%% file: main.tex
\documentclass[12pt,reqno]{amsart}

\usepackage[left=1in, right=1in, top=1.1in,bottom=1.1in]{geometry}
\usepackage{array}

\usepackage[title]{appendix}
\usepackage{booktabs}

\usepackage{graphicx}
\graphicspath{ {./fig/}}

\usepackage[T1]{fontenc}
\usepackage{lmodern}
\usepackage{amsthm,amssymb,amsmath,amsfonts,mathrsfs,amscd}
\usepackage{bm}
\usepackage{enumitem}
\usepackage{stmaryrd}
\usepackage{multirow}

\usepackage[font={scriptsize}]{caption}

\usepackage{hyperref}
\hypersetup{
  colorlinks   = true,
  citecolor    = green,
  linkcolor    = red
}

\theoremstyle{remark}

\theoremstyle{remark}

\DeclareMathOperator{\var}{var}

\newcommand{\cX}{\mathcal{X}}
\newcommand{\cW}{\mathcal{W}}
\newcommand{\E}{\mathbb{E}}

\newcommand{\bu}{\bm{u}}

\newcommand{\argmin}{\operatornamewithlimits{argmin}}

\title{Dynamics of Liquidity Surfaces in Uniswap v3}
\author[J. Risk]{Jimmy Risk}
\author[S.-N. Tung]{Shen-Ning Tung}
\author[T.-H. Wang]{Tai-Ho Wang}
\date{\today}

\address{Jimmy Risk \newline
Department of Mathematics and Statistics, \newline
Cal Poly Pomona\newline
3801 W Temple Ave, Pomona CA 91768
}
\email{jrisk@cpp.edu}

\address{Shen-Ning Tung \newline
Department of Mathematics, \newline
National Tsing Hua University \newline
Hsinchu, Taiwan
}
\email{tung@math.nthu.edu.tw}

\address{Tai-Ho Wang \newline
Department of Mathematics \newline
Baruch College, The City University of New York \newline
1 Bernard Baruch Way, New York, NY10010
}
\email{tai-ho.wang@baruch.cuny.edu}

\keywords{Automatic market making, Decentralized exchange, Decentralized finance}

\begin{document}

\begin{abstract}
This paper presents a comprehensive study on the empirical dynamics of Uniswap v3 liquidity, which we model as a time-tick surface, $L_t(x)$. Using a combination of functional principal component analysis (FPCA) and dynamic factor methods, we analyze three distinct pools over multiple sample periods. Our findings offer three main contributions: a statistical characterization of automated market maker liquidity, an interpretable and portable basis for dimension reduction, and a robust analysis of liquidity dynamics using rolling window metrics. For the 5 bps pools, the leading empirical eigenfunctions explain the majority of cross-tick variation and remain stable, aligning closely with a low-order Legendre polynomial basis. This alignment provides a parsimonious and interpretable structure, similar to the dynamic Nelson-Siegel method for yield curves. The factor coefficients exhibit a time series structure well-captured by AR(1) models with clear GARCH-type heteroskedasticity and heavy-tailed innovations.

\end{abstract}

\maketitle

\input{Intro}
\input{Background}

\input{Data_Methodology}

\input{Analysis}

\input{Conclusion}

\bibliographystyle{alpha}
\bibliography{Ref}

\input{Appendix}

\end{document}

%% file: Intro.tex

\section{Introduction}
Decentralized finance (DeFi) \cite{Harvey2021defi, Gobet2023DecentralizedFB} is rapidly transforming financial markets, with automated market makers (AMMs) playing a crucial role in this evolution. As a revolutionary force, decentralized exchanges (DEXs) \cite{Capponi2021AdoptionDB} have emerged, attracting millions of users to DeFi projects and offering a way to bypass the constraints of traditional financial systems.

Introduced by Uniswap Labs, Uniswap marked a pivotal shift within the DEX landscape by replacing traditional order-book mechanisms with its innovative AMM design. Version 1 enabled censorship-resistant spot trading, allowing swaps between Ether (ETH), Ethereum's native asset, and any ERC-20 token (e.g., USDC, WBTC). Version 2 \cite{Adams2020UniswapV2} generalized to any two ERC-20 tokens and added an on-chain oracle (a time-weighted average price). Version 3 \cite{Adams2021UniswapV3} introduced concentrated liquidity, allowing liquidity providers (LPs) to allocate capital only within a chosen price interval. Fees accrue from swaps executed when the current price lies within the LP's defined range.

Liquidity is a fundamental aspect of efficient markets, measuring the ease with which an asset can be exchanged without appreciable price impact \cite{Bouchaud2018Microscope}. In centralized venues, liquidity is extensively studied through limit-order books (LOBs); queueing and stochastic partial differential equation (SPDE) models \cite{ContStoikov2010LOB, Huang2015QRM, ContMuller2021SPDE} effectively capture depth dynamics and resilience. By contrast, empirical work on AMM liquidity has so far centered on static range-allocation or fee-tier optimization.  For example, Heimbach, Schertenleib, and Wattenhofer document that realized LP pay-offs hinge on tick-width and position lifetime \cite{heimbach2022risks}.  Fan et al.~\cite{Fan2021Strategic} formalize the dynamic liquidity provision problem and focus on a general class of strategies to maximize LP earnings, and follow-up work provides a convex stochastic optimization problem that quantifies the optimal interval width for a belief-driven LP, providing implications for contract design \cite{Fan2022Differential}.  Related is work on optimal adaptive allocation through deep reinforcement learning \cite{zhang2023adaptive}, equilibrium liquidity distribution via mean-field games \cite{bayraktar2024dex}, and predictable-loss-aware optimal strategies derived from dynamic continuous time models \cite{cartea2024decentralized}.  

On the more statistical side, panel regressions connect liquidity spreads and depth to gas costs, volatility, and fee revenue \cite{zhu2024drives}; researchers use linear models to forecast lookahead volume \cite{miori2023defi}; and statistical tests assess price discovery (across pools) and market efficiency \cite{alexander2025price}.  Complementary work measures impermanent loss and LP risk \cite{loesch2021impermanent} and clusters trader archetypes via graph embeddings \cite{miori2024clustering}.  

In summary, liquidity plays a crucial role in understanding larger aspects of DeFi.  A better understanding of the liquidity surface $L_t(x)$ aids comprehensive risk assessment, capital efficiency, and the development of sophisticated trading strategies.  As mentioned, Uniswap v3 specifically generates a two-dimensional \emph{liquidity surface} \(L_t(x)\) indexed by block-time \(t\) and tick distance \(x\) that represents the quantized price range within the Uniswap v3 architecture. In other words, each trading pair has its own temporal landscape, enabling empirical liquidity analysis, inference, and modeling.  Surprisingly, we find that a thorough statistical description of the $L_t(x)$ as a dynamic liquidity surface in $(t,x)$ is still missing.  This work is intended to fill that gap.

In particular, we apply methods in dynamic factor models and functional principal component analysis (FPCA) to obtain a low-rank factor representation.  This has been applied to find low-rank structures in other dynamic functional financial objects, such as yield curves \cite{Litterman1991CommonFactors, Christensen2011Affine, Oprea2022UsePCA}, implied volatility surfaces \cite{Avellaneda2020PCA}, and forward curves \cite{Shen2009ModelingForecasting, Hays2012FunctionalDFM}.  

To motivate this approach, consider the well-known Nelson–Siegel (NS) factor model as introduced by Charles Nelson and Andrew Siegel (1987) \cite{nelson1987parsimonious}.  This was used as a parsimonious parametric framework to fit the term structure of interest rates (yield curves). The NS functional form uses three factors that are combined to approximate the cross-sectional shape of yields across maturities in a flexible but low-dimensional way \cite{Diebold2008Global}:
\begin{equation}\label{eq:NS-equation}
    \hat{y}_t(\tau) = \beta_{1,t} \cdot 1 + \beta_{2,t} \cdot \frac{1-e^{-\lambda \tau}}{\lambda \tau} + \beta_{3,t} \Big(\frac{1-e^{-\lambda \tau}}{\lambda \tau} - e^{-\lambda \tau}\Big),
\end{equation}
where where $\hat{y}_t(\tau)$ is the model-implied yield at time $t$ and maturity $\tau$, and $\beta_{1,t}$, $\beta_{2,t}$, $\beta_{3,t}$ are time-varying coefficients (factors) while $\lambda$ is a decay parameter that governs the shape. While the original model \cite{nelson1987parsimonious} considered fixed coefficients (typically re-estimated), this \emph{dynamic factor model} was a later extension by Diebold and Li (2006) \cite{DieboldLi2006Forecasting} to incorporate time-varying factors. This extends the cross-sectional specification to time-varying factors with their own dynamics.  Despite its simplicity, this Dynamic Nelson–Siegel (DNS) model provides an excellent fit to the term structure at each point in time and often yields forecasts that outperform more theoretically complex models in out-of-sample interest rate forecasting. By projecting the entire yield curve onto three persistent factors and then modelling those factors with low-order autoregressions, Diebold and Li achieved a practical and accurate forecasting method for yields \cite{DieboldLi2006Forecasting}.  Christensen et al.~(2011) later connect with existing no-arbitrage models for yield curves to add a no-arbitrage link to DNS without sacrificing parsimony \cite{Christensen2011Affine}.

One of the remarkable aspects of the Nelson–Siegel basis is that its three latent factors naturally align with the principal components (PCs) observed in empirical yield curve data \cite{DieboldLi2006Forecasting}. Most of the variation in the yield curve can be explained by these three orthogonal factors, colloquially known as level, slope, and curvature, which persistently match the factors in the NS model. Specifically, a principal component analysis on a panel of interest rates across maturities reveals the first PC to be a \textit{level shift} affecting all maturities, the second PC a \textit{slope factor} that tilts the curve by raising short rates and lowering long rates or vice versa, and the third PC a \textit{curvature factor} that bends the curve (hump-shaped, affecting medium maturities more than short or long ones).  This aligns with findings by Litterman and Scheinkman (1991) \cite{Litterman1991CommonFactors}, who, through eigendecomposition of the covariance matrix of zero-return vectors, showed that the first factor (level) explains 89-90\% of variance and is nearly constant across maturities; the second factor (slope/twist) explains about 8\% and exhibits a monotone sign change; and the third factor (curvature/butterfly) explains 1-2\% and affects medium maturities more. This low-rank structure is also found in other financial surfaces. For instance, Cont and da Fonseca (2002) \cite{ContDaFonseca2002Dynamics} observed a similar ``level-slope-curvature'' for implied volatility surfaces as functions of moneyness and time to maturity.

\subsection{Contributions}
In this study, we focus on the empirical dynamics of the liquidity surface $(t,x) \mapsto L_t(x)$ to answer three core questions: What is the statistical nature of these fluctuations? Can they be quantified? Can they be modelled in a parsimonious way? Our investigation begins nonparametrically, finding a stable low-rank structure when PCA is performed over various time windows. We also find that the nonparametric eigenmode basis consistently aligns well with the Legendre polynomial basis, providing an initial attempt at a persistent structure akin to the NS decomposition \eqref{eq:NS-equation}. By leveraging both decompositions, we identify explicit and persistent stochastic structures in the temporal components.

Our specific contributions are as follows:
\begin{itemize}
    \item We provide a preliminary view of the surface $(t,x) \mapsto L_t(x)$ for three key liquidity pairs (Ethereum ETH-USDC 5bps (basis points), Ethereum ETH-USDC 30bps, and Arbitrum ARB-USDC 5bps) across three distinct time periods, accompanied by a high-level statistical analysis.  
    \item We demonstrate that the infinite-dimensional dynamics (in $x$) can be effectively analyzed through a lower-dimensional principal component projection, yielding temporal coefficients and eigenmodes over tick distance.
    \item We identify persistent autoregressive [order 1] (AR(1))–generalized autoregressive conditional heteroskedasticity (GARCH) time series structures, with heavy-tailed innovations for the temporal coefficients.
    \item We show that the Ethereum ETH-USDC 5bps and Arbitrum ARB-USDC 5bps pools maintain a stable eigenmode structure that closely aligns with the Legendre polynomial basis, a conclusion we quantify using subspace distance metrics.
    \item We use the Legendre polynomials as an interpretable tool for the liquidity surface and illustrate the effect of its orthogonal components through volatility shocks.
    \item We verify the robustness of our statistical findings by replicating the analysis with various preprocessing schemes.
\end{itemize}

The rest of this work is outlined as follows: Section \ref{sec:background} covers the background, with \ref{sec:background-uniswap} detailing the Uniswap v3 framework and the meaning of a liquidity surface, and \ref{sec:background-statistical} covering the statistical methods used. Section \ref{sec:data} provides details on the datasets and preprocessing used. The follow-up Section \ref{sec:empirical-analysis} performs the statistical analysis, including PCA and Legendre basis decompositions, assessing stability, persistent stochastic structures, and the effects of shocks. Lastly, Section \ref{sec:discussion-conclusion} finalizes the discussion and provides avenues for future work.

%% file: Background.tex
\section{Background}\label{sec:background}

\subsection{Uniswap v3 Liquidity Provision}\label{sec:background-uniswap}
Uniswap v3 introduced a revolutionary approach to decentralized liquidity provision, allowing liquidity providers (LPs) to concentrate their capital within specific price ranges. This mechanism fundamentally redefines the relationship between an LP's token holdings and their contribution to the market, moving beyond the uniform liquidity distribution of earlier Automated Market Maker (AMM) versions.

\subsubsection{Concentrated Liquidity in Detail} \label{sec:CL}
\begin{figure}[!ht]
\centering
    \includegraphics[width=0.9\linewidth]{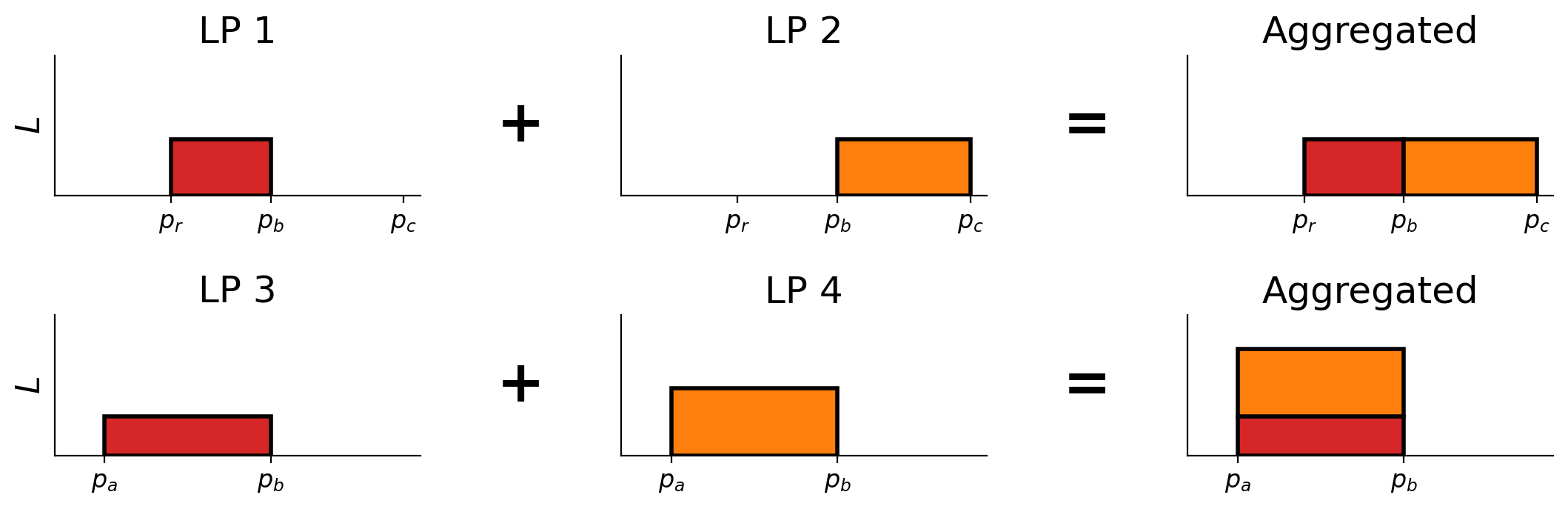}
    \caption{Aggregated Liquidity}
    \label{fig:additivity}
\end{figure}

Concentrated liquidity is the cornerstone of Uniswap v3's capital efficiency. Unlike previous iterations, where liquidity was spread evenly across the entire price curve, Uniswap v3 empowers LPs to strategically allocate capital within a user-defined price range $[p_l, p_r)$. This innovative feature hinges on several interconnected concepts:

\begin{enumerate}
    \item \textbf{Liquidity Level $L$:} This parameter quantifies the amount of liquidity an LP provides within their chosen range. Conceptually, it represents the virtual reserves contributed by the LP, analogous to the constant-product formula, $L = \sqrt{xy}$, used in Uniswap v2. An LP's liquidity level directly determines its share of trading fees and its exposure to price movements as long as the market price remains within its defined range.
    \item \textbf{Liquidity Provision Range $[p_l, p_r)$:} This is the specific price interval an LP selects for capital deployment. When the current market price $p$, as reported by the AMM, resides within this designated range, the LP's assets become active for swaps, following a predetermined bonding curve \cite{Adams2021UniswapV3}. The precise selection of this range is paramount, as it dictates the periods during which the LP's capital actively participates in facilitating trades.
    \item \textbf{Aggregated Liquidity:}
    Uniswap v3 achieves its distinctive \textit{aggregated} liquidity profile by seamlessly combining contributions from numerous individual LPs, each providing liquidity within their own defined price ranges (see Figure \ref{fig:additivity}). Denote the $i$-th LP's position by its liquidity level $L_i$ and its chosen price range $[p_{\ell,i}, p_{r,i})$. An individual LP's liquidity $L_i$ is only active when the current market price $p$ falls within their specified range. The total liquidity, $L_{\text{total}}(p)$, available in the Uniswap v3 pool at any given price $p$ is then the sum of the liquidity levels of all individual LP positions whose active ranges encompass that price:
    $$
    L_{\text{total}}(p) = \sum_{i=1}^N L_i 1_{[p_{\ell,i}, p_{r,i})}(p).
    $$
    This summation illustrates how the distinct liquidity contributions from individual LPs are effectively combined at each price point. This process creates a potentially non-uniform liquidity distribution across the entire price spectrum, directly influenced by where LPs choose to concentrate their capital \cite{TungWang2024CLMM}. Consequently, an LP's token holdings and fee earnings are dynamically determined by its individual liquidity level $L_i$ and how the market price $p$ interacts with its active range $[p_{\ell,i}, p_{r,i})$.
\end{enumerate}

\subsubsection{Comparison with Traditional Limit Order Books}
Uniswap v3 represents a significant evolution in AMM design, effectively bridging the gap between constant function market makers (CFMMs) \cite{AngerisChitra2020ImprovedPO} and traditional electronic limit order books (LOBs). While CFMMs like Uniswap v2 offer simplicity, they lack the expressiveness of LOBs, which maintain a list of outstanding buy and sell orders at specified prices \cite{Gould2013LimitOB}. Uniswap v3's concentrated liquidity feature allows liquidity provider (LP) positions to function akin to limit orders in LOBs, thereby approximating more complex demand curves without the full state complexity of a traditional order book \cite{MilionisM2023ComplexityApproximation}.

Despite sharing the goal of enabling trades at preferred price levels, Uniswap v3's narrow-range LP positions and traditional LOB limit orders exhibit fundamental differences in their underlying mechanisms and operational characteristics. These distinctions are detailed below:

\begin{enumerate}
    \item \textbf{Price Specification:} A primary distinction lies in how the price is specified. In Uniswap v3, LPs define a price \textit{range}, providing continuous liquidity across that interval. Conversely, LOBs rely on specific, \textit{discrete price points} for each individual order.
    \item \textbf{Liquidity Provision Mechanism:} Liquidity provision differs fundamentally. Within a defined range, Uniswap v3 liquidity is continuous and formula-driven, always available according to the bonding curve until fully utilized. In contrast, LOB liquidity is offered in discrete, fixed quantities at specific price levels.
    \item \textbf{Operational Mechanisms and Trade Execution:} Uniswap v3 operates as an AMM, utilizing a constant-product formula (adjusted for the specified price range) to facilitate swaps. Trades execute instantly against the available liquidity within an LP's range until that liquidity is depleted. In contrast, LOBs function by matching submitted buy and sell orders, executed through a greedy matching process at the specified price or better.
    \item \textbf{Behavior with Market Price Movements:} The behavior of positions under market price fluctuations also diverges. If the market price moves outside of a Uniswap v3 LP's defined range, the position becomes inactive, holding its current asset composition until the price re-enters the range. This often necessitates active management and potential position withdrawal by the LP to realize intended gains or avoid impermanent loss. In LOBs, limit orders remain in the book until they are either filled or explicitly cancelled by the user, irrespective of broader market price movements.
\end{enumerate}

\subsection{Statistical Methods}\label{sec:background-statistical}
Suppose for $x \in \mathcal{X} = [-1,1]$ and $t \geq 0$ we observe $\{y_t(x)\}$.  To connect with the previous discussion, $y_t(x) = \log L_t(x)$, and tick $x$ has been standardized to $[-1,1]$ (see Section \ref{sec:preprocessing} for details).   
The idea is to decompose 
\begin{equation}\label{eq:decomposition}
    y_t(x) = m(x) + \sum_{k=1}^K \beta_{t,k} u_k(x) + r^{(K)}_t(x)
\end{equation}
into $K$ \emph{factors} with \emph{basis} (or loadings) $\{u_k(x)\}_{k=1}^K$ and \emph{coefficients} (or factor scores) $\{\beta_{t,k}\}_{k=1}^K$, and $r^{(K)}_t(x)$ is the \emph{residual truncation error}. Through this decoupling, one can separately analyze the temporal dynamics through the time series $(\beta_{t,k})_{t \geq 0}$ and spatial structure through the $(u_k(x))_{x \in \mathcal{X}}$. Loosely, the goals are
\begin{itemize}
    \item $K$ small enough and an interpretable structure on the $\{u_k\}_{k=1}^K$ so that the model remains parsimonious and interpretable;
    \item some persistence so that certain properties (e.g., stable basis, distributions of coefficient processes) are stable across time windows;
    \item $K$ large enough for an appropriate basis $\{u_k\}_{k=1}^K$ so that $r^{(K)}_t(x)$ is (approximately) idiosyncratic.
\end{itemize}
It is important to note that the first and last goals are often difficult to achieve simultaneously.

\subsubsection{Dynamic Factor Extraction}\label{sec:dynamic-factor-extraction}
In practice, we observe a data matrix $Y$ with entries $\{y_t(x_m)\}$ for $t = 1, \ldots, T$ and $m = 1, \ldots M$. We assume the $x_m$ are equally spaced. Each row of $Y$ corresponds to the observed time-$t$ curve, $\mathbf{y}_t = \big[y_t(x_1), \ldots, y_t(x_M)\big]$. Given the lack of prior information about the statistical and functional dynamics of $y_t(x)$, a nonparametric approach such as principal component analysis (PCA) is appropriate. For this reason, we assume the data are centered unless otherwise noted; otherwise, replace $\mathbf{y}_t$ with $\mathbf{y}_t - \overline{\mathbf{y}}$ where $\overline{\mathbf{y}} = \frac{1}{T} \sum_{t=1}^T \mathbf{y}_t$.

The following approach applies dynamic factor models \cite{stock2011dynamic}, which are both conceptually and mathematically similar to functional principal component analysis (FPCA) \cite{ramsay2005functional}. The former is often employed in econometric settings where there is no necessary structure on the $M$ coordinates.  The latter considers all objects in $L^2(\cX)$ rather than $\mathbb{R}^M$, thus imposing structure in the $x$ coordinate.  Our equally-spaced and dense setup recovers the same subspace as FPCA in the limit; see Section \ref{sec:basis-discussion-appendix} for details. Note that despite having an enticing name, \emph{dynamic FPCA} as popularized by Hormann et al.~\cite{hormann2015dynamic} is not appropriate as it replaces \eqref{eq:decomposition} with a non-factorable double sum (one infinite), improving theoretical properties but hindering the purpose of parsimonious interpretability. 

Denote the sample $M \times M$ covariance matrix as
\begin{equation}\label{eq:PC-decomp}
    \hat{\Sigma} := \frac{1}{T} Y^\top Y = U \Lambda U^\top,
\end{equation}
where $U \Lambda U^\top$ is its eigendecomposition (principal component) form.  Here, $\Lambda = \text{diag}(\lambda_1, \ldots, \lambda_M)$ are the eigenvalues of $\hat{\Sigma}$ and $U$ has orthonormal columns consisting of $\hat{\Sigma}$'s orthonormal eigenvectors $\bu_1, \ldots, \bu_M$ with $\bm{u}_k = [u_k(x_1), \ldots, u_k(x_M)]^\top$.  For simplicity, assume $T > M$ and full rank, so that the eigenvalues can be ordered $\lambda_1 > \lambda_2 > \cdots > \lambda_M > 0$.

For a fixed $K \leq M$, PCA yields the optimal rank-$K$ least-squares reconstruction: 
\begin{equation}\label{eq:pca-optimal}
\{\beta_{t,k}\}, U_K = \argmin_{\{\tilde{\beta}_{t,k}\}, \tilde{U}_K} \sum_{t=1}^T \sum_{m=1}^M \Big(y_{t}(x_m) - \sum_{k=1}^K \tilde{\beta}_{t,k} \tilde{u}_k(x_m)\Big)^2,
\end{equation}
where $\beta_{t,k} = \langle \mathbf{y}_t, \mathbf{u}_k\rangle$ under the Euclidean inner product on $\mathbb{R}^M$.  This solution is also unique (up to sign) if we assume ordered eigenvalues and orthonormality on the $U_K$.  If $\mathbf{y}_t$ is weakly stationary with $\var(\mathbf{y}_t) = \Sigma = U \Lambda U^\top$, then the same holds true at the population level:
\begin{equation}
    \{\beta_{t,k}\}, U_K = \argmin_{\{\tilde{\beta}_{t,k}\}, \tilde{U}_K} \E\Big\|\mathbf{y}_t - \sum_{k=1}^K \tilde{\beta}_{t,k} \tilde{\mathbf{u}}_k\Big\|^2.
\end{equation}
In summary, the optimal rank-$K$ representation in \eqref{eq:decomposition} is the one produced by the covariance (according to squared error, and under expectation in the population setting).  

\emph{Remark}: If $\{y_t(x)\}$ is weakly stationary, $\hat{\Sigma}$ is a consistent estimator of the population covariance. If it is not, the decomposition and optimality of \eqref{eq:PC-decomp} and \eqref{eq:pca-optimal} still hold, but the interpretations change (see Section \ref{sec:preprocessing}).

We shift the analysis to the truncated 
\begin{equation}\label{eq:y-truncated}
    y_t^{(K)}(x_m) = \sum_{k=1}^K \beta_{t,k} u_k(x_m), \qquad m = 1, \ldots, M
\end{equation}
so that $t$ and $x$ have been decoupled.  The interpretation is that each $k$th \emph{factor} consists of the pair $\big(\beta_{t,k}, u_k(x)\big)$, and is one piece that contributes orthogonally to $y_t^{(K)}(x)$, in order of decreasing variance explained.  Often one chooses $K$ via the cumulative proportion of variance explained, where the (cumulative) proportion of variance explained by factor $k$ is given by
\begin{equation}\label{eq:pve}
    \text{PVE}_j = \frac{\lambda_j}{\sum_{j=1}^M \lambda_j}, \qquad \text{CPVE}_K = \sum_{j=1}^K \text{PVE}_j,
\end{equation}
so that $\text{CPVE}_M=1$. If one finds that $K$ is suitable to achieve a desired threshold (say, 95\%), these $K$ components are individually analyzed. The $u_k(x)$ offer insight into a static structure of $y_t(x)$ in the $x$ coordinate.  Under weak stationarity, these are time-invariant (asymptotically in the sample case). The scores $\{\beta_{t,k}\}$ capture temporal dynamics (e.g. AR(1) for $k$ or vector AR(1) (VAR(1)) across $k$). Statistical properties of the $\beta_{t,k}$ transfer to the associated structural effect $u_k(x)$ when reconstructing $y_t(x)$. Thus, the \emph{stochastic properties of the $k$th factor}, such as mean reversion, heavy tails, or heteroskedasticity, can be analyzed through its corresponding $\beta_{t,k}$.

While PCA provides a data-driven basis, we also explore the alignment of this basis with a fixed, interpretable basis, such as the Legendre polynomials. Details on these comparisons and the methodology for evaluating off-grid points are provided in Appendix \ref{sec:basis-discussion-appendix}.

\subsubsection{Forecasting}  One obtains forecasts naturally by forecasting the $\beta_{t,k}$ through traditional time series methods (e.g.~$\text{AR}(1)$ or $\text{VAR}(1)$ in discrete time).  Let $\mathcal{F}_t = \sigma\big(\{y_s(x_i)\}_{i=1}^M, s \leq t\big)$ be the \emph{filtration} containing all relevant information up to time $t$ and assume the loadings $u_k(x) \in \mathcal{F}_t$, either estimated and fixed or by using a fixed basis. Define
\[
\hat\beta_{t+h\mid t,k}:=\E[\beta_{t+h,k}\mid\mathcal F_t],\qquad
\hat y^{(K)}_{t+h\mid t}(x):=\E\!\big[y^{(K)}_{t+h}(x)\mid\mathcal F_t\big].
\]
Then by linearity
\[
\hat y^{(K)}_{t+h\mid t}(x)=\sum_{k=1}^K \hat\beta_{t+h\mid t,k}\,u_k(x).
\]
Further, $\hat{y}_{t+h\mid t}(x) = \hat{y}^{(K)}_{t+h\mid t}(x)$ if we assume the remainder satisfies  $\E[r_{t+h}^{(K)}(x) | \mathcal{F}_t]=0$.  Consequently, prediction intervals and other distributional quantities can be obtained.  Note that basis drift affects optimal forecasts, but it is usually mild for short horizons.  In practice, the basis is periodically updated.

%% file: Data_Methodology.tex
\section{Data and Methodology}\label{sec:data}
This section details the specific Uniswap v3 liquidity data used in this study, the necessary preprocessing steps, and how the data's characteristics inform our modelling approach. Our primary focus is on understanding liquidity behavior relative to the current market price through in-sample examination.

\subsection{Uniswap v3 Liquidity Data Characteristics}
Our analysis begins with raw liquidity data sourced from an internal API provided by Teahouse Finance, a decentralized finance (DeFi) asset-management platform specializing in Uniswap. The goal is to understand the aggregated liquidity dynamics over time and the discretized price space.  Seminal work in Uniswap v3 provides some preliminary stylized facts pertinent to our modelling approach \cite{Fan2021Strategic, Fan2022Differential}:
\begin{itemize}
    \item A large fraction of the liquidity mass lies near the current price, meaning we typically observe $L_t(x)$ to be relatively large near $x=0$. Here, $x$ represents the normalized logarithmic pool price, with $x=0$ corresponding to the current pool price. Note that there may still be peaks further out.
    \item The shape of the liquidity surface can vary significantly depending on LP risk preferences and chosen price interval configurations. 
    \item Market volatility significantly influences LP behavior. In high-volatility regimes, LPs tend to widen their liquidity ranges, which shifts liquidity mass outwards and consequently lowers the central peak. Conversely, periods of low volatility are often characterized by a tighter, steeper concentration of liquidity around the mid-price.    
\end{itemize}

One goal is to verify these to the best of our ability and explain them through the dynamic factor lens.  Throughout, we add additional stylized facts that arise through the statistical analysis.

\subsubsection{Datasets Utilized}
Our analysis incorporates data from a variety of Uniswap v3 pools to assess commonalities, generalizability, and improve robustness of takeaways:
\begin{itemize}
    \item \textbf{Ethereum Mainnet 5bps ETH-USDC pool:} This pool serves as our primary focus due to its consistently high trading volume and deep liquidity, providing a rich dataset for in-depth analysis.
    \item \textbf{Ethereum Mainnet 30bps ETH-USDC pool:} Included to investigate how different fee tiers (and thus different `tickSpacing` values) impact liquidity distribution and dynamics, allowing for a comparison against the 5bps pool.
    \item \textbf{Arbitrum 5bps ARB-USDC pool:} Included as a crucial robustness check to assess if observed features generalize across distinct blockchain environments, specifically Arbitrum (an Ethereum Layer-2 scaling solution), which has different block times and transaction costs.
\end{itemize}
The times and block numbers are displayed in Table \ref{tab:datasets}.  Block spacing was chosen as 115200 for ARB5 and 2400 for ETH to obtain an (approximate) 8-hour time spacing.

\begin{table}[!ht]
    \centering
    \begin{tabular}{l|l|c|ccc|c}
        \multicolumn{1}{c|}{Dataset} & \multicolumn{1}{c|}{Pool} & \multicolumn{1}{c|}{bps} & \multicolumn{3}{c|}{Time Range} & \multicolumn{1}{c}{Spacing} \\
         &  &  && Start & End &  \\ \hline
        \multirow{2}{*}{ARB5} & \multirow{2}{*}{ARB--USDC} & \multirow{2}{*}{5}
            & Date--time: & Aug 23, 2023 & Nov 26, 2024 & 8.39 h \\
         &  & 
            & Block Number: &\texttt{124163200} & \texttt{278646400} & \texttt{115200} \\ \hline
        \multirow{2}{*}{ETH5, ETH30} & \multirow{2}{*}{ETH--USDC} & \multirow{2}{*}{5, 30}
            & Date-time: & Aug 1, 2021 & Nov 26, 2024 & 8.26 h \\
         &  &
            & Block Number: &\texttt{12940529} & \texttt{21274529} & \texttt{2400} \\
    \end{tabular}
    \caption{Full dataset ranges. Dates are rounded to calendar days (UTC). The Spacing column reports time spacing (hours) on the date row and block step on the block row. ETH5 and ETH30 share the same blocks and times.  Spacing in hours is computed as the average hourly time between the clock times at the exact block-number intervals.}
    \label{tab:datasets}
\end{table}

\subsection{Applying Dynamic Factor / Functional PCA Models}
We give details on our implementation of Uniswap v3 liquidity into the framework in Section \ref{sec:background-statistical}.  Time $t$ is \emph{block number}, a discrete index advanced by consensus when a new block is appended to the chain. Block numbers provide a consistent, immutable, and sequential temporal proxy for ordering events. In general, $t$ is stochastically related to its corresponding clock time $c(t)$.  For example, Ethereum is partitioned into slots of 12 seconds, and at most one block can be proposed per slot.  One typically observes $c(t) - c(t-1)$ to be approximately 12 seconds, but it can sometimes be greater due to skipped or empty blocks and other factors \cite{kapengut2023event}. Arbitrum is slightly different and has an approximate gap of 0.25 seconds per block \cite{kalodner2018arbitrum}. Thus, the liquidity surface can be indexed by either the discrete block time $t$ or clock-time $c(t)$. When there is no ambiguity, we may interchange $t$ or $c(t)$ for clarity. 

Next, Uniswap v3's price space is discretized into \textit{ticks}. The price at an integer tick index $i$ is defined by $P(i) = 1.0001^i$, meaning each tick represents a 0.01\% (1 basis point) price movement. Liquidity ranges for LPs are set between specific \emph{tick-spacing multiples} $s$. The parameter $s$ varies by pool fee tier (e.g., a 0.05\% fee tier uses $s=10$) and directly determines the minimum width and granularity of liquidity provision ranges. The total liquidity available at any given price $p$ is the sum of individual LP contributions whose active ranges encompass that price, as previously defined in Section \ref{sec:CL}. This aggregation results in the non-uniform liquidity distribution we aim to model.  

To analyze liquidity dynamics relative to the current market price, we employ a relative tick coordinate system, analogous to how liquidity in LOBs is often described relative to the best bid/ask \cite{ContMuller2021SPDE}. Essentially, this normalizes the current pool price tick to 0, which aids in capturing the functional form of liquidity, where behavior is assumed to be more consistently described when referenced to the current AMM price rather than absolute price levels. Given the vast range of possible Uniswap v3 ticks, this transformation simplifies analysis by focusing on relative behavior. For background on price variations in financial markets, see \cite{cont2001empirical}.

To simplify our modelling approach, we analyze the liquidity at a fixed number of relative ticks around the current market price (See Appendix \ref{sec:lp-surface-appendix} for details). This method allows us to focus on the shape of the liquidity surface close to the price, effectively transforming a variable-length dataset into a fixed-length functional form suitable for analysis.

\subsubsection{Further Preprocessing}\label{sec:preprocessing}

We make a few miscellaneous comments about preprocessing in the subsequent analysis.

\begin{itemize}
    \item We work with the logarithmically transformed $y_t(x) = \log L_t(x)$ as it matches with how $x$ is related to the log price (since $\log P(i) = i \log 1.0001$), and also helps stabilize the associated probability distribution.  Furthermore, this is common practice in the analysis of financial curves to ensure positivity.  
    \item We work with the undifferenced log-liquidity data.  This is common in situations where interpretable factors are paramount.  For example, the dynamic Nelson-Siegel framework for yield curves is applied to undifferenced data to directly capture the level, slope, and curvature of the term structure \cite{DieboldLi2006Forecasting}, despite unit root behavior in many samples. Here, the focus is on providing an interpretation of the inherent shape and evolution of the liquidity surface rather than time differences, obtaining an interpretable and parsimonious decomposition.  Evidence and implications of nonstationarity are discussed throughout the analysis.
\end{itemize}



%% file: Analysis.tex
\section{Empirical Analysis}\label{sec:empirical-analysis}
This section presents the empirical analysis of Uniswap v3 liquidity surfaces. The analysis is broken into two main types.  The first uses the full data in Table \ref{tab:datasets} and works over overlapping rolling windows to assess how certain phenomena persist in time.  The second focuses on three specific windows with a more in-depth statistical analysis and interpretation in each, comparing across windows and datasets.  The windows are described in Table \ref{tab:dataset-windows}.  These windows serve to localize the principal component decomposition to illustrate how within-window results provide valuable insights and are aligned across datasets (up to block-number variability).

\begin{table}[!ht]
    \centering
    \begin{tabular}{c|l|rr|rr}
         & & \multicolumn{2}{c|}{Arbitrum} & \multicolumn{2}{c}{Ethereum Mainnet} \\
        Window & & \multicolumn{2}{c|}{ARB--USDC} & \multicolumn{2}{c}{ETH--USDC} \\
        & & \textbf{Start} & \textbf{End} & \textbf{Start} & \textbf{End} \\ \hline
        \multirow{2}{*}{1} & Date & Sep 1, 2023 & Jan 22, 2024 & Sep 1, 2023 & Jan 13, 2024 \\
                           & Block      & \texttt{126928000} & \texttt{173008000} & \texttt{18038129} & \texttt{18998129} \\ \hline
        \multirow{2}{*}{2} & Date & Feb 1, 2024 & Jun 16, 2024 & Feb 1, 2024 & Jun 14, 2024 \\
                           & Block      & \texttt{176233600} & \texttt{222313600} & \texttt{19130129} & \texttt{20090129} \\ \hline
        \multirow{2}{*}{3} & Date & Jul 1, 2024 & Nov 12, 2024 & Jul 1, 2024 & Nov 12, 2024 \\
                           & Block      & \texttt{227497600} & \texttt{273577600} & \texttt{20210129} & \texttt{21170129} \\
    \end{tabular}
    \caption{Training windows for each dataset. Each window contains $T=400$ observations spaced at approximately 8-hour intervals. Dates are rounded to calendar days (UTC). Ethereum 5bps and 30bps share the same windows and blocks.}
    \label{tab:dataset-windows}
\end{table}

\paragraph{Hyperparameter choices.} The main analysis utilizes $M=201$ ticks, so $x=0$ is the current price and $x \in \{-1.0, \ldots, -0.01\}$ and $\{0.01, \ldots, 1.0\}$ are the 100 (log) prices on either side.  Additionally, the data are provided using \texttt{blockNumber} as the time unit. We use a block-number spacing so that $y_t(x)$ and $y_{t-1}(x)$ are approximately 8 hours apart; this corresponds to 115200 blocks for ARB5 and 2400 for ETH.  This analogously matches a typical 8-hour daily trading window in centralized exchanges.  We initially tried a 2-hour spacing, but found that movement in the liquidity surface was too sparse to model reliably.  Note that liquidity data are updated continuously, so there are no breaks during nighttime or during holidays.  Additionally, whenever PCA is performed, it uses $T=400$ rows of data (approximately 4.5 months), which provides enough data for analysis but not too long a window to over-generalize.  

Appendix \ref{sec:alt-preprocessing} performs analysis of alternative choices of $M$ and also $T=800$.  Comparative details are discussed in the main text when relevant.

\subsection{Initial Inspection of Log-Liquidity Surface}

\begin{figure}[ht]
    \centering
    \includegraphics[width=0.9\linewidth]{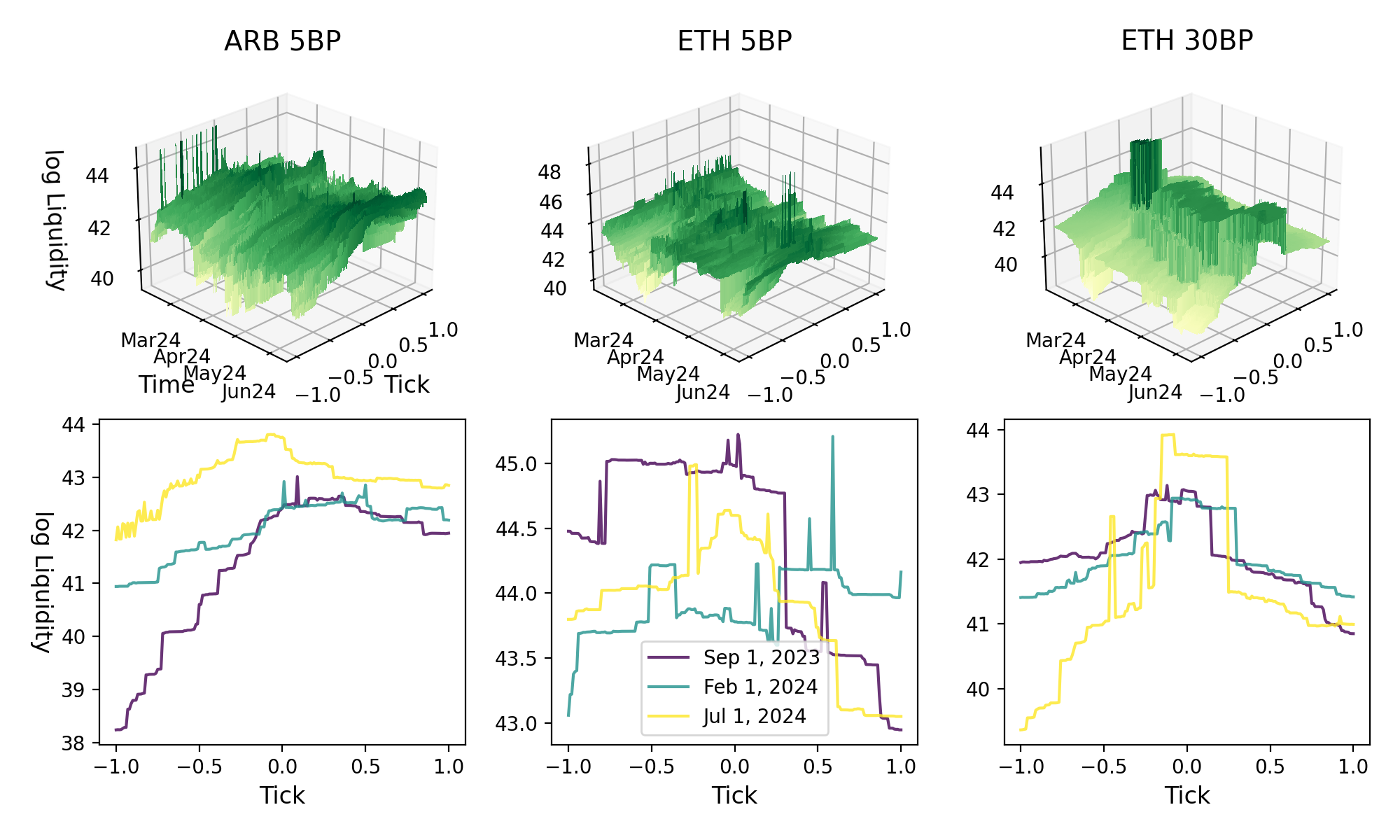}
    \caption{Top panel: raw log-liquidity surfaces $y_t(x)$ over time $t$ and relative tick $x$ according to Window 2. Bottom: cross-sectional plots of $y_{t_0}(x)$ against $x$ with $t_0$ being the first time in its respective window.}
    \label{fig:raw-liquidity}
\end{figure}

As an initial inspection, Figure \ref{fig:raw-liquidity} displays a 3D surface plot of $(t,x) \mapsto y_t(x)$ (top panel) for the second window of each dataset, and a fixed-time cross-sectional plot of $y_{t_0}(x)$ vs $x$, where $t_0$ is the first time in each window (bottom panel).  The first and third windows are left out of the top panel for space reasons.

The 3D surfaces generally show liquidity to be unimodal, though the mode may shift over time, and additional temporary modes can appear due to stochastic fluctuations. This is clearly present in the snapshots for ARB5 and ETH30, and both appear roughly concentrated around the current market price ($x=0$). ARB5 has some asymmetric tilting across time points, and ETH30 has shifts in variability, with the latest date displayed (Jul 1, 2024) having noticeably lower values toward the edges, and some spikes toward the center.  The ETH5 pool exhibits a flatter profile (top panel) and has several spikes in its cross-sections, making the general pattern less discernible. These irregularities are also visible in its surface plots, with some also appearing in ARB5 at various points.  

\begin{figure}[ht]
    \centering
    \includegraphics[width=0.9\linewidth]{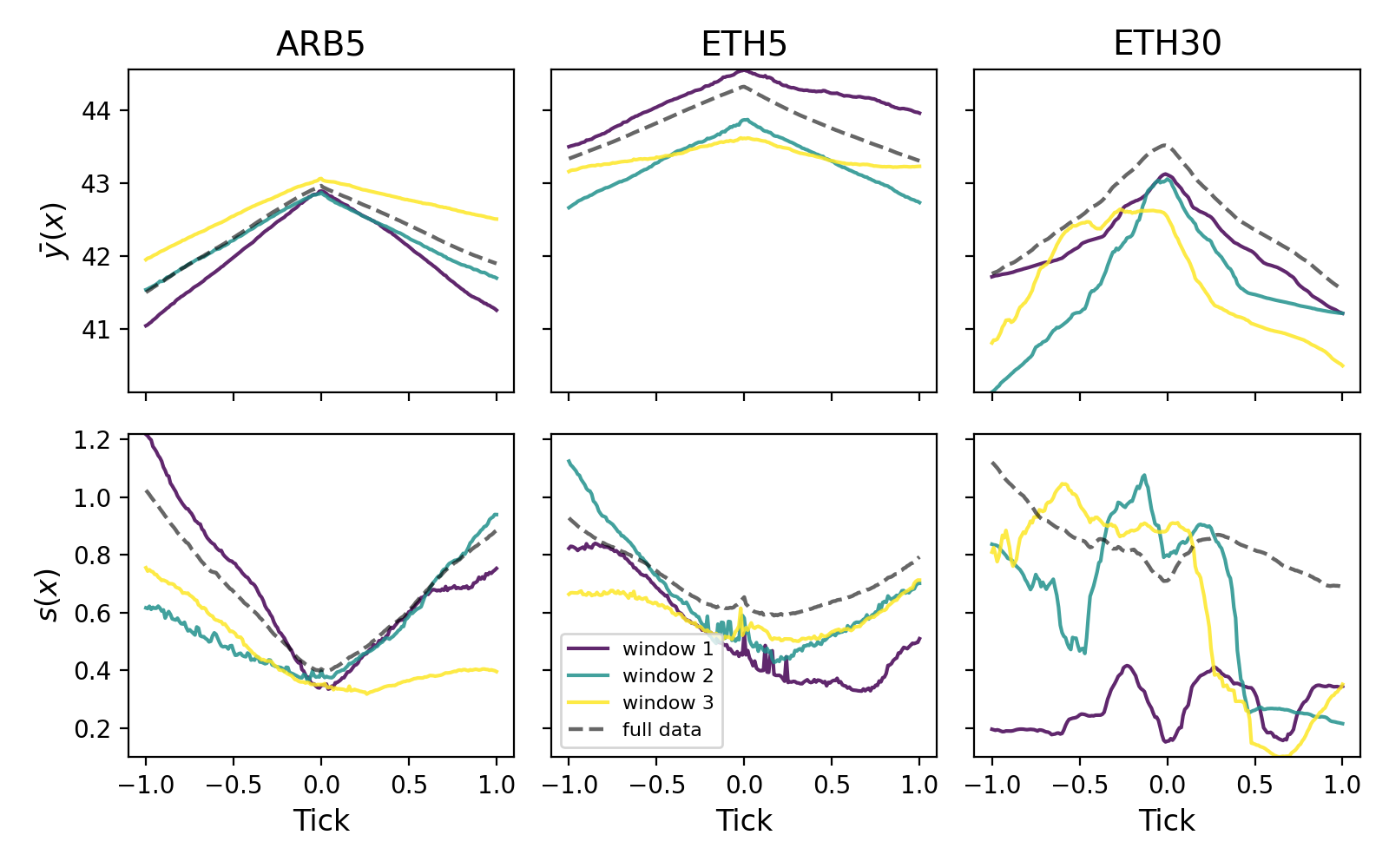}
    \caption{Sample mean and standard deviation functions taken over its corresponding window, plotted against relative tick $x$, across all datasets and windows in Table \ref{tab:dataset-windows}.  The dashed black lines are averages taken over the full dataset in Table \ref{tab:datasets}.}
    \label{fig:sample-mean-var}
\end{figure}

To better detect how liquidity profiles shift within a 4.5-month period, we compute the sample mean and standard deviation functions of the log-liquidity:
\begin{equation}
    \overline{y}(x) = \frac{1}{T} \sum_{t=1}^T y_t(x), \qquad s(x) = \sqrt{s^2(x)}, \quad s^2(x) = \frac{1}{T} \sum_{t=1}^T \big(y_t(x) - \overline{y}(x)\big)^2
\end{equation}
over each window and also the full dataset. These are plotted in Figure \ref{fig:sample-mean-var}.

Consistent with Figure \ref{fig:raw-liquidity}, all mean functions in Figure \ref{fig:sample-mean-var} display a tent-like shape centered at $x=0$, generally consistent with the mean over the full data.  ARB5 and ETH5 have notably stable shapes in the mean (although there are shifts), with some flattening occurring in Window 3 for both (more exaggerated for ETH5).  ETH30 has a similar shape for the first two windows (and full average), but has a less discernible ``tent" shape in the third window, with a more flattened mode whose location is shifted toward $x = -0.25$.

The standard deviation functions of the three datasets vary significantly. While ARB5 and ETH5 share some similarities, such as a valley shape (which can be tilted or smoothed) with higher variance at the edges, they also show distinct characteristics. The ETH5 dataset, in particular, has a spike in variance at $x=0$, which then rapidly drops before increasing again towards the edges. The asymmetry in variance for both ARB5 and ETH5, with higher variability around negative ticks, may be due to LPs anticipating price movements. In contrast, the ETH30 dataset exhibits a noticeably different and less stable structure. This is evidenced by the lack of a clear, persistent 'tent' shape and significant shifts in its standard deviation function over different time periods, reflecting more complex liquidity dynamics in this higher-fee pool.

\subsection{Persistence of Variation Explained}\label{sec:persistence-of-variation-explained}

\begin{figure}[ht]
    \centering
    \includegraphics[width=0.975\linewidth]{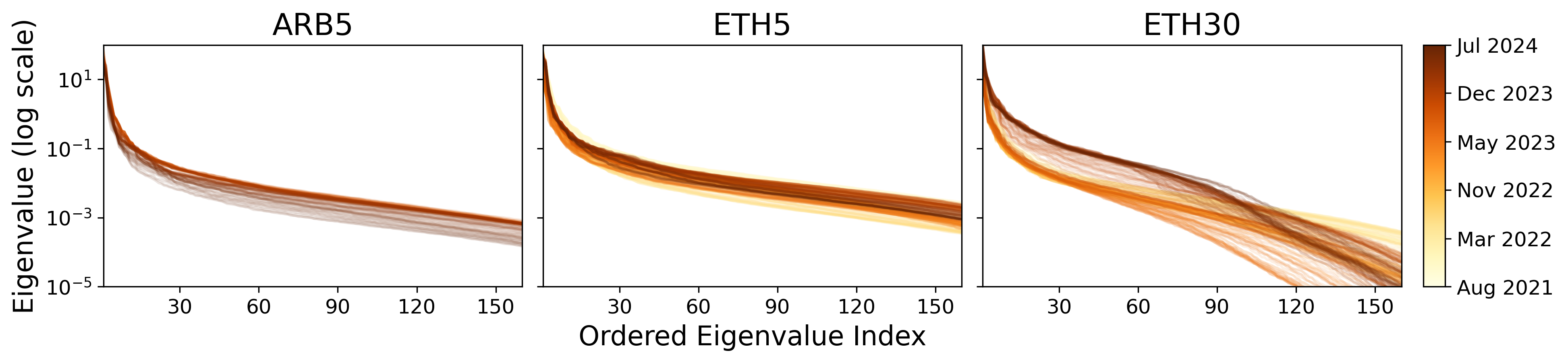}    
    \includegraphics[width=0.975\linewidth]{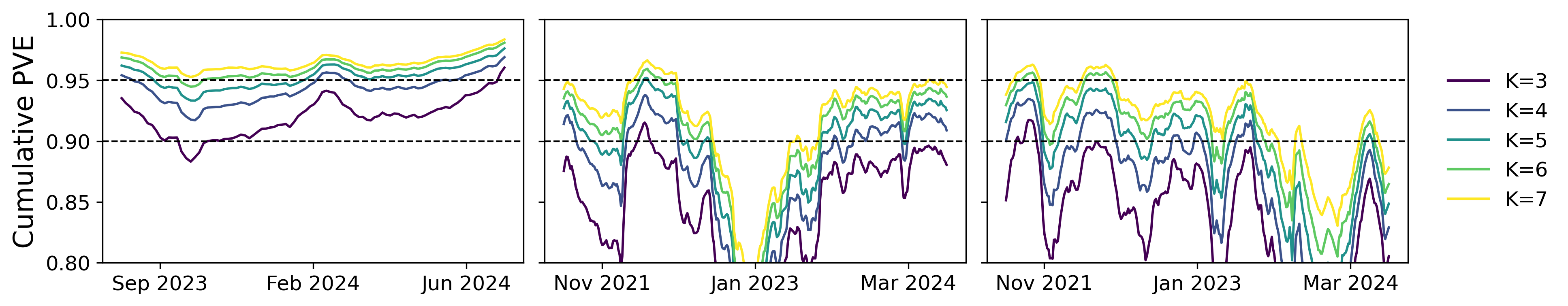}\\

    \caption{Rolling-window principal component analysis (PCA) of liquidity surfaces. Top: ordered eigenvalues (log-scale) for each window of length $T=400$, color-coded by the window start date $t_j$ (lighter = earlier, darker = later). Bottom: cumulative proportion of variance explained (CPVE; Eq.~\eqref{eq:pve}) versus $t_j$ for fixed ranks $K \in \{3,4,5,6,7\}$. Columns correspond to ARB5, ETH5, and ETH30.}
    \label{fig:eigenvalue-CPVE}
\end{figure}

With the noticeably different shapes across windows, we perform rolling window analysis to see how eigenvalues and proportion of variance explained vary across time.  To clarify the timing and meaning of \textit{rolling window}, we denote $t_j = 0, 10, 20, \ldots$ for $j=1, 2, \ldots$ as the start dates of a rolling window, where $t_j=0$ is the same as $t=0$ for the full dataset in Table \ref{tab:datasets}.  Then PCA is performed over the rolling liquidity surface $(y_{t_j}(x), y_{t_{j}+1}(x), \ldots, y_{t_j+T-1}(x))$ for each $j=1, 2, \ldots$.  Thus, each contains its own decomposition and summary statistics.  As before, $T=400$ and $M=201$ so that each surface shares $400-10=390$ times with the previous in sequence.  Useful summary statistics over the windows are shown in Figure \ref{fig:eigenvalue-CPVE}.

The top panel shows the ordered eigenvalues on the log-scale, colored by $t_j$.  The lighter colors (toward yellow) are at earlier dates, and the darker (brown) are later.  All three pairs experience a rapid decay up to the 25th or 30th eigenvalue.  ARB5 and ETH5 consistently stabilize as linear afterward, with stable slopes and slightly shifted intercepts over time periods.  Moving to ETH30, it has a similar shape to the other two prior to 2023 (yellow and light orange), but exhibits a persistent regime shift in late 2023, with a slower decay both in the higher ranks and later ranks (up to approximately the 90th rank).  The stability across dates suggests that a lower rank structure is suitable for ARB5 and ETH5, while ETH30 may be too unstable.  

The bottom panel of the figure looks specifically at the proportion of variance explained over the rolling windows for ranks $K=3, 4, 5, 6, 7$.  The $y$-axis is the cumulative PVE as in Equation \eqref{eq:pve} and the $x$-axis is the rolling window start date.  A rank-5 structure appears adequate for ARB5 to provide 95\% of variance explained, nearly uniform over the times considered; a rank 3 structure satisfies 90\% cumulative PVE.  ETH5 has a similar conclusion for $K=5$ aside from a period of high volatility in Summer 2022 to Summer 2023.  ETH30 is generally more volatile, requiring five components to explain 90\% of variance up to Summer 2023; seven is generally not adequate up to 95\%.  Additionally, the regime shift observed in the top panel (late 2023) is present for ETH30: its cumulative PVE drops but does not stabilize as of the end date of data collection.

In summary, ARB5 and ETH5 are generally stable in eigenvalue decay.  ARB5 and ETH5 share many similarities in their cumulative PVE, although ETH5 experiences a drift to require more PCs to attain a certain PVE around Jan 2023.  Note that it is possible that ARB5 could also experience this phenomenon, but it is unobservable due to a lack of data. ETH30 has temporal irregularities across both panels.  

Appendix \ref{sec:alt-preprocessing} provides PVE plots for the choices of $M=101, 51$, and $11$ ticks in comparison to $M=201$; see Figure \ref{fig:prop-var-over-M}.  There is little difference in moving from $M=101$ to $M=201$, which justifies the choice of $M=201$, providing a balance between a sufficient and necessary amount of data.

\subsection{PCA: Basis Functions And Their Interpretations}
This section performs the PC decomposition in Section \ref{sec:dynamic-factor-extraction} for each dataset and window in Table \ref{tab:dataset-windows}.  This yields the coefficient series $(\beta_{t,k})_{t = 1}^T$ and PCA basis $u_k(x), x \in \{-1.0, -0.99, \ldots, 0.99, 1.0\}$ for $k=1, \ldots, K$.

We choose $K=5$ as components to analyze, as this choice is sufficient for 90--95\% of cumulative PVE for ARB5 and ETH5 over the windows considered.  It also appears that for these windows, no reasonable choice of $K$ will work for ETH30, but its analysis is considered a useful case study.  We begin by analyzing the basis functions $u_k(x)$ to assess commonalities and interpretations.  Subsection \ref{sec:time-series-analysis-PCA} analyzes the time series scores to attach a temporal meaning to the coefficient dynamics.

\begin{figure}[!ht]
    \centering
    \includegraphics[width=0.975\linewidth]{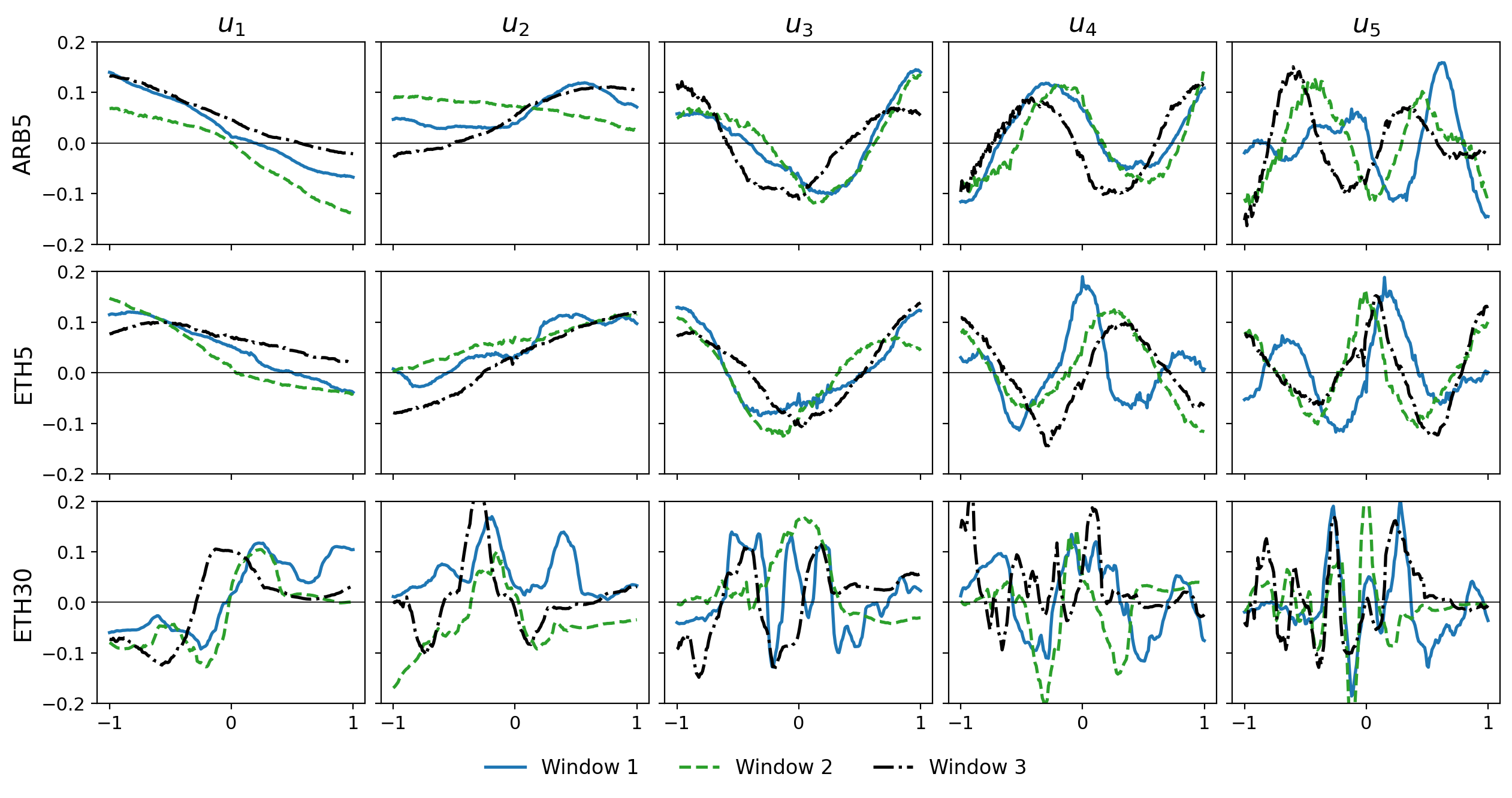}
    \caption{First five PCA basis functions for each dataset.  Rows: dataset, columns: basis function index.  Three windows are displayed per subpanel.  Note that these are unique up to a sign change.}
    \label{fig:PCA-basis-functions}
\end{figure}

Figure \ref{fig:PCA-basis-functions} displays the first five basis functions for each dataset and window. As a baseline, ARB5 demonstrates a generally consistent shape and interpretation across windows:
\begin{itemize}
    \item $u_1(x)$: Exhibits a \textit{slope effect}, consistently tilting the liquidity surface. A shock in this factor increases liquidity toward $x=-1$ and simultaneously decreases it toward $x=+1$.  This mode explains the largest proportion of variance.
    \item $u_2(x)$: Represents a \textit{mixed level/slope effect}: liquidity generally increases as these basis functions are strictly positive (aside from part of window 3). The asymmetry adds a slope, where one side is increased more than the other.
    \item $u_3(x)$: Characterized as a \textit{curvature effect}: acts as a bell curve or symmetric beta distribution.  A positive shock in this factor will decrease the liquidity toward the center and increase it at the edges.
    \item $u_4(x)$ and $u_5(x)$: Show increasing complexity with four and five inflection points, respectively, contributing to finer details of the liquidity surface.
\end{itemize}

Note that within each window, a more pronounced "level" effect can be obtained by a linear combination of $u_1$ and $u_2$, offsetting their slopes.  Comparing Window 3 with Windows 1 and 2 illustrates the idea of the same effect but in a rotated basis, where an approximate level effect can be found regardless of window, but each requires a different way to linearly combine the basis functions.

The ETH5 basis functions are also consistent across windows.  The summary is highly similar to that of ARB5; we highlight a few differences:
\begin{itemize}
    \item $u_1(x)$: Similar in shape to ARB5, but now mostly positive.  A positive shock in this factor will increase liquidity in the $x=-1$ direction, with less of an effect toward the $+1$ direction. 
    \item $u_2(x)$: Slightly more irregular than ARB5.  Modes for Windows 1 and 2 are positive, whereas Window 3 crosses $y=0$ closer to $x=0$.
\end{itemize}
The remaining three basis functions are still similar, albeit with some slight irregularities and less discernable shapes. Across both ARB5 and ETH5, each of $u_k$, $k=3, 4, 5$ tends to have $k-2$ inflection points, consistent with a polynomial representation. One exception is Window 1 for ETH5, which seems to have a higher polynomial order for its fourth and fifth basis functions, indicating a more complex functional shape during that period. 

A consistent interpretation of ETH30 is challenging.
\begin{itemize}
    \item $u_1(x)$: Often a logistic type shape.  A positive shock in this factor would increase liquidity sharply for positive $x$ and decrease it for negative, although the interpretation shifts slightly in Window 3 and in Window 2 appears to be making that transition.
    \item $u_2(x)$: In all windows, a positive shock would increase liquidity in a neighborhood of negative $x$ close to $x=0$. Otherwise, there are not many similarities across windows.
\end{itemize}
Higher order basis functions become rapidly uninterpretable, aside from factor 3 for window 2, which is surprisingly a nice mound shape.  There seems to be a severe instability over time for ETH30's basis; this is further analyzed in Section \ref{sec:basis-stability}.

\subsection{PCA: Coefficient Time Series}\label{sec:time-series-analysis-PCA}

For each principal component time series score associated with the factors above, we provide a time series analysis in two steps: (1) a lightweight overview for general observations (directly below), and (2) a second pass that targets specifics found in the first pass, including considering specific heteroskedastic models and innovation distributions (Section \ref{sec:detailed-time-series}).

\subsubsection{Arbitrum ETH-USDC 5bps (ARB5)} 
\begin{figure}[!ht]
    \centering
    \begin{tabular}{ccc}
        $\qquad$Window 1 (2023–2024) & Window 2 (2024) & Window 3 (2024) \\
        \includegraphics[width=0.31\linewidth]{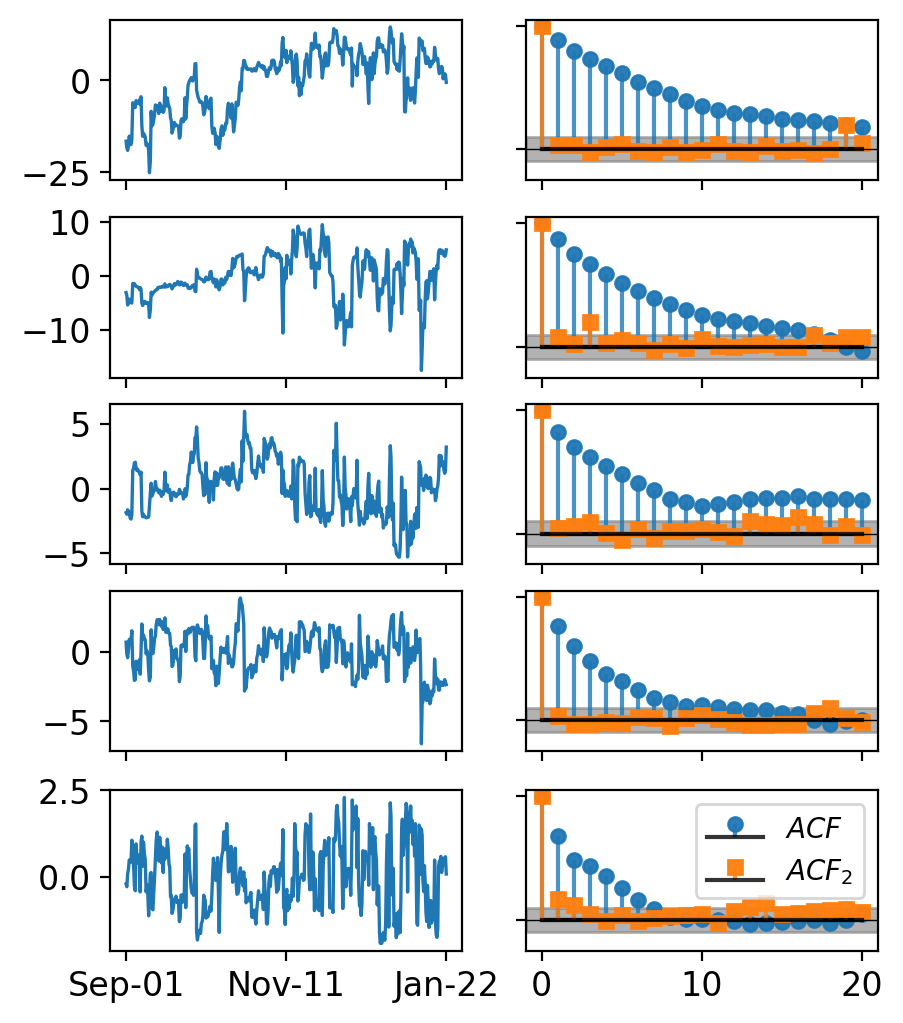} & $\quad$\includegraphics[width=0.31\linewidth]{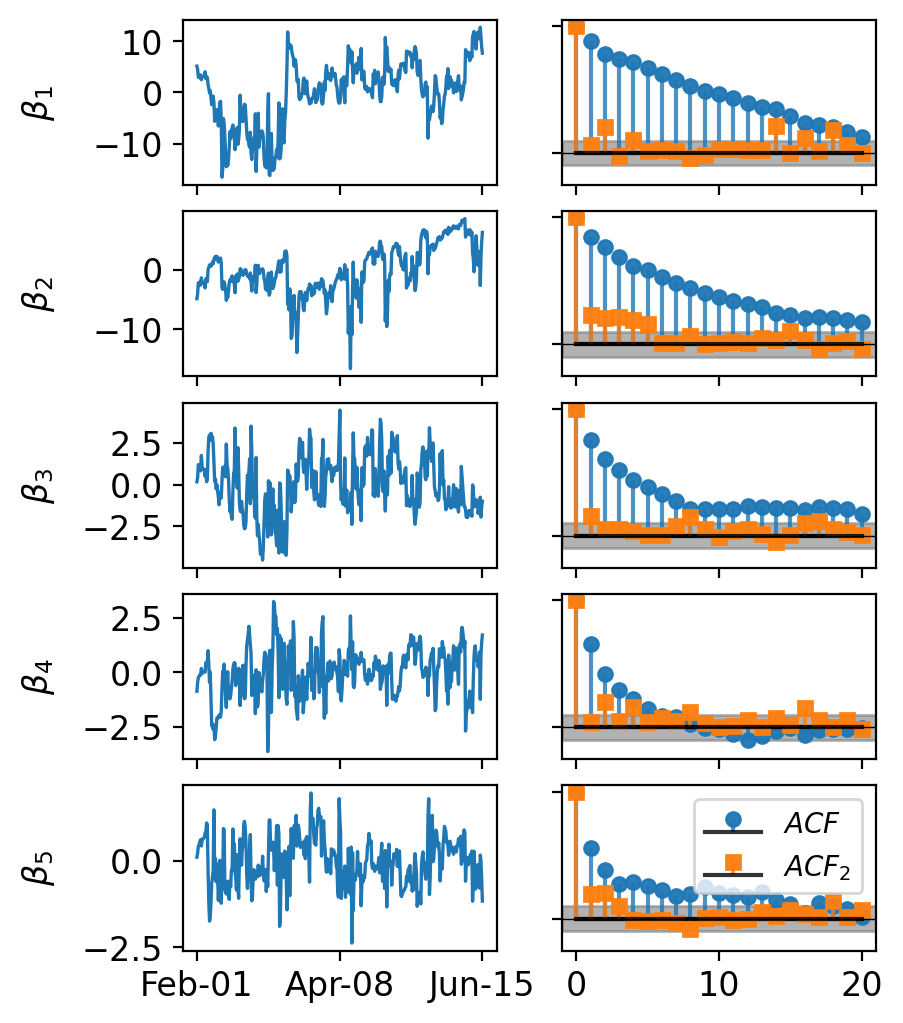} & $\quad$\includegraphics[width=0.31\linewidth]{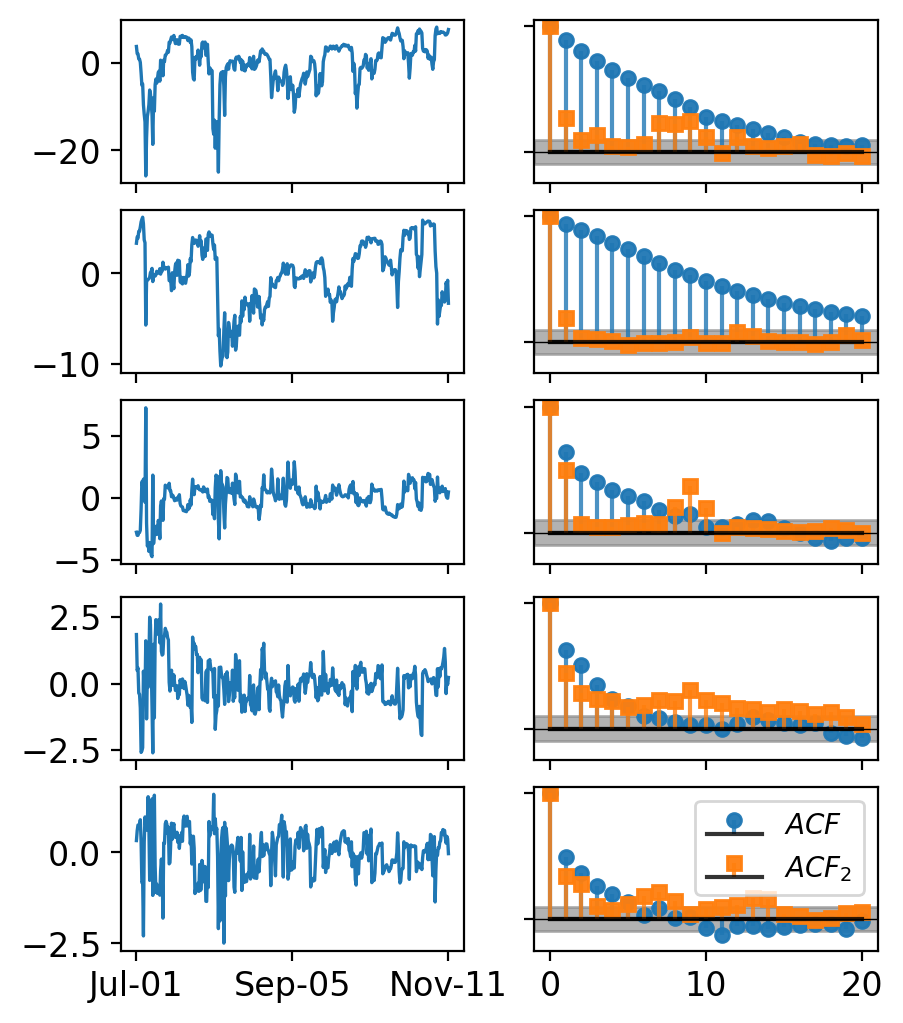}
    \end{tabular}
    \caption{Time series (left) and autocorrelation functions of $\beta_{t,k}$ (right, blue) for ARB5. $\text{ACF}_2$ (orange) refers to the autocorrelation of the squared AR(1) residuals $\hat{e}^2_t$. The gray bar indicates the 95\% pointwise cutoff for significance from zero autocorrelation. Each ACF equals 1 at lag 0.}
    \label{fig:time-series-ARB5}
\end{figure}

Figure \ref{fig:time-series-ARB5} presents the time series of the first five principal component scores ($\beta_{t,k}$) and their autocorrelations (ACFs) for the ARB5 dataset. Table \ref{tab:arb5-time-series} summarizes key statistics for the $\beta_{t,k}$, in particular, its proportion of variance explained, augmented Dickey-Fuller (ADF) test $p$-value, sample standard deviation $\text{sd}(\beta_{t,k})$ (over the window), and mean reversion time $\tau_k = -1/\log |\phi|$ when fitted to an AR(1) model.  The ADF test assesses a null of a unit root against a trend stationary alternative.  Note that we are not performing any formal hypothesis testing, although a larger $p$-value is suggestive of a unit root. For a benchmark comparison, we fix $p>0.1$ as a cutoff to indicate potential unit root behavior (larger values moreso).  These are always visually cross-checked with the time series and its ACF.  Note that $\tau_k$ and $\text{sd}(\beta_{t,k})$ are in reference to an 8-hour period of time: $\tau_k/3$ is the daily reversion time.  The standard deviation $\text{sd}(\beta_{t,k})$ generally does not scale with the sampling interval, so there is no straightforward rule to obtain a daily version.

\begin{table}
\begin{center}\footnotesize
\begin{tabular}{l|rrrr|rrrr|rrrr}
\toprule
\multirow{2}{*}{$k$} & \multicolumn{4}{c}{Window 1} & \multicolumn{4}{c}{Window 2} & \multicolumn{4}{c}{Window 3} \\
\cmidrule{2-5} \cmidrule{6-9} \cmidrule{10-13}
 & PVE$_k$ & ADF $p$ & $\text{sd}(\beta_{t,k})$ & $\tau$ & PVE$_k$ & ADF $p$ & $\text{sd}(\beta_{t,k})$ & $\tau$ & PVE$_k$ & ADF $p$ & $\text{sd}(\beta_{t,k})$ & $\tau$ \\
\midrule
1 & 0.702 & 0.019 & 8.323 & 9.156 & 0.612 & 0.141 & 6.325 & 8.499 & 0.653 & 0.011 & 5.353 & 9.047 \\
2 & 0.184 & 0.000 & 4.257 & 7.295 & 0.274 & 0.050 & 4.228 & 5.914 & 0.260 & 0.033 & 3.376 & 16.839 \\
3 & 0.042 & 0.149 & 2.035 & 5.803 & 0.043 & 0.000 & 1.671 & 3.599 & 0.034 & 0.000 & 1.227 & 2.259 \\
4 & 0.024 & 0.000 & 1.523 & 3.776 & 0.018 & 0.000 & 1.079 & 2.428 & 0.014 & 0.000 & 0.792 & 2.137 \\
5 & 0.009 & 0.000 & 0.945 & 2.591 & 0.007 & 0.003 & 0.679 & 1.753 & 0.008 & 0.000 & 0.601 & 1.421 \\

\bottomrule
\end{tabular}
\end{center}
\caption{Summary statistics for principal component time series of ARB5 for each window. $\tau$ is mean reversion time $-1/\log|\phi|$ from AR(1) fit (in units of 8 hours).}
\label{tab:arb5-time-series}
\end{table}

There is a substantial jump (18--28\%) in explained variance from the first to the second factor compared to the second to third (3--4\%). The decay shape of the ACFs is also persistent across windows: the fourth and fifth factor series typically exhibit rapid decay, while the first two are much slower.  This is apparent in the table, with rapidly decaying mean reversion times, especially for $k=3, 4, 5$.  The standard deviations also show a similar decay at higher modes.  Across windows, the standard deviations and mean reversion times are fairly similar, with some deviations per mode and window.  The most consistent is the reversion time of the first factor series, with $\tau \approx 8.5$--$9.2$ ($2.8$--$3$ days).

The ADF $p$-values suggest that most series are stationary, with the exception of Factor 3 in Window 1 (figure: slow tapering of ACF and non-zero level is indicative of regime shift) and Factor 1 in Window 2 (linear decay in ACF like random walk).

The gray bar in the ACF plots represents the standard $1.96/\sqrt{n}$ asymptotic 95\% pointwise cutoff for significant autocorrelation. The orange dots display the ACF of the squared residuals ($\hat{e}_t^2$) when the series is fitted to an AR(1).  These suggest the presence of heteroskedasticity (GARCH effects) in nearly all series, some more pronounced than others.

\subsubsection{First Pass Commonalities and Takeaways Across Datasets}

There are many commonalities across datasets.  As to not be repetitive with details, the analogous figures and tables for ETH5 and ETH30 are given in Appendix \ref{sec:tab-fig-eth5-eth30-appendix} (specifically, Figure \ref{fig:time-series-ETH5} and Table \ref{tab:ETH5-time-series} for ETH5, and \ref{fig:time-series-ETH30} and \ref{tab:eth30-time-series} for ETH30), with a brief summary below.

ETH5 shares many similarities with ARB5, with some differences. Its fourth and fifth components have a significantly lower mean reversion time compared to the first three. For components 1 and 2, the times are mostly larger (varying from 0.75--3 times larger).  The proportion of variances explained is also generally lower for the first two, and higher for the later components, although it is not extreme.  The majority of coefficient series for ETH5 appear stationary, with one unit root in each window (component 1 in Windows 1 and 2, and component 2 in Window 3).  The series standard deviations are similar in magnitude to ARB5.

ETH30 is less regular.  There are now 1--3 nonstationary series in each window (evidence of unit root according to the ADF $p$), and visual inspection of the ACF verifies a slow linear decay in many of the series, some clearly nonstationary.  The proportions of variance explained also increase in later components; for example, $\text{PVE}_4, \text{PVE}_5$ in Windows 1 and 2 increase by 2x compared to ETH5 (which already increased compared to ARB5), mirroring the discussion surrounding Figure \ref{fig:eigenvalue-CPVE}.

A strong heteroskedastic effect is common across all three datasets, as evidenced by $\text{ACF}_2$ in the figures, which has several significant pieces (often at lower lags). The magnitude of $\text{ACF}_2$ across lags is remarkably similar across datasets and windows, appearing strongest in Window 3 and for the ARB5 dataset. Overall, several common features of the principal component time series emerge across all three datasets. The ACFs of the factor scores generally exhibit an exponential decay, characteristic of AR(1) processes. Additionally, significant autocorrelations in the squared residuals strongly suggest the presence of GARCH effects, indicating time-varying volatility in the principal components, a finding that is almost uniform across all components, datasets, and windows. For ARB5 and ETH5, most windows appear to have at most one series with a unit root, which is more pronounced for ETH5. In contrast, ETH30 consistently shows 1-3 series per window with this effect, and it is generally stronger. Finally, across all datasets, a notable shock in the first coefficient series is observed in Window 3 around August. From the start of Window 2, the first two coefficient series in both ETH5 and ETH30 share a common pattern of dropping from a neutral or positive value to a negative one, followed by a jump in the opposite direction in March. ARB5 shares a similar, though less pronounced, shift in its first coefficient series.

\subsubsection{Detailed Time Series Models}\label{sec:detailed-time-series}

To further investigate the statistical properties of the score series, we perform model selection across mean, volatility, and distribution specifications using the Bayesian Information Criterion (BIC) \cite{kass1995bayes}. To ease notation, we drop the subscript $k$ in $\beta_{t,k}$ until needed.  For a given series $(\beta_{t})_{t=1}^T$ and model $\mathcal{M}$ (broadly containing distributional assumptions, parameters, etc.), write
\begin{equation}
\beta_{t} = \mu_t(\mathcal{F}_{t-1}; \mathcal{M}) + \varepsilon_t,\qquad
\varepsilon_t = \sigma_t(\mathcal{F}_{t-1}; \mathcal{M})\, z_t,\qquad
z_t \overset{\text{iid}}{\sim} F(\mathcal{M}),    
\end{equation}
where $\mathcal{F}_{u}=(\beta_{s}, \sigma_{s}, \varepsilon_{s})_{s \leq u}$. The GARCH(1,1) model yields the volatility recursion
\begin{equation}\label{eq:GARCH}
\sigma_{t}^2 = \omega + \alpha \varepsilon_{t-1}^2 + \gamma \sigma_{t-1}^2, \qquad \omega > 0, \alpha, \gamma \geq 0. 
\end{equation}
The condition $\alpha + \gamma < 1$ yields a stationary covariance.  The ARCH(1) model is simply a GARCH(1,0), meaning it drops the $\sigma_{t-1}^2$ term.  

For a given time series model $\mathcal{M}$ with $d$ estimated parameters, let $\hat\ell$ be the maximized log-likelihood with sample size $T$. The Bayesian information criterion (BIC) is defined as
\begin{equation}
\text{BIC} = -2\hat\ell + d \log T.
\end{equation}
BIC balances goodness of fit $-2\hat\ell$ with a complexity penalty $d \log T$.  When comparing models on the same dataset, the model with lower BIC is preferred.  More specifically, if two models $\mathcal{M}_1$ and $\mathcal{M}_2$ have equal prior probabilities, the posterior odds admit the approximation \cite{kass1995bayes}
\[
\frac{\mathbb{P}\big(\mathcal{M}_1 \mid (\beta_{t})_{t=1}^T)\big)}{\mathbb{P}\big(\mathcal{M}_2 \mid (\beta_{t})_{t=1}^T\big)}
\approx \exp\left(\frac{1}{2}\big(\text{BIC}_2 - \text{BIC}_1\big)\right).
\]
By monotonicity, one often relates these odds with the difference $\Delta_{1,2}\text{BIC} = \text{BIC}_2 - \text{BIC}_1$ as a measure of preference toward $\mathcal{M}_1$ (a lower $\text{BIC}_1$ gives larger $\Delta_{1,2}\text{BIC}$, increasing its odds).  Kass and Raftery \cite{kass1995bayes} give four evidence cutoffs, for $\Delta_{1,2}\text{BIC}$ which are $(0, 2)$, $[2, 6)$, $[6, 10)$, and $[10, \infty)$, corresponding respectively to labels ``not worth a bare mention'', ``positive'', ``strong'', and ``very strong''. Thus, to compare across many $\mathcal{M}_j, j = 1, \ldots, J$ for the same $(\beta_{t})_{t=1}^T$, it is sufficient to compare each model relative to that with the smallest BIC.

The subsequent analysis considers two model selection procedures using BIC as a comparative metric.  The following is performed for all factor score series within each window and dataset:
\begin{enumerate}
    \item Using an AR(1) mean (no trend or intercept), fit a collection of heteroskedastic models and distributions.  This is done first because the current analysis strongly suggests an AR(1) style component with apparent heteroskedasticity but imprecise details.
    \item Using volatility models based on step (1), double-check if the mean specification is correct.
\end{enumerate}

For the first step, the idea is to check some overarching properties and commonalities across datasets, windows, and factors.  Since there are no specific hypotheses to be tested, we sweep over a large collection of volatility specifications and innovation distributions that can assess a variety of properties.  All modelling is done with the \texttt{arch} package \cite{sheppard2021arch} in Python and utilizes the majority of available volatility processes: constant volatility, ARCH(1), GARCH(1,1), EGARCH(1,0,1), EGARCH(1,1,1), GJR-GARCH(1,1,1), and TARCH(1,1,1).  See Appendix \ref{sec:appendix-time-series-model-details} for full details of the seven models tested.  
In summary, this collection of volatility models can identify certain traits based on the data's model preference:
\begin{itemize}
    \item not heteroskedastic (homoskedastic): constant volatility;
    \item no volatility clustering: ARCH(1);
    \item volatility clustering: all GARCH variants;
    \item asymmetry effects: EGARCH(1,1,1), GJR-GARCH(1,1,1), TARCH(1,1,1);
    \item multiplicative volatility dynamics: EGARCH (both 1,0,1 and 1,1,1);     
    \item outlier sensitivity: TARCH is less sensitive than (GJR-)GARCH to extreme values; EGARCH further compresses through logarithm.
\end{itemize}
We also consider all of the available default distributions for the driver of the innovations, $z_t$.  These are the normal, $t$, skew-$t$, and generalized error distribution (GED).  Heavy tails are a hallmark of the $t$-distribution.  For the GED, we restrict its shape parameter to be greater than 2 as an assessment for platykurtic behavior ($t$ will check for leptokurtic).  The skew-$t$ informs whether there is an instantaneous asymmetry effect (compare with the volatility asymmetry, which is state-dependent).

Parameters are estimated by MLE using \texttt{arch} (Python).  Appendix \ref{sec:tab-fig-eth5-eth30-appendix} shows a heatmap comparing all $3 \cdot 3 \cdot 5=45$ series across all $4 \cdot 6 = 24$ combinations of volatility and distributions.  The results are overwhelmingly consistent in favor of some GARCH effect with heavy-tailed innovations ($t$-distribution).  More specifically:
\begin{enumerate}
    \item GARCH or GARCH variants always (100\%) produced the lowest BIC, when compared to ARCH(1) or a constant volatility.
    \item The $t$ or skew-$t$ distribution was always (100\%) preferred to the normal or platykurtic GED.
    \item The top three choices were EGARCH(1,0,1)-$t$ ($26/45=57.8\%$), EGARCH(1,1,1)-$t$ ($6/45 = 13.3\%$), and TARCH(1,1,1)-skew $t$ ($5/45=11.1\%$).
    \item There were 11 cases where there was strong or very strong evidence against EGARCH(1,0,1)-$t$ ($\Delta \text{BIC} > 6$).  Except for one case with GARCH(1,1)-$t$, they were either EGARCH(1,1,1)-$t$ or a skew-$t$ paired with GARCH(1,1), EGARCH(1,0,1), or TARCH(1,1,1).  Out of those 11 cases, 7 of them occurred in ARB5.
\end{enumerate}
Aside from the last bullet, there were no clear patterns where preference was for or against a certain model according to dataset or time window.

In general, conditional heteroskedasticity with heavy tails is unequivocal.  The common winner is EGARCH(1,0,1)-$t$, suggesting symmetric volatility with multiplicative growth and heavy tails.  When asymmetry is preferred, it is mixed between the innovations (skew-$t$) or volatility process (TARCH(1,1,1) or EGARCH(1,1,1)) or both. The results also suggest a preference toward regularizing large shocks, since EGARCH generally wins, and TARCH is always preferred over GJR-GARCH.

To check if AR(1) is appropriate, we compare mean models using EGARCH(1,0,1) volatility with $t$-distributed noise based on the last step.  This simply checks different AR($p$) orders, specifically $p=1,2,3$, as well as an ARMA($1,1$) model for a potential moving average effect.  Note that $p=3$ could have contextual relevance as three lag units equate to one day (so it could be related to a regional trading time, like the operational time differences of America, Europe, Asia).  A BIC heatmap comparing choices is provided in Appendix \ref{sec:tab-fig-eth5-eth30-appendix}.  Toward Occam's razor, consider the cases where there was a BIC difference of $>6$ against AR(1):  This occurred in $8/45 = 17.8\%$ of the series considered, and ARMA(1,1) was never preferred.  An AR(3) model was only preferred in ETH5 Window 1, for $k=2,3,4$.  The remaining 5 cases of AR(2) were scattered across datasets, windows, and component number.  Thus, AR(1) is a reasonable choice according to BIC, and is further supported by the ACF decay rates in Figure \ref{fig:time-series-ARB5}.

\subsection{Basis Stability}\label{sec:basis-stability}
\begin{figure}[!ht]
    \centering
    \includegraphics[width=0.975\linewidth]{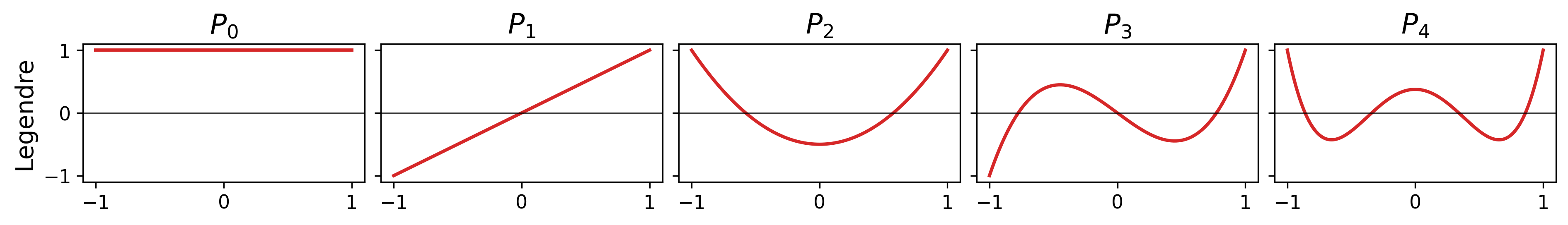}
    \caption{First five Legendre polynomials: $P_0(x) = 1$, $P_1(x) = x$, $P_2(x) = \frac{1}{2}(3x^2-1)$, $P_3(x) = \frac{1}{2}(5x^3-3x)$, $P_4(x) = \frac{1}{8}(35 x^4 - 30x^2 + 3)$.}
    \label{fig:legendre}
\end{figure}

With many commonalities existing both in the statistical properties of the scores and the PCA basis functions, we move toward assessing subspace stability and working with a common basis. We use the same rolling window approach as in Section \ref{sec:persistence-of-variation-explained}, now keeping track of the full sequence of $T \times K$ matrices $(U_{t})_{t \geq 0}$ over each rolling window start time $t$, where $U_t$ is the matrix whose columns are the orthogonal eigenfunctions in the PCA decomposition as computed over the length $T$ window $[t, t+T)$.  We measure basis stability using the projection distance in Equation \eqref{eq:projection-distance}.   To specifically measure how much a subspace is drifting, we use two reference bases:
\begin{itemize}
    \item $d(U_t, U_0)$, the subspace distance between the basis $U_t$ from PCA data over $[t, t+T)$ and $U_0$, the basis estimated at inception $[0, T)$, and
    \item $d(U_t, U_{\mathcal{L}})$, where $\mathcal{L}$ are the Legendre polynomials.
\end{itemize}
To motivate the latter, recall that Figure \ref{fig:PCA-basis-functions} showed similarities with polynomials of increasing degree, relating to terms of its inflection points.  Additionally, rotating the first two basis functions together formed a type of ``intercept plus slope.''  The Legendre polynomials $\{P_n\}_{n=0}^{\infty}$ form an orthogonal basis over $[-1,1]$ (when equipped with the Lebesgue measure), where $P_n$ is of degree $n$.  The first five are displayed in Figure \ref{fig:legendre}.  As intended, they share many similarities when compared with the empirical bases in Figure \ref{fig:PCA-basis-functions}.  

\begin{figure}[!ht]
    \centering
    \includegraphics[scale=0.8]{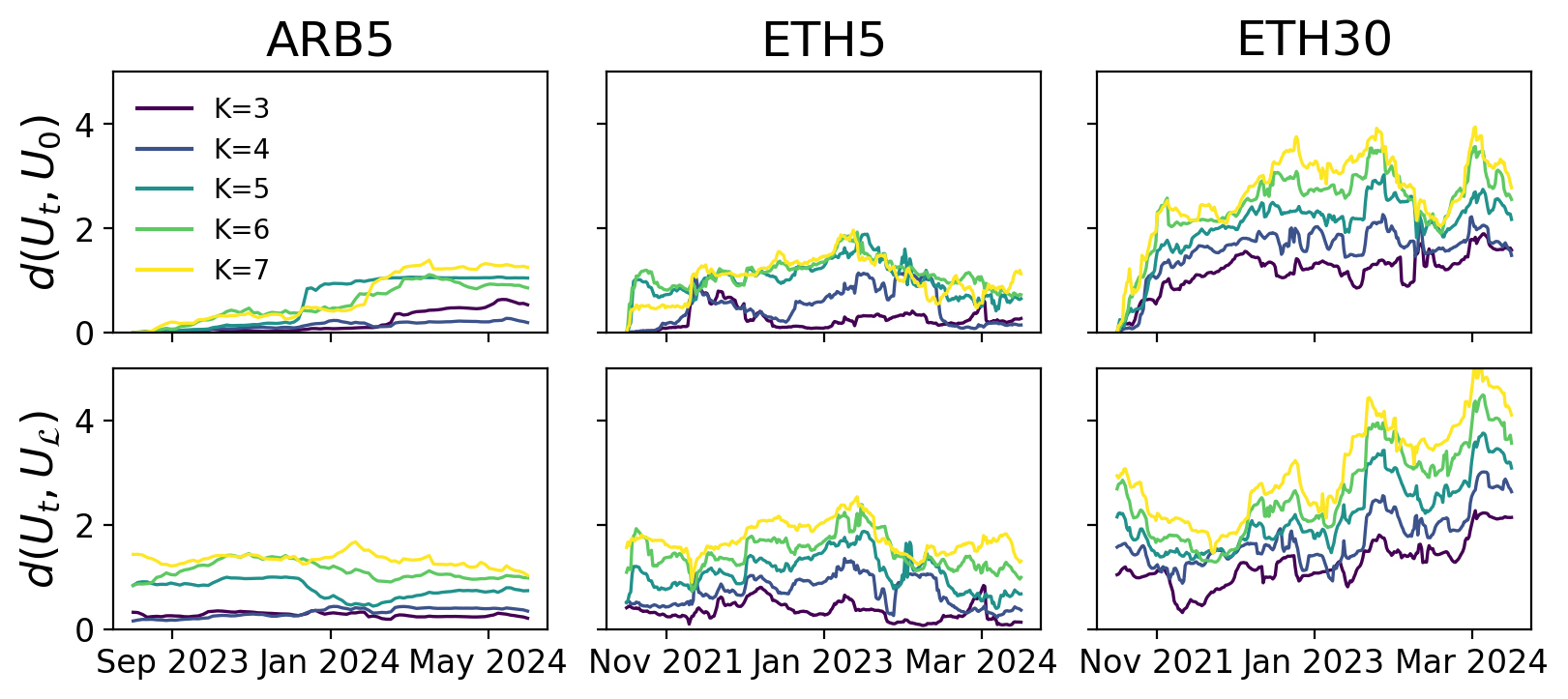}
    \caption{Distance between subspaces.  $U_t$ refers to the eigenmode basis performing PCA over times $\{t, t+1, \ldots, t+T-1\}$, and $U_{\mathcal{L}}$ refers to the Legendre basis. $x$-axis values are equally spaced, corresponding to the rolling window start time. $K$ refers to the dimension of the basis used.}
    \label{fig:eigenmode-stability}
\end{figure}

Figure \ref{fig:eigenmode-stability} displays the subspace distance \eqref{eq:projection-distance} for $K=3, 4, 5, 6, 7$ over the rolling window start date $t$.  The Legendre coefficients were determined as in Equation \eqref{eq:beta-general-basis} using Simpson's rule.  The top panel shows $d(U_t, U_0)$, the amount by which the basis has drifted since the initial eigenbasis was obtained at $t=0$, and the bottom panel shows $d(U_t, U_{\mathcal{L}})$, comparing the empirical eigenbasis against the fixed Legendre basis.

For a reference benchmark, recall from Section \ref{sec:basis-discussion-appendix} that $0 \leq d \leq K < M-K$, and the expected distance between two random $K$-dimensional subspaces is $K(1-K/M)$, or 2.955, 3.920, 4.876, 5.821, 6.756 for $K=3,4,5,6,7$ respectively.  

Looking at the figure in order of datasets, ARB5 generally indicates high stability.  All $K$ exhibit relatively small $d(U_t, U_0)$ throughout.  Notably, $K=4$ is more stable than $K=3$ relative to inception, suggesting information from higher modes is retained even far into the future.  A possible breakpoint appears in higher dimensions around Jan 2024, with a jump for $K=5$ that later propagates to $K=6$ and $K=7$.  Alignment with the Legendre basis is tight, as $d(U_t, U_{\mathcal{L}})$ remains low across time and $K$, similar in magnitude to $d(U_t, U_0)$ for later $t$.  The fourth-dimensional subspaces are consistently of low distance, and even up to the seventh dimension appear quite stable.

ETH5 is qualitatively similar, but is generally noisier with more fluctuations.  Interestingly, $d(U_t, U_0)$ shows transient drift that later reverts, a property hallmark of processes that are weakly stationary with slow mean reversion.  ETH5 is also generally close to $U_{\mathcal{L}}$, with a similar drop in similarity toward middle dates.  Interestingly, all three-dimensional subspaces for ETH5 are highly close even during periods of higher drift.

ETH30's nonstationarity is fully present in Figure \ref{fig:eigenmode-stability}.  The subspace distances increase rapidly a few months from inception.  Each distance is slowly drifting higher, potentially stabilizing to a value close to the random subspace expected distance (e.g.~2.955 for $K=3$ and $6.756$ for $K=7$).

Appendix \ref{sec:alt-preprocessing} repeats the rolling subspace distances in Figure \ref{fig:eigenmode-stability} considering different rolling window sizes ($T=200, 800$ compared to $T=400$) and approximate subsampling intervals ($4$ and $16$ hours compared to $8$).  The overall takeaways are the same regardless of these choices, with an increase in stability when $T=800$ is used.

In summary, the bases for ARB5 and ETH5 are highly stable and comfortably align with the fixed Legendre basis up to at least the fourth dimension. This finding is significant because it provides a clear, interpretable, and natural starting point for modeling liquidity surfaces, a property that the ETH30 pool does not share. The ability to use a fixed, orthogonal basis like the Legendre polynomials simplifies analysis and enables consistent interpretation across different time periods. The next section illustrates its use and shows that its basis functions offer a natural interpretation of the liquidity surface.

\subsection{Legendre Basis for Liquidity Surface}\label{sec:Legendre}
\begin{figure}[!ht]
        \centering
        \includegraphics[width=1\linewidth]{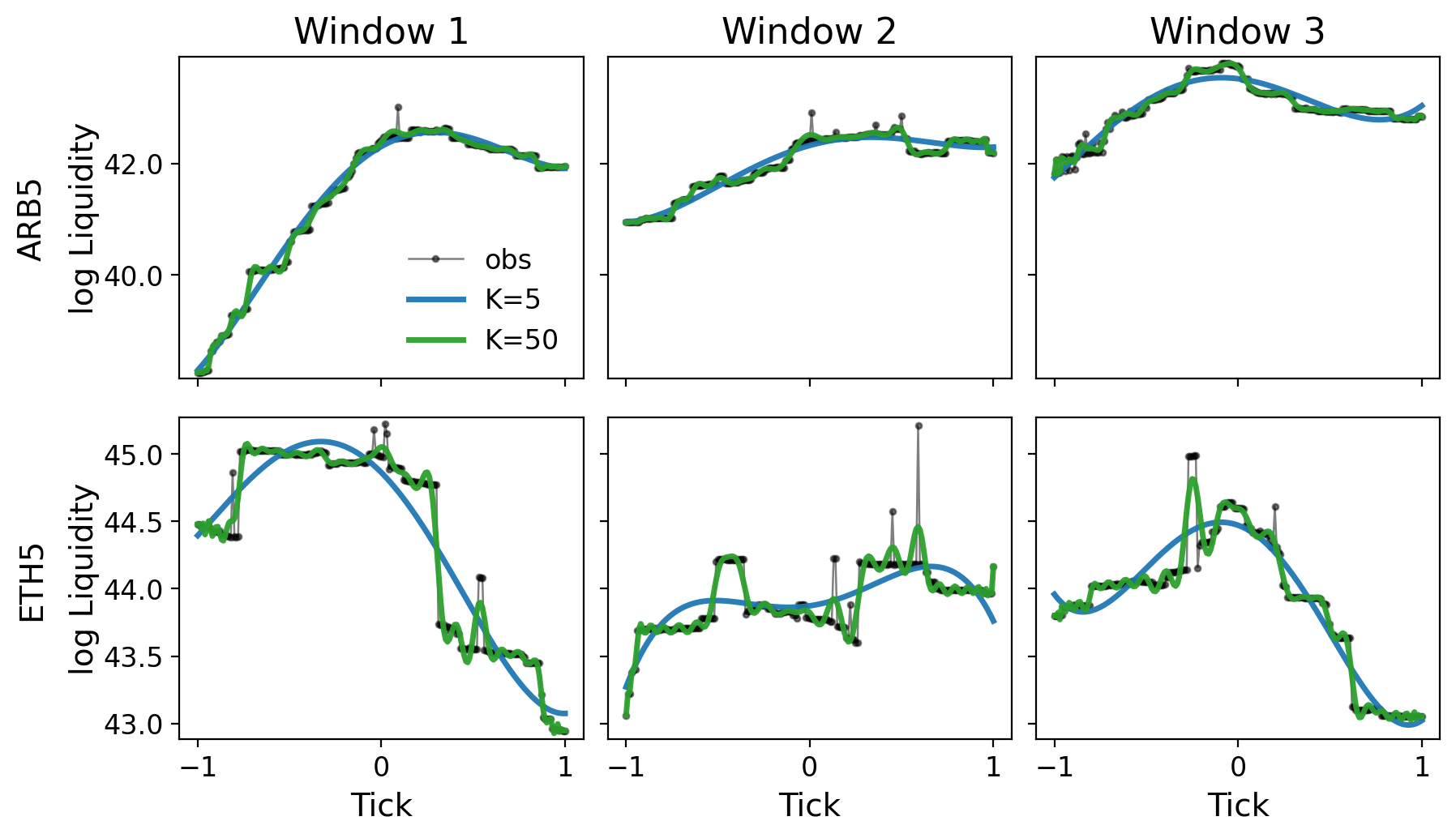}
        \caption{Cross-sections comparing Legendre reconstructions using $K=5$ versus $K=50$ over all windows and ARB5 and ETH5.  Each cross-section uses the first $t$ in that window.}
        \label{fig:reconstruction-5-vs-50}
\end{figure}

For ARB5 and ETH5, and each window, we fit to the Legendre basis and compare against the raw data.  For indexing consistency, enumerate $k=1, 2, \ldots$ and write $\psi_k(x) = P_{k-1}(x)$ as the Legendre polynomial of degree $k-1$.  Specifically, this means to consider the rank $K$ reconstruction
\begin{equation}\label{eq:legendre-reconstruction}
    \hat{y}^{(K)}_t(x) = \sum_{k=1}^K \beta_{t,k} \psi_k(x), \qquad y_t(x) = \hat{y}^{(K)}(x) + r^{(K)}_t(x), \qquad x \in [-1, 1],
\end{equation}
where the $\beta_{t,k}$ are determined as in Equation \eqref{eq:beta-general-basis} using Simpson's quadrature rule, and $r^{(K)}_t$ is the remainder.  Since non-zero intercepts in the $\beta_{t,k}$ provide useful insight, no demeaning is performed prior to the fit.  Figure \ref{fig:reconstruction-5-vs-50} shows the Legendre fit for ARB5 and ETH5, separately fitted over datasets and windows for the first cross-section in each window (analogous to the cross-sections in Figure \ref{fig:raw-liquidity}). This is done for $K=5$ (blue curve) and $K=50$ (green), with the observed $y_t(x_m)$ overlayed (grey points).

The $K=50$ reconstruction cleanly interpolates except in cases of large jumps or potential outlier spikes (particularly evident in the ETH5 case), providing a low bias but high variance fit.  With $K=5$, it is a smoother, low-variance fit.  Visually, it is highly accurate for ARB5 and captures ETH5's coarse shape. Note that neither is objectively better; larger $K$ lowers approximation error but risks fitting noise; $K=M$ interpolates in-sample, a typical bias-variance tradeoff.  The remainder of the text focuses on the first five components ($k=1,\dots,5$) since (1) orthogonality means taking $K>5$ does not change their interpretation or coefficients, (2) empirically, $K=5$ explains $\approx90$–$95\%$ of variance across windows under the PCA eigenbasis, and the Legendre basis is closely aligned.

\subsubsection{Factor Specifics and Effects of Shocks}
In contrast with PCA eigenfunctions, the Legendre basis yields a fixed collection of functions across time.  One may interpret the $k$th basis function in the following way:
\begin{itemize}
    \item $\psi_1(x)=1$ (\textit{level}):  A positive shock raises liquidity uniformly over prices, representing liquidity being added everywhere.
    \item $\psi_2(x)=x$ (\textit{slope/tilt}). Increases liquidity toward $x=-1$ while decreasing toward $x=+1$ (or vice versa), matching the empirical ``slope effect.”
    \item $\psi_3(x)=\frac{1}{2}(3x^2-1)$ (\textit{curvature}): A positive shock would lift the edges toward $x=\pm 1$ and depress at the current price $x=0$, the same as ``flattening" the liquidity surface, which has concentration at zero.  A shock in the opposite direction enforces a bell shape.
    \item $\psi_4(x)=\frac{1}{2}(5x^3-3x)$ (\textit{inhomogeneous tilt}).  This has the same tilt effect as $\psi_2(x)$ at $x=0$ and $\pm 1$, but a reversed effect toward the two inflection points $x = \pm 1/\sqrt{5} \approx \pm 0.4472$.  A positive shock leaves $x=0$ untouched, increases liquidity near $x= -1/\sqrt{5}, 1$, and decreases liquidity near $x= 1/\sqrt{5}, -1$.
    \item $\psi_5(x)=\frac{1}{8}(35x^4 - 30x^2 + 3)$ (\textit{mid-level repositioning}). Increasing the corresponding weight shifts liquidity symmetrically on $x \approx \pm \sqrt{3/7}$ toward the middle and edges.  Effectively, liquidity near (but not at) the middle would shift to either the far edges or toward $x=0$.
    \end{itemize}
The interpretations become unwieldy as $K$ increases.   Note that the lower order pieces hold interpretation even if $K>5$ by orthogonality of the $\psi_k(x)$.

\begin{figure}[!ht]
    \centering
    \includegraphics[width=1\linewidth]{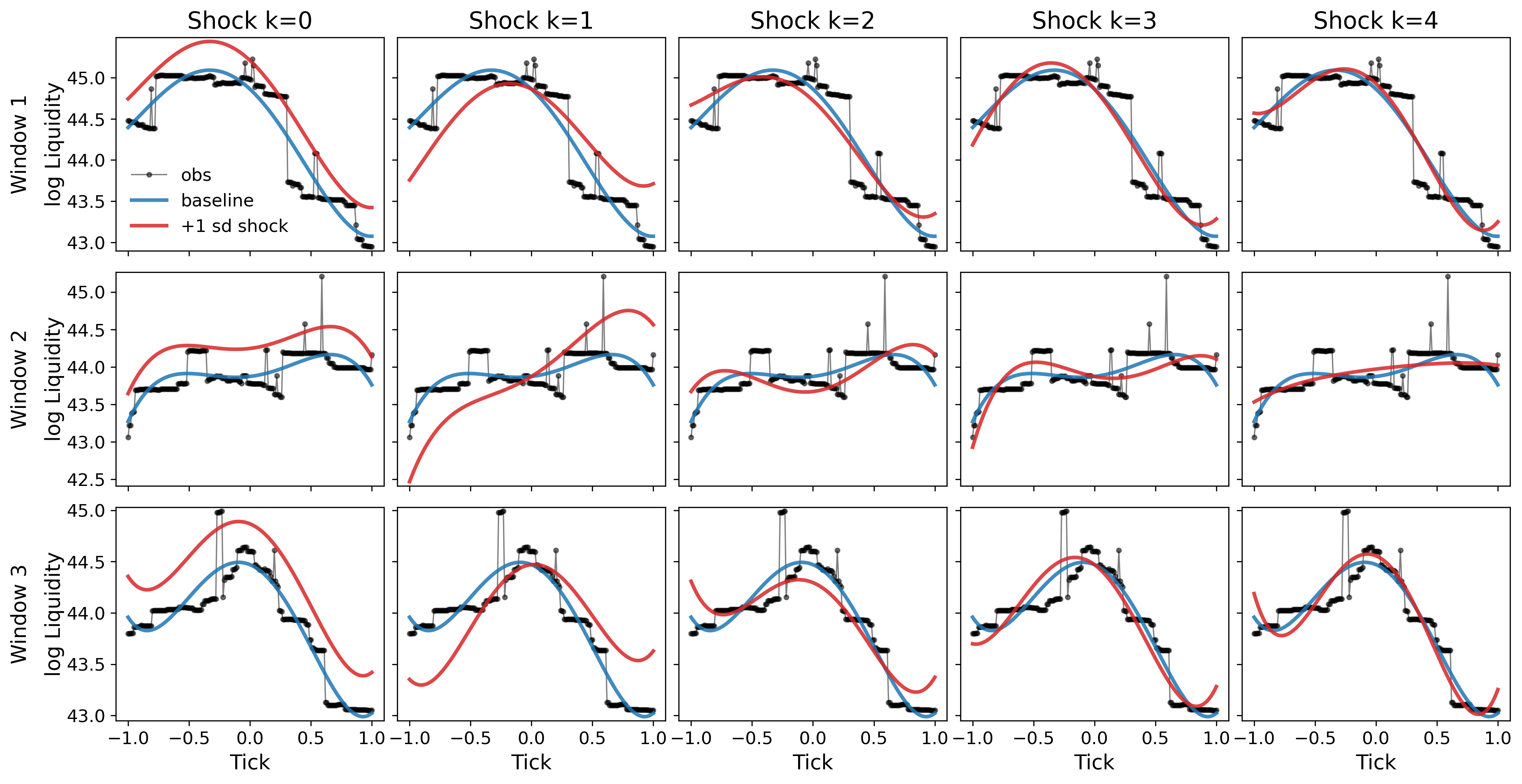}
    \caption{ETH5: effect onto the full cross-section $y_t(x)$ of a 1 standard deviation shock on the $k$th component, $k=1, \ldots, 5$, where $t$ is the beginning of the window considered.}
    \label{fig:eth5-shock}
\end{figure}

To better understand these interpretations, Figure \ref{fig:eth5-shock} works with ETH5 and reveals the effect of a positive ``shock'' to each factor, displaying the first cross-section in each window (mirroring the Figure \ref{fig:raw-liquidity}).  Specifically, $\beta_{t,k}$ is replaced with $\beta_{t,k} + \text{sd}(\beta_{t,k})$ in the reconstruction Equation \eqref{eq:legendre-reconstruction}.  This is done separately for each $k$ to observe its individual effect on the curve.  The sample standard deviation is used for simplicity to illustrate the effect of a ``typical'' shock.  The figure shows the observed $y_{t_0}(x_m)$ (grey points) as well as the Legendre projected versions $\hat{y}_t^{(5)}(x)$ (blue curve) and its shocked version (red curve).  The ARB5 version is Figure \ref{fig:arb5-shock} in Appendix \ref{sec:tab-fig-legendre-appendix}, showing the same patterns.

The observations match the bulleted discussion above, as expected.  Notably, a shock of $k=1$ increases the liquidity curve uniformly, $k=2$ tilts it, and $k=3$ distributes mass away from the current price at $x=0$.  The higher order effects are less obvious (likely because of a smaller standard deviation), although a close inspection reveals $k=4$ providing the inhomogeneous tilt.  This is especially apparent in Window 2 where $x=0$ is left untouched, $x<0$ shows a decrease at $x=-1$ and increase at $x=-1/\sqrt{5}$, and the reverse holds for $x>0$.  For $k=5$, liquidity increases at both the far tails and center, redistributing mass near $x= \pm \sqrt{3/7}$ (local minima of $\psi_5(x)$).


\subsubsection{Legendre Time Series}\label{sec:legendre-time-series}

We briefly repeat the time series analysis as done in Section \ref{sec:time-series-analysis-PCA}.
\begin{figure}[!ht]
    \centering
    \begin{tabular}{ccc}
        $\qquad$Window 1 (2023–2024) & Window 2 (2024) & Window 3 (2024) \\
        \includegraphics[width=0.31\linewidth]{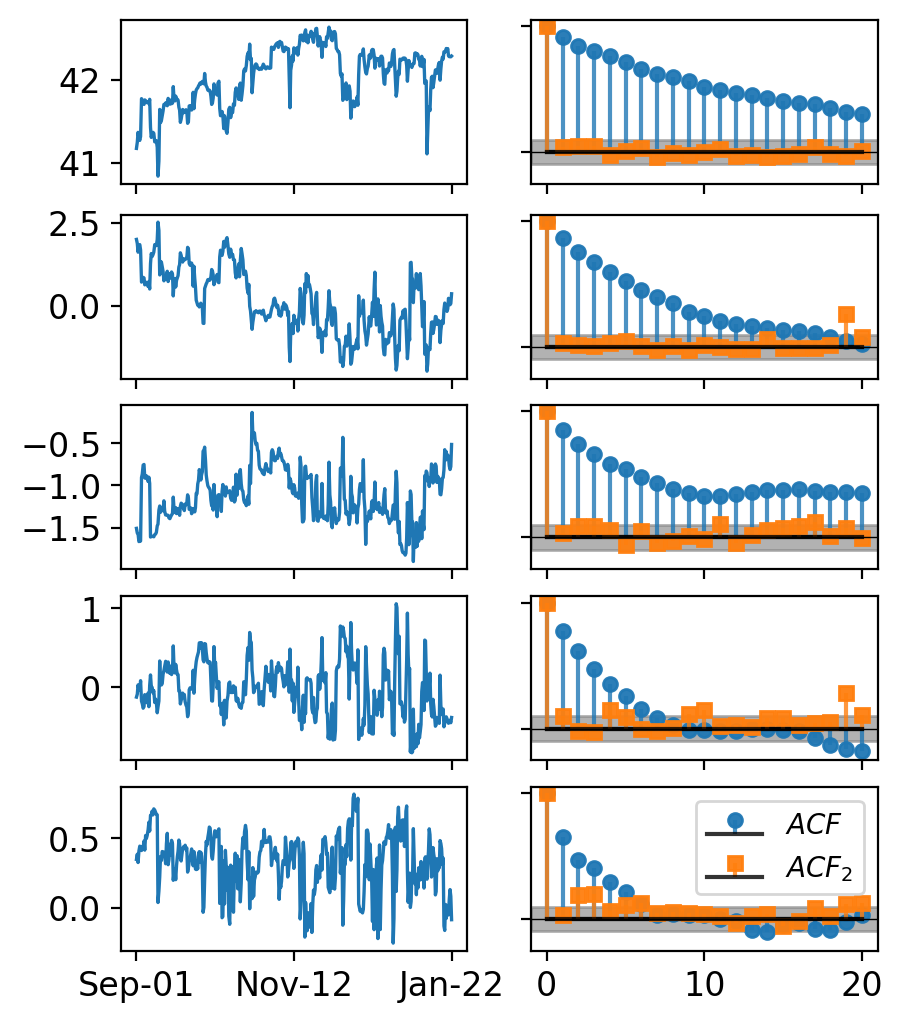} & $\quad$\includegraphics[width=0.31\linewidth]{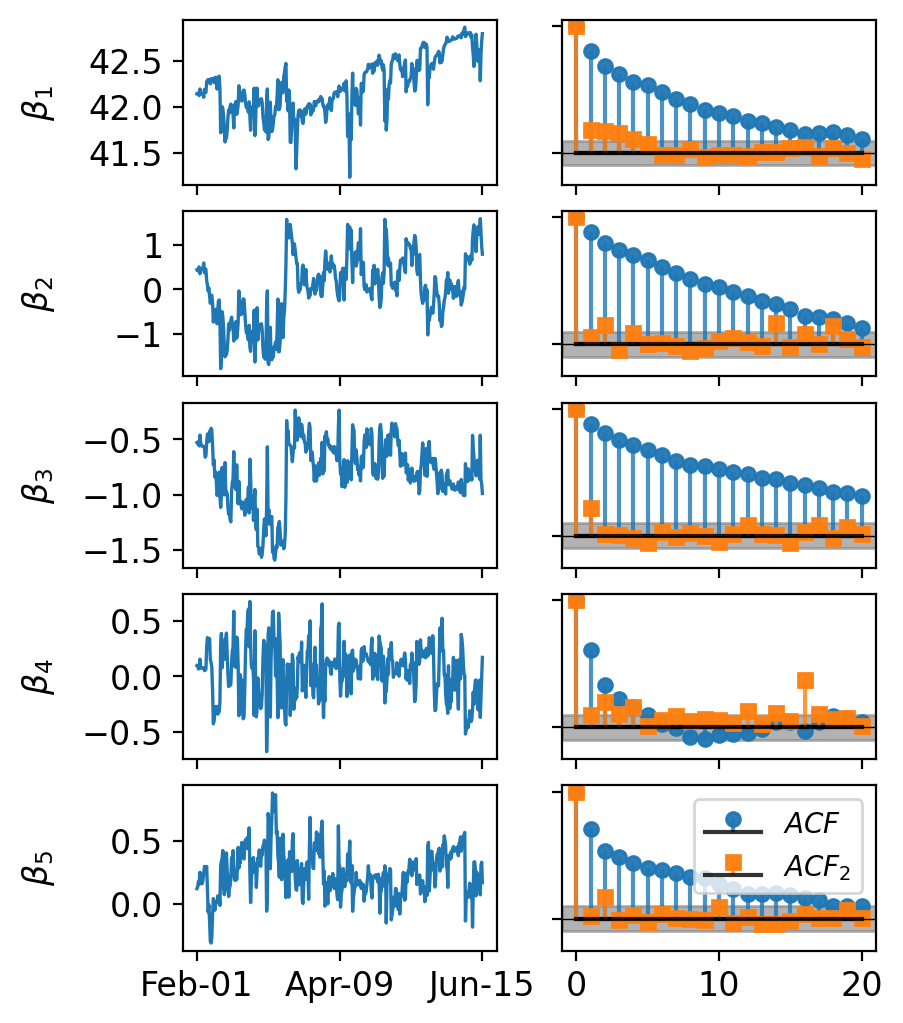} & $\quad$\includegraphics[width=0.31\linewidth]{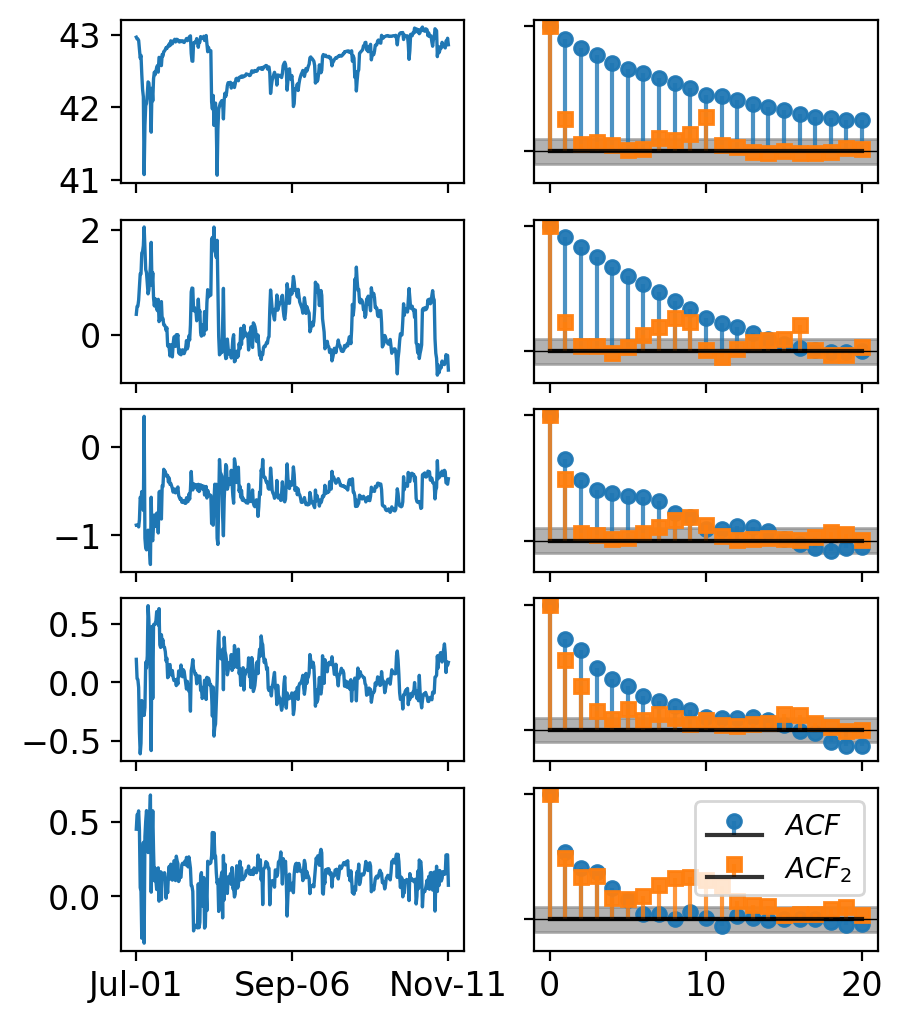}
    \end{tabular}
    \caption{Legendre Fit: Time series (left) and autocorrelation functions of Legendre coefficients $\beta_{t,k}$ (right, blue) for ARB5. $\text{ACF}_2$ (orange) refers to the autocorrelation of the squared AR(1) residuals $\hat{e}^2_t$. The gray bar indicates the 95\% pointwise cutoff for significance from zero autocorrelation. Each ACF equals 1 at lag 0.}
    \label{fig:legendre-time-series-ARB5}
\end{figure}

\begin{figure}[!ht]
    \centering
    \begin{tabular}{ccc}
        $\qquad$Window 1 (2023–2024) & Window 2 (2024) & Window 3 (2024) \\
        \includegraphics[width=0.31\linewidth]{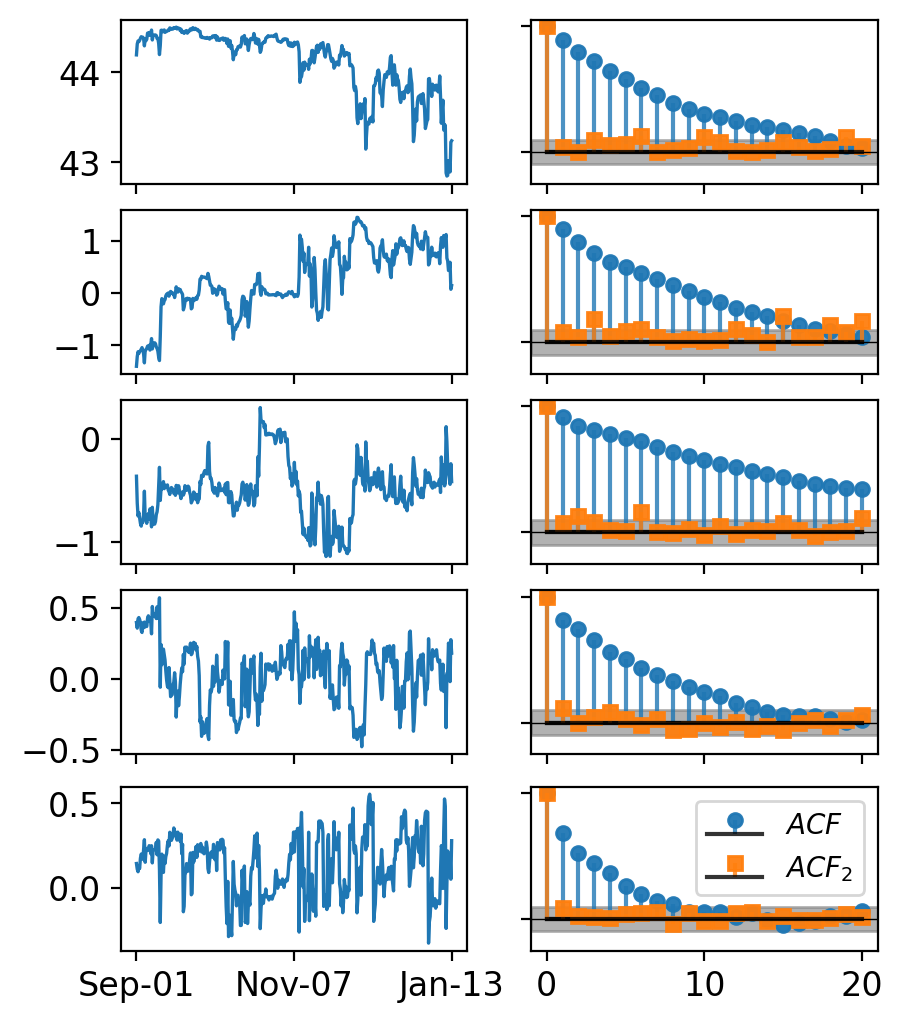} & $\quad$\includegraphics[width=0.31\linewidth]{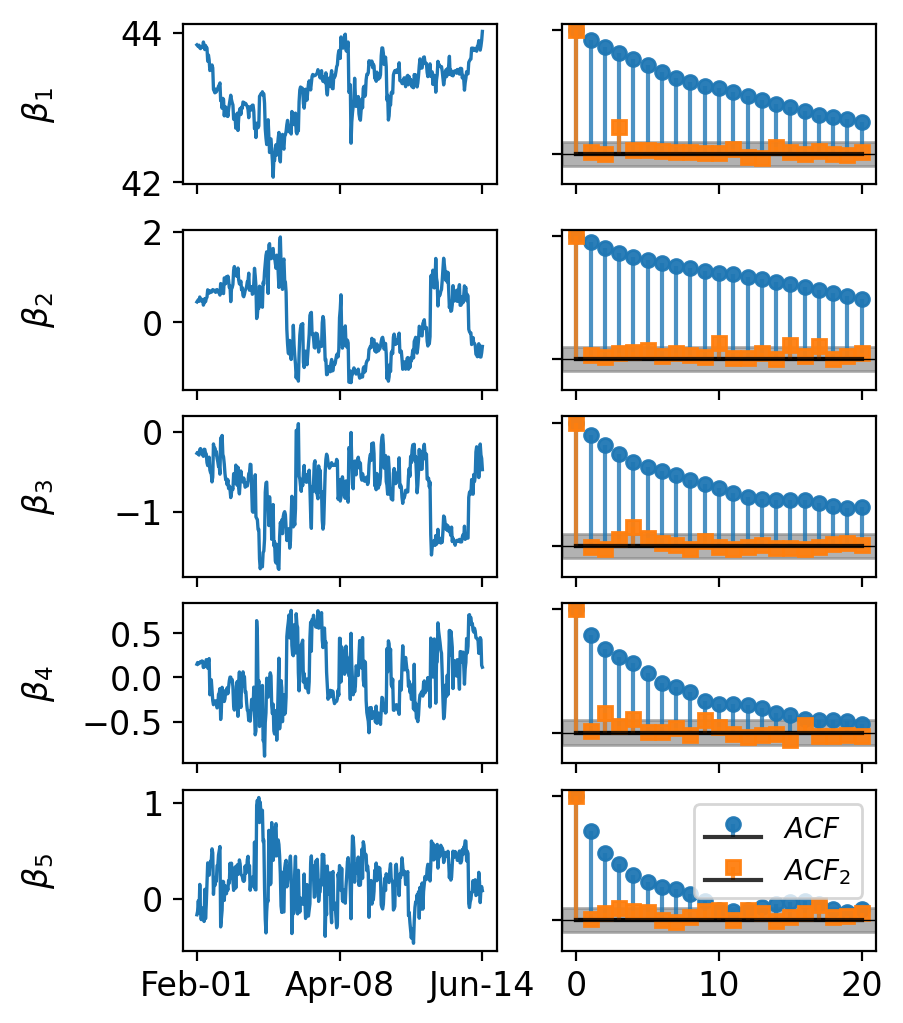} & $\quad$\includegraphics[width=0.31\linewidth]{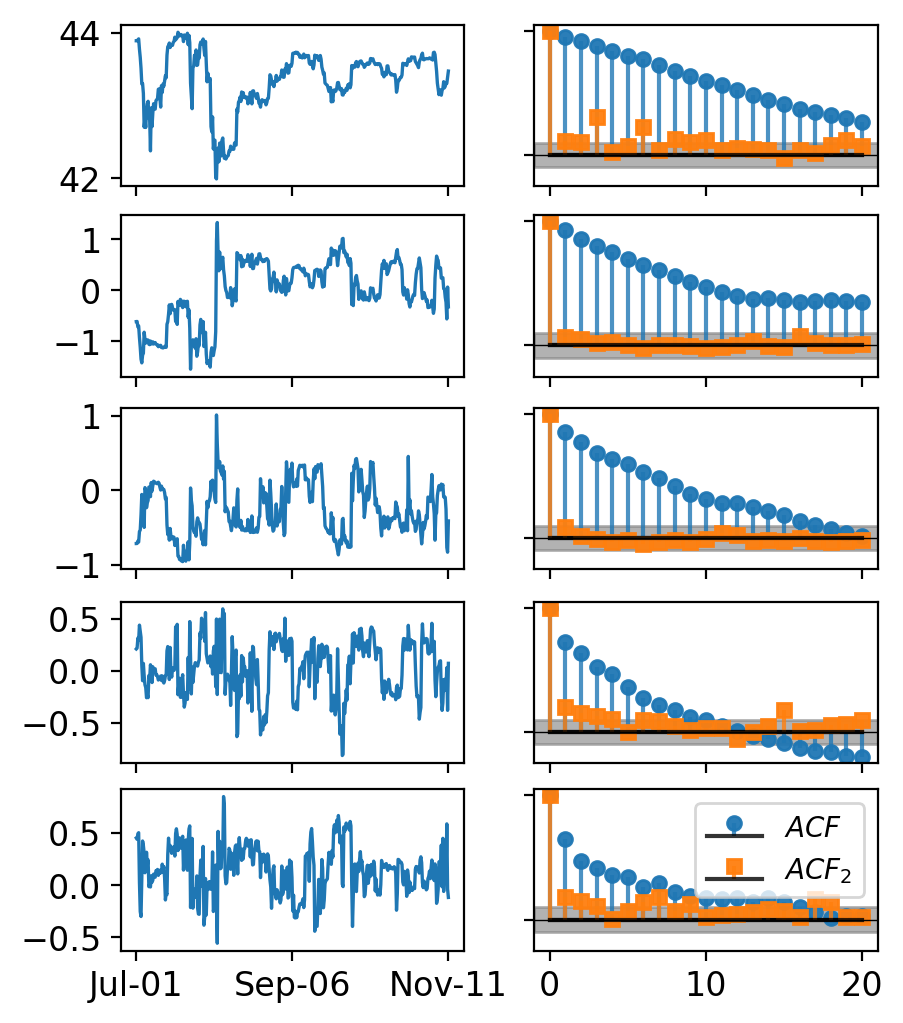}
    \end{tabular}
    \caption{Legendre Fit: Time series (left) and autocorrelation functions of Legendre coefficients $\beta_{t,k}$ (right, blue) for ETH5. $\text{ACF}_2$ (orange) refers to the autocorrelation of the squared AR(1) residuals $\hat{e}^2_t$. The gray bar indicates the 95\% pointwise cutoff for significance from zero autocorrelation. Each ACF equals 1 at lag 0.}
    \label{fig:legendre-time-series-ETH5}
\end{figure}

\begin{table}
\begin{center}\footnotesize
\begin{tabular}{l|rrrr|rrrr|rrrr}
\toprule
\multirow{2}{*}{$k$} & \multicolumn{4}{c}{Window 1} & \multicolumn{4}{c}{Window 2} & \multicolumn{4}{c}{Window 3} \\
\cmidrule{2-5} \cmidrule{6-9} \cmidrule{10-13}
 & ADF $p$ & $\overline{\beta_{t}}_{,k}$ & $\text{sd}(\beta_{t,k})$ & $\tau$ & ADF $p$ & $\overline{\beta_{t}}_{,k}$ & $\text{sd}(\beta_{t,k})$ & $\tau$ & ADF $p$ & $\overline{\beta_{t}}_{,k}$ & $\text{sd}(\beta_{t,k})$ & $\tau$ \\
\midrule
1 & 0.030 & 42.044 & 0.346 & 11.437 & 0.225 & 42.235 & 0.311 & 4.555 & 0.022 & 42.651 & 0.315 & 9.051 \\
2 & 0.009 & 0.094 & 0.948 & 7.387 & 0.054 & 0.039 & 0.743 & 8.399 & 0.002 & 0.249 & 0.544 & 11.378 \\
3 & 0.075 & -1.125 & 0.308 & 6.678 & 0.001 & -0.793 & 0.281 & 7.831 & 0.000 & -0.522 & 0.185 & 2.350 \\
4 & 0.000 & -0.028 & 0.347 & 3.890 & 0.000 & 0.049 & 0.231 & 2.034 & 0.000 & 0.036 & 0.174 & 3.115 \\
5 & 0.000 & 0.327 & 0.210 & 2.390 & 0.000 & 0.250 & 0.183 & 2.908 & 0.000 & 0.145 & 0.125 & 1.603 \\
\bottomrule
\end{tabular}
\end{center}
\caption{Legendre Fit: Summary statistics for Legendre coefficients of ARB5 for each window. $\tau$ is mean return time $-1/\log|\phi|$ from AR(1).}
\label{tab:legendre-arb5-time-series}
\end{table}

\begin{table}
\begin{center}\footnotesize
\begin{tabular}{l|rrrr|rrrr|rrrr}
\toprule
\multirow{2}{*}{$k$} & \multicolumn{4}{c}{Window 1} & \multicolumn{4}{c}{Window 2} & \multicolumn{4}{c}{Window 3} \\
\cmidrule{2-5} \cmidrule{6-9} \cmidrule{10-13}
 & ADF $p$ & $\overline{\beta_{t}}_{,k}$ & $\text{sd}(\beta_{t,k})$ & $\tau$ & ADF $p$ & $\overline{\beta_{t}}_{,k}$ & $\text{sd}(\beta_{t,k})$ & $\tau$ & ADF $p$ & $\overline{\beta_{t}}_{,k}$ & $\text{sd}(\beta_{t,k})$ & $\tau$ \\
\midrule
1 & 0.652 & 44.128 & 0.349 & 9.569 & 0.118 & 43.278 & 0.372 & 16.322 & 0.023 & 43.362 & 0.397 & 23.758 \\
2 & 0.069 & 0.235 & 0.634 & 11.443 & 0.033 & 0.026 & 0.804 & 17.784 & 0.184 & -0.013 & 0.607 & 14.919 \\
3 & 0.031 & -0.499 & 0.274 & 10.886 & 0.000 & -0.722 & 0.402 & 9.717 & 0.000 & -0.257 & 0.350 & 6.550 \\
4 & 0.000 & 0.042 & 0.210 & 5.029 & 0.001 & 0.001 & 0.343 & 4.205 & 0.000 & 0.031 & 0.260 & 3.125 \\
5 & 0.000 & 0.135 & 0.176 & 2.609 & 0.000 & 0.214 & 0.261 & 3.012 & 0.000 & 0.146 & 0.230 & 2.357 \\
\bottomrule
\end{tabular}
\end{center}
\caption{Legendre fit: Summary statistics for Legendre coefficients time series of ETH5 for each window. $\tau$ is mean return time $-1/\log|\phi|$ from AR(1).}
\label{tab:legendre-eth5-time-series}
\end{table}

Many of the takeaways in these figures coincide with conclusions from the PCA discussion.  We list some notable remarks:
\begin{itemize}
    \item The same spikes in midsummer 2024 are present in Window 3, for both ARB5 and ETH5.
    \item The time series and ACFs carry many of the same properties as the PCA version.  A particular example is $k=3$, ETH5, Window 1, where the PCA and Legendre scores are nearly indistinguishable.  
    \item Heteroskedasticity is the biggest thing that seems to be adjusted, sometimes amplified or shifted across $k$ (see e.g.~ARB5 Window 3)
\end{itemize}

The tables remove the proportion of variance explained, which are not directly applicable to a fixed basis expansion, and add $\overline{\beta_t}_{,k}$, the average over that period.  
\begin{itemize}
    \item The first coefficient is an intercept.  The sample mean $\overline{\beta_t}_1$ is large and dominates the other $\overline{\beta_t}_{,k}$, appropriate as it represents the level effect of liquidity.
    \item The average $\overline{\beta_t}_{,3}$ for the curve effect is uniformly negative across windows and datasets and large relative to its standard deviation.  This makes sense because increasing the curve effect flattens it out.  A negative baseline is needed to obtain the bell curve shape.
    \item Conversely, the average mid-level repositioning $\overline{\beta_t}_{,5}$ is uniformly positive.  This counteracts the pure ``quadratic'' shape from the negative $\overline{\beta_t}_{,2}$, creating a tail effect akin to a bell curve.  It also has the fastest mean reversion time (approximately 1 day, sometimes faster for ARB5), suggesting a transient effect.
    \item The slope effects $(k=2, 4)$ always have large $\text{sd}(\beta_{t,k})$ relative to their averages.  This suggests an insignificant average slope effect, i.e.~the liquidity surface can drift from asymmetry but will revert toward it.  The mean reversion time for $k=2$ is quite high with $\tau \approx 7.5$--$18$.  In contrast, the cubic $k=4$ has $\tau \approx 2$--$5$.
    \item Mean reversion times decrease rapidly as $k$ increases, particularly for $k=4$ and 5.
    \item We still observe 0--1 unit roots across the first five coefficient series within each window/dataset.  The ones with ADF $p>0.1$ are the level effect $(k=1)$, common across ARB5 Window 2, and ETH5 Windows 1 and 2, and the slope effect $(k=2)$ for ETH5 Window 3.  These look like a random walk/unit root (slow decay).  For Window 3, it looks like a regime shift by the ACF (quick decay then level off at a nonzero value), and the switch in the time series.  
\end{itemize}

\paragraph{\emph{Detailed Time Series Analysis}} We repeat the investigation of the time series as in Section \ref{sec:detailed-time-series} for the Legendre time series.  Rather than looking for broad patterns across all datasets and windows, the focus is shifted toward each basis function as they carry a fixed meaning in this context.

Figure \ref{fig:legendre-heteroskedastic-BIC-heatmap} shows a heatmap comparing all $3 \cdot 2 \cdot 5=30$ series across all $4 \cdot 6 = 24$ combinations of volatility and distributions.  In aggregate, the patterns are very similar to the PCA case, favoring a GARCH effect with heavy tails.  The $t$ or skew-$t$ distribution are always preferred to the normal or platykurtic GED, and the choices still include EGARCH(1,0,1)-$t$ ($12/30=40\%$), TARCH(1,1,1)-skew $t$ ($6/30=20\%$), with GARCH(1,1)-$t$ the third most preferred now ($4/30=13.3\%)$.  Interestingly, there are now two cases for an ARCH(1)-$t$, but both have a $\Delta\text{BIC} < 1$ compared to the EGARCH(1,0,1)-$t$.  With fixed factor interpretations across windows, it is important to point out some specific observations:
\begin{figure}[ht]
    \centering
    \includegraphics[width=0.95\textwidth]{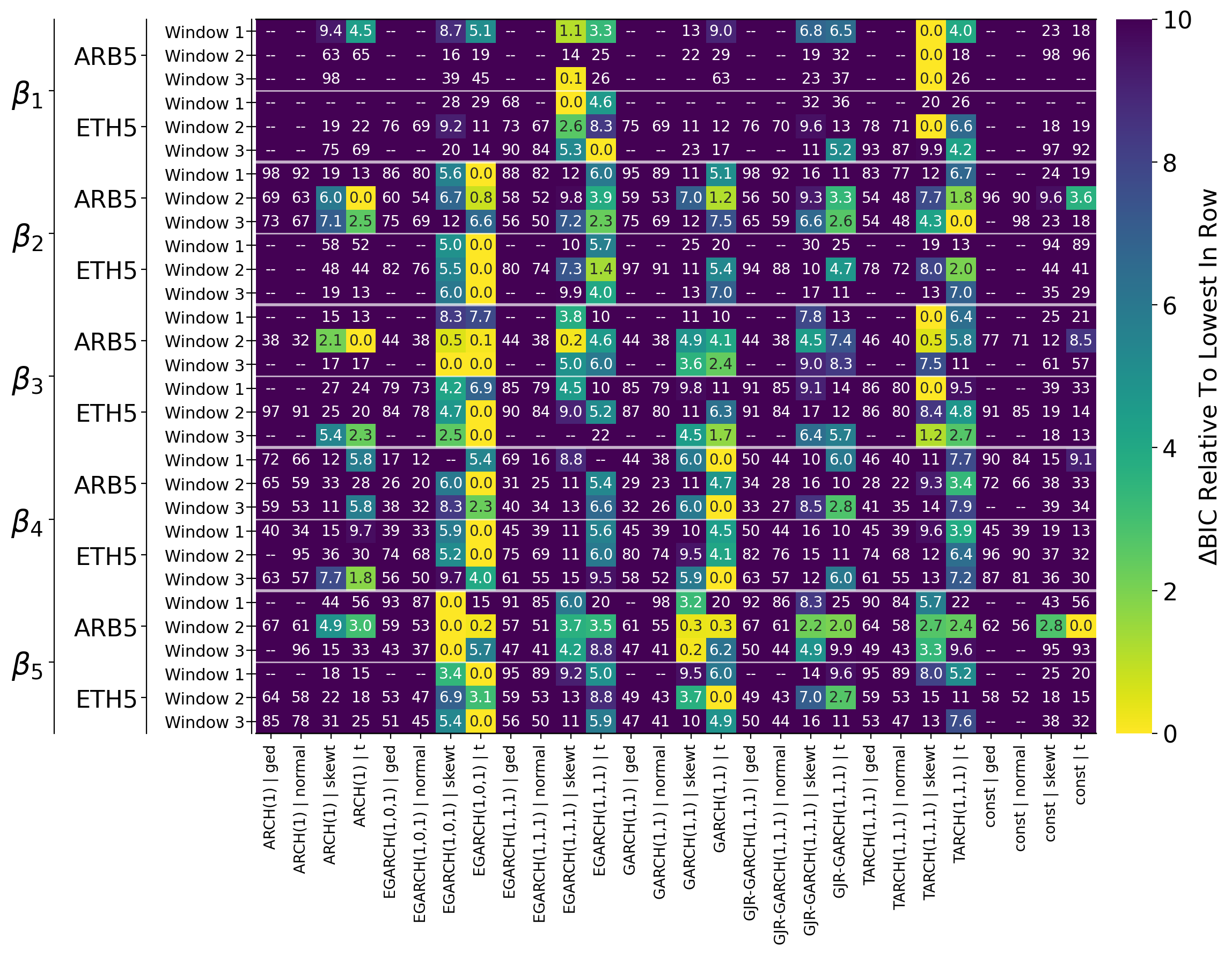}
    \caption{Legendre Coefficients: Heatmap of BIC scores according to heteroskedasticity and distribution assumptions.  Reported is the $\Delta$ BIC relative to the lowest (best) for that series.  All use an AR(1) mean.  Any value of ``--" had $\Delta BIC > 10$.}
    \label{fig:legendre-heteroskedastic-BIC-heatmap}
\end{figure}

\begin{enumerate}
    \item For $\beta_{1,t}$, the coefficients associated with the level effect $P_0(x)=1$, there is unanimous preference for the asymmetric TARCH(1,1,1) or EGARCH(1,1,1) over EGARCH(1,0,1) (with 5/6 preferring skew-$t$ and the other a symmetric-$t$).  All of these have $\Delta\text{BIC}>6$ compared to the best EGARCH(1,0,1) in that row, aside from ARB5 Window 1 ($\Delta \text{BIC} = 5.1$).  Thus, the level effect exhibits high asymmetry, both in the volatility (state dependent) and innovations (instantaneous shocks).
    \item EGARCH(1,0,1) or GARCH(1,1) is always preferred for $\beta_{4,t}$ and $\beta_{5,t}$ over asymmetric EGARCH(1,1,1) or TARCH(1,1,1).   The skew-$t$ wins for ARB5, and $t$ for ETH5.  Thus, these high-order effects prefer symmetric volatilities, with potential skew in the innovations.
    \item $\beta_{2,t}$ and $\beta_{3,t}$ are split across all choices mentioned above.
\end{enumerate}

We repeat the mean analysis, using EGARCH(1,0,1) volatility with $t$-distributed noise.  A detailed BIC heatmap is in Figure \ref{fig:legendre-mean-BIC-heatmap} in Appendix \ref{sec:tab-fig-legendre-appendix}.  The takeaways are similar to those of the PCA discussion, with a general preference toward AR(1) with a couple of exceptions.  This time, the first two factors $(k=1, 2)$ are consistently AR(1) (10/12 give lowest BIC, and the other two have $\Delta BIC = 1.8, 4.7$).  The higher-order factors are more mixed, but still with a strong preference toward AR(1). AR(3) was never preferred.

%% file: Conclusion.tex
\section{Discussion and Conclusion}\label{sec:discussion-conclusion}
This research explored the application of \textit{dynamic factor models} and \textit{functional principal component analysis (FPCA)} to model and analyze liquidity profiles in Uniswap v3. Our findings demonstrate the potential of these methodologies to effectively capture the complex dynamics in liquidity distributions across price ranges. By treating liquidity profiles as functional data, we can leverage tools for dimensionality reduction and forecasting that maintain the structure of the original data.

Our empirical analysis highlights a \textit{persistent low-rank factor representation} within the Uniswap v3 liquidity surfaces of the 5 bps pools. For all datasets and time periods considered, the factor time series coefficients exhibited \textit{autoregressive (AR) dynamics}, clear \textit{generalized autoregressive conditional heteroskedasticity (GARCH) effects}, and \textit{heavy tails}. This was consistent for both the empirical (PCA) bases and a Legendre basis. A statistical verification using the Bayesian Information Criterion (BIC) overwhelmingly favored GARCH variants over constant volatility models, confirming a preliminary observational analysis. The models also consistently preferred heavy-tailed innovation distributions, such as the Student's $t$-distribution, over the normal distribution, suggesting that liquidity surfaces are prone to occasional extreme shocks. Additionally, there appears to be approximately one unit root present in the collection of factor time series for the 5 bps pools, a commonality in other functional time series in finance, such as yield curves. In summary, heavy tails, conditional heteroskedasticity, and AR(1)-style mean-reversion are unequivocal features of (log)-liquidity surfaces.

A key finding is the \textit{stability of the empirical PCA basis} for the 5 bps pools (Ethereum ETH-USDC and Arbitrum ARB-USDC). This low-rank structure is highly stable across rolling time windows and also aligns consistently with a low-order Legendre polynomial basis. This alignment is a critical discovery, as it provides a fixed, interpretable set of factors, akin to the level, slope, and curvature factors found in the Nelson-Siegel model for yield curves, which can be used for consistent analysis and modeling. This finding suggests that the fundamental "shape" of liquidity surfaces for these pools can be parsimoniously represented by a collection of orthogonal factors with persistent distributional properties. This sets the stage for accurate forecasting, uncertainty quantification, and analysis of economic shocks.

In contrast, the Ethereum 30 bps ETH-USDC pool revealed a highly nonstationary structure with significant temporal irregularities. Its PCA basis drifted substantially over time, requiring a higher number of principal components to explain the same proportion of variance observed in the 5 bps pools. This stark difference suggests that the higher 30 bps fee tier may attract liquidity providers with different patterns, resulting in a less stable and less consistent liquidity surface.

\subsection{Stylized Facts of Uniswap v3 Liquidity Surfaces}
The Legendre factor model can be used to empirically verify the stylized facts in the seminal work \cite{Fan2021Strategic, Fan2022Differential}. The statistical analysis of the log-liquidity surface confirms several stylized facts from existing literature and reveals new ones.

\subsubsection{Verification of Known Stylized Facts}
\begin{itemize}
    \item We confirm the claim that ``\emph{we typically observe $L_t(x)$ to be relatively large near $x=0$.}'' The average coefficients $\overline{\beta_t}_k$ for the asymmetric terms $(k=2,4)$ in Tables \ref{tab:legendre-arb5-time-series} and \ref{tab:legendre-eth5-time-series} are both effectively zero across all time windows, meaning asymmetry averages out over time. We also uniformly observe that $\overline{\beta_t}_3 < 0$ (significantly so) and $\overline{\beta_t}_5>0$, both of which contribute to a larger liquidity mass around $x=0$.
    \item We provide empirical evidence for the stylized fact that ``\emph{The shape of the liquidity surface can vary significantly depending on LP risk preferences and chosen price interval configurations.}'' and that ``\emph{Market volatility significantly influences LP behavior}''. Our findings show that in high-volatility regimes, LPs tend to widen their liquidity ranges, shifting liquidity mass outwards and lowering the central peak. Conversely, in low-volatility periods, liquidity is characterized by a tighter, steeper concentration around the mid-price. According to our model, this phenomenon is captured by the "flattening effect" or $\psi_3(x)$ with coefficients $\beta_{t,3}$. A higher average $\beta_{t,3}$ value corresponds to a flatter liquidity surface, providing clear evidence for risk-averse LP behavior.
\end{itemize}

\subsubsection{New Stylized Facts}
Our statistical analysis reveals a succinct set of new facts about the liquidity surface dynamics.

\begin{itemize}
    \item \textbf{Low-Rank Structure:} Dynamic factor analysis shows that 3 to 7 orthogonal components are sufficient to explain 90\% to 95\% or more of the variation in the log-liquidity surface. This percentage occasionally drops to 80\% during periods of high volatility.
    \item \textbf{Persistent Time Series:} The time series associated with these factors exhibit AR(1)-like behavior when stationary. There is overwhelming evidence of GARCH(1,1) volatility and heavy-tailed innovations, and these facts hold consistently across the pools studied.
    \item \textbf{Basis Alignment:} For the 5 bps pools (ARB-USDC and ETH-USDC), the empirical PCA bases strongly align with the basis of Legendre polynomials. This consistency across time windows allows for the use of a parsimonious, fixed-factor model. The distributional facts mentioned above also hold for the Legendre coefficient series.
    \item \textbf{Nonstationarity:} Approximately one unit root is present in the collection of coefficient time series for the 5 bps pools, as found in both the PCA and Legendre bases. However, for the 30 bps ETH-USDC pool, the coefficient series are highly nonstationary, and there is no consistent basis, indicating a more complex and less stable liquidity structure.
\end{itemize}

\subsection{Future Work}
Our findings present several compelling avenues for future research, building upon the foundational framework we have established. An immediate next step is to examine whether the trends observed in the ETH-USDC 5bps and 30bps pools hold for other token pairs. While a full analysis of every possible pair is beyond the scope of this study, it would be valuable to empirically verify if less dominant pools are, in general, more nonstationary and less stable. We hypothesize that this is the case because the lower volume may attract liquidity providers who deploy more static or passive positions. This makes the liquidity dynamics more susceptible to price movements, which are themselves highly nonstationary.

From a similar perspective, it is worth investigating if the basis stability, particularly the alignment with the Legendre basis, holds for a wider array of dominant pools, such as those involving memecoins, stablecoin-stablecoin pairs, or the BTC-ETH pair. We anticipate that the stylized facts established in this study will remain consistent for high-volume pairs. More generally, extending the analysis to various token pairs and blockchain networks beyond Ethereum and Arbitrum would provide valuable insights into the generalizability of our findings and the potential impact of network-specific factors on liquidity dynamics.

From a modelling perspective, our work can be extended by exploring alternative orthogonal polynomial bases over $[-1,1]$, such as the weighted Jacobi polynomials. These polynomials are orthonormal according to the weighted measure $w(x) = (1-x)^\alpha (1+x)^\beta$, where $\alpha, \beta > -1$. The standard Legendre basis is a special case where $\alpha=\beta=0$. This offers additional flexibility, as the weights can be chosen to emphasize different parts of the liquidity curve. For example, setting $\alpha=\beta>0$ places more weight toward the center of the curve, while $\alpha\neq\beta$ introduces an asymmetry, favoring one side of the price more than another. The optimal choice of basis ultimately depends on the practitioner's specific needs and the economic interpretation desired.

Moreover, the Jacobi polynomials are orthogonal eigenfunctions for the Jacobi differential operator:
$$
\mathcal{L} = (1-x^2) \frac{d^2}{dx^2} + (\beta - \alpha - \left\{\alpha + \beta + 2)x\right\}\frac{d}{dx}
$$
associated with the eigenvalues $\lambda_n = -n(n+\alpha+\beta+1)$ for $n$ being nonnegative integers. It is thus tempting to use a parametric dynamical model for the evolution of the (log) liquidity profile in the form of a stochastic partial differential equation (SPDE) driven by a finite-dimensional Brownian motion
\[
dy_t(x) = \left\{\mathcal{L}y_t(x) - V(t,x) y_t(x) + h_t(x)\right\} dt + \sum_{r=1}^R \sigma_t^m(x) dW^r_t \quad \mbox{ for } x \in [-1,1], 
\]
where the differential operator $\mathcal{L}$ depicts the local diffusion-convection behavior of the liquidity profile $y_t$, $V \geq 0$ corresponds to the percentage cancellation rate of liquidity at site $x$, and $h \geq 0$ captures the incoming flow rate of liquidity at $x$, and $R$ is the number of Brownian drivers. The Jacobi polynomials, being the eigenfunctions for the Jacobi operator $\mathcal{L}$, help decompose the above SPDE into an infinite-dimensional system of SDEs in eigenmodes. As in the empirical analysis done in Section \ref{sec:Legendre} using the Legendre bases and the fact that the eigenvalues $\lambda_n$ diverge to negative infinity as $n\to\infty$, one would expect that the SDEs with the first few (approximately 5 to 7) eigenmodes will dominantly capture the dynamics of the (log) liquidity profile. The SPDE with the Jacobi differential operator thus provides a succinct and financially intuitive, easy-to-interpret model for the liquidity surface. We leave the details of this line of research for a future follow-up study.


Forecasting is a natural and crucial extension of this work, especially with the use of a fixed, interpretable basis. We propose a suitable forecasting recipe using a \textit{vector autoregression (VAR) model} with a \textit{multivariate GARCH process}. Consider the factor coefficient vector $\bm{\beta}_t = [\beta_{t,1}, \ldots, \beta_{t,K}]^\top$. An appropriate model is
\begin{equation}\label{eq:var-garch}
    \bm{\beta}_{t+1} = \bm{a} + \bm{A} \bm{\beta}_t + \bm{\epsilon}_{t+1}, \qquad \bm{\epsilon}_{t+1} | \mathcal{F}_t \sim t_\nu(0, \bm{H}_{t+1}), 
\end{equation}
Here, $\bm{a}$ is an intercept vector, $\bm{A}$ contains the autoregression parameters, and $\bm{H}_t$ follows a multivariate GARCH specification. This approach provides a full distributional forecast of the liquidity surface. This model would allow for:
\begin{itemize}
    \item Full simulation of future paths for computing loss quantiles and conditional variance.
    \item Simulation of economic shocks by adjusting the current state or model parameters.
    \item Fast computation of look-ahead error metrics suitable for rolling forecast error calculations.
\end{itemize}
In this forecasting framework, the number of factors $K$ would need to be determined beforehand, either by selecting a fixed, conservative number (e.g., $K=5$) or based on a specific reconstruction metric. A thorough forecasting study using this methodology would be a valuable contribution to the field.

\section*{Acknowledgements}
S. N. T. is grateful for the financial support from the National Science and Technology Council of Taiwan under grant 114-2115-M-007-012-MY3, "Mathematical Foundation of Automated Market Makers."

%% file: Appendix.tex
\appendix

\section{Basis Discussion}\label{sec:basis-discussion-appendix}
Values at $x \notin \{x_1, \ldots, x_M\}$ are not allowable in \eqref{eq:y-truncated} without regularity: infinitely many $L^2$ functions interpolate the grid exactly but differ for $x \in (x_{m-1}, x_m)$. Thus, off-grid evaluation requires some smoothness or structure on $u_k$ (or on $y_t$) and a numerical scheme.  

One approach is to choose an orthogonal basis $\{\psi_k\}_{k \geq 1}$ of $L^2([-1,1])$, for example Legendre, Fourier, or wavelets. The optimal $L^2$ projection coefficients are $\tilde{\beta}_{t,k} = \int_{-1}^1 y_t(x) \psi_k(x)dx$.  Since this is intractable, as no off-grid evaluations are allowed, one typically approximates $\tilde{\beta}_{t,k} \approx \beta_{t,k}$ via a quadrature rule to determine weights $(w_m)_{m=1}^M$, e.g.~trapezoidal or Simpson's rule (both work in the equally spaced case).  In particular,
\begin{equation}\label{eq:beta-general-basis}
    \beta_{t,k} = \sum_{m=1}^M w_m y_t(x_m) \psi_k(x_m)
\end{equation}
or for a fixed $K$ in vector form
\begin{equation}
    \bm{\beta}_t = (\Psi^\top W \Psi)^{-1} \Psi^\top W \mathbf{y}_t,
\end{equation}
where $\Psi$ is $M \times K$ with $m,k$ entry $\psi_k(x_m)$, and $W = \text{diag}(w_1, \ldots, w_M)$. This can be connected with the general FPCA approach, which assumes $y_t\in L^2([-1,1])$ is stationary with mean $m$ and covariance kernel $C: [-1,1]^2 \rightarrow \mathbb{R}$, with $C$ symmetric, positive definite, square-integrable, and continuous. Analogous to PCA, Mercer's Theorem yields $C(x,x')=\sum_{k=1}^\infty\lambda_k\phi_k(x)\phi_k(x')$ with orthonormal $\{\phi_k\}$.  The Karhunen-Lo\`eve expansion subsequently provides $y_t(x)=m(x)+\sum_{k=1}^\infty\eta_{t,k}\phi_k(x)$, which is the $L^2$-optimal basis when truncated to $K$ terms.
 With equally spaced grids $\{x_m\}_{m=1}^M$, the eigenpairs of the sample covariance $\hat{\Sigma}$ converge (in operator norm) to those of the covariance operator as $M, T \rightarrow \infty$ (relative to the quadrature rule). 
 
In summary, a fixed basis $\{\psi_k\}_{k \geq 1} \subset L^2([-1,1])$ allows full evaluation of \eqref{eq:y-truncated} at off-grid points. Although this is suboptimal compared to the Mercer basis, it still provides a consistent reconstruction in the limit, assuming sufficient regularity.  Additionally, if the two bases align closely, one has the added benefit of a fixed interpretation if the $\psi_k(x)$ provide it.

\paragraph{\emph{Comparing Subspace Alignment}} 
In practice, a data-dependent basis is not fixed over time due to sampling or estimation error, or nonstationarity.  One may also want to use an alternative data-dependent basis, such as diagonalizing the \emph{long-run} covariance (found to perform well in forecasting tasks \cite{hissinaga2024interest}, but suboptimal for low-rank representation of $y_t(x)$ in \eqref{eq:pca-optimal}), or a fixed basis $\{\psi_k\}_{k \geq 1}$ for a consistent interpretation, like Nelson-Siegel for yield curves, but still want it to be ``close'' to the empirical bases.  Thus, it is useful to have a method to compare across bases.

Note that in any $K$-rank reconstruction \eqref{eq:y-truncated}, all $K \times K$ orthogonal rotations of factors yield the same reconstruction. Thus, when two bases reconstruct the same $y$, it means they span the same subspace.  The degree to which these subspaces do not align can be compared through \emph{projection distance}.  Specifically, let $U_1, U_2 \in \mathbb{R}^{M \times K}$ be two rank-$K$ matrices with orthogonal columns, each representing a particular basis, for example $U_K$ from the PCA decomposition above, or $\Psi$ from the fixed orthonormal basis.  A standard metric to compare subspaces is the projection distance:
\begin{equation}\label{eq:projection-distance}
    d_S(U_1, U_2) = \frac{1}{2}\| P_1 - P_2\|^2_F, \qquad P_j = U_j U_j^\top \in \mathbb{R}^{M \times M},
\end{equation}
where $\|\cdot\|_F^2$ is the Frobenius norm.  It follows that $0 \leq d_S(U_1, U_2) \leq \min(K, M-K)$.  The subspaces are the same if and only if $d_S=0$.

Thus, subspace drift across windows $\cW_1$ and $\cW_2$ can be measured by comparing $d_S(U_{\cW_1}, U_{\cW_2})$.  As mentioned, one can also assess alignment with a fixed $\Phi$.  For example, if $d_S(U_{\cW}, \Phi)$ is generally small over all windows $\cW$, then the basis $\Phi$ is relatively stable for the data-generating process.  

One benchmark stricter than the maximal distance $\min(K, M-K)$ is one according to a random subspace: if $U \in \mathbb{R}^{M \times K}$ corresponds to a random $K$-rank projection (Haar-distributed random matrix), then $E[d_S(U_j, U)] = K(1-\frac{K}{M})$ (see Section 3.3 in Meckes (2019) \cite{meckes2019random}).  Values below this indicate more meaningful alignment than compared to a random subspace.

\section{Liquidity Surface Details and Rank-Standardized Coordinates} \label{sec:lp-surface-appendix}
To set notation, denote the \emph{raw liquidity surface} as $\{\mathcal{L}_t(i)\}_{t \ge 0, i \in s\mathbb{Z}}$, where $\mathcal{L}_t(i)$ is the raw step function corresponding to liquidity at time $t$ and tick $i$.  Let $s$ be the pool's tick-spacing (e.g.\ $s=10$ for the 0.05\% tier) and $\tilde i_t$ the current price tick at time $t$.  To simplify the modelling approach, we rank-standardize ticks, focusing on locations where liquidity changes across ticks.

Define the jump (event) ticks
\[
E_t := \{i\in s\mathbb{Z} : \mathcal{L}_t(i)\neq \mathcal{L}_t(i-s)\},
\]
i.e., the boundary ticks of the piecewise-constant $\mathcal{L}_t$ (flat interiors excluded). Order jumps around $\tilde i_t$ using a \emph{rank} index $r\in\mathbb{Z}$:
\[
\cdots < i_t^{(-2)} < i_t^{(-1)} < i_t^{(0)} := \tilde i_t < i_t^{(1)} < i_t^{(2)} < \cdots,
\]
where, for $r>0$ (resp.\ $r<0$), $i_t^{(r)}$ is the $|r|$-th jump strictly to the right (resp.\ left) of $\tilde i_t$.

Fix an odd $M\ge3$ and let $j^* := (M+1)/2$. We retain the anchor and the $(M-1)/2$ nearest jumps on each side. For $j=1,\dots,M$ define the orderings $i_{t,j}:= i_t^{(j-j^*)}$
and the rank-standardized coordinates
\[
x_j := \frac{2(j-j^*)}{M-1}\in[-1,1],\qquad j=1,\dots,M.
\]
Note that $x_{j^*}=0$ corresponds to the current price. By construction, $\{x_j\}_{j=1}^M$ is a fixed, equally spaced grid. Assume $M$ is small enough that at each $t$ there are at least $(M-1)/2$ jumps on each side of $\tilde i_t$. The corresponding liquidity surface to be modelled over $x \in [-1, 1]$ is $L_t(x_j) := \mathcal{L}_t(i_{t,j})$.

This setup simplifies modelling but has two caveats: (i) to forecast future liquidity in absolute ticks, one must jointly model price (as in standard LOB settings), and (ii) it tracks jump locations (where the piecewise curve moves). In practice, jumps typically occur at each step near the current price, in which case $x_j$ is simply an affine transform of the raw tick $j$.

\section{Additional Tables and Figures for ETH5 and ETH30 Per Window (PCA)}\label{sec:tab-fig-eth5-eth30-appendix}

\begin{figure}[!ht]
    \centering
    \begin{tabular}{ccc}
        $\qquad$Window 1 (2023)& Window 2 (2023)& Window 3 (2023--2024)\\
        \includegraphics[width=0.31\linewidth]{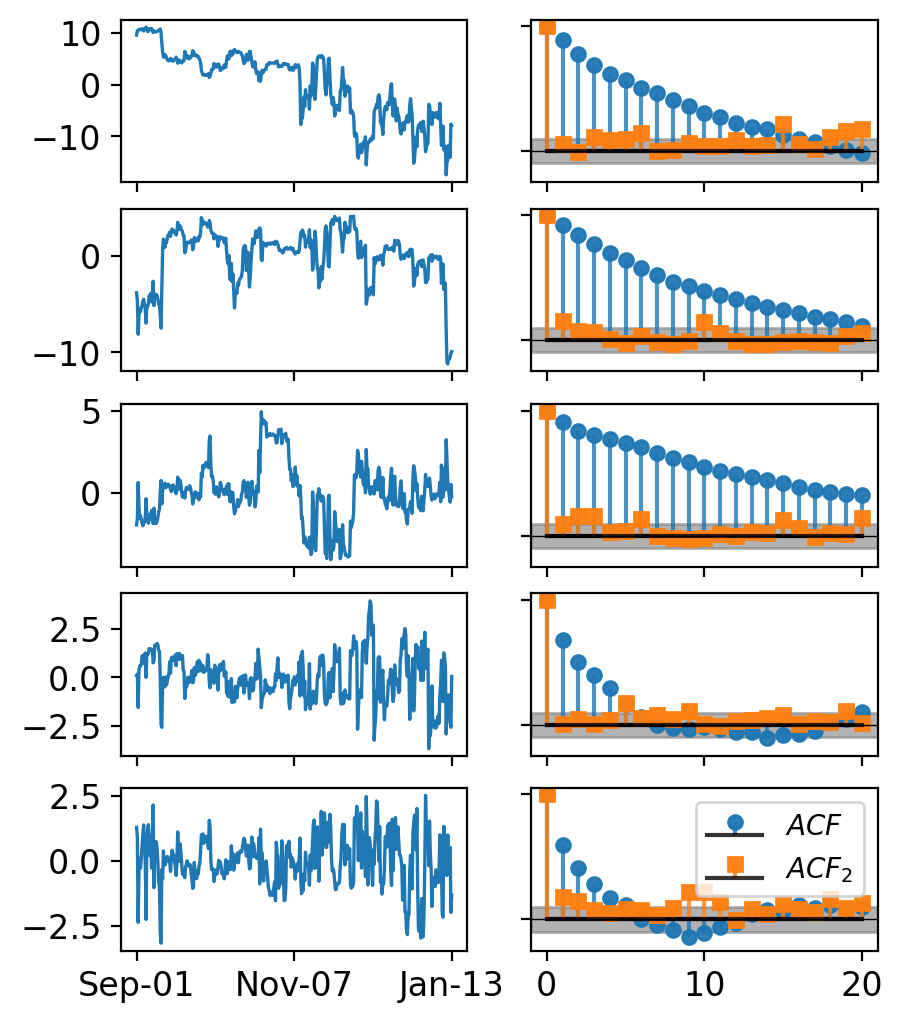} & $\quad$\includegraphics[width=0.31\linewidth]{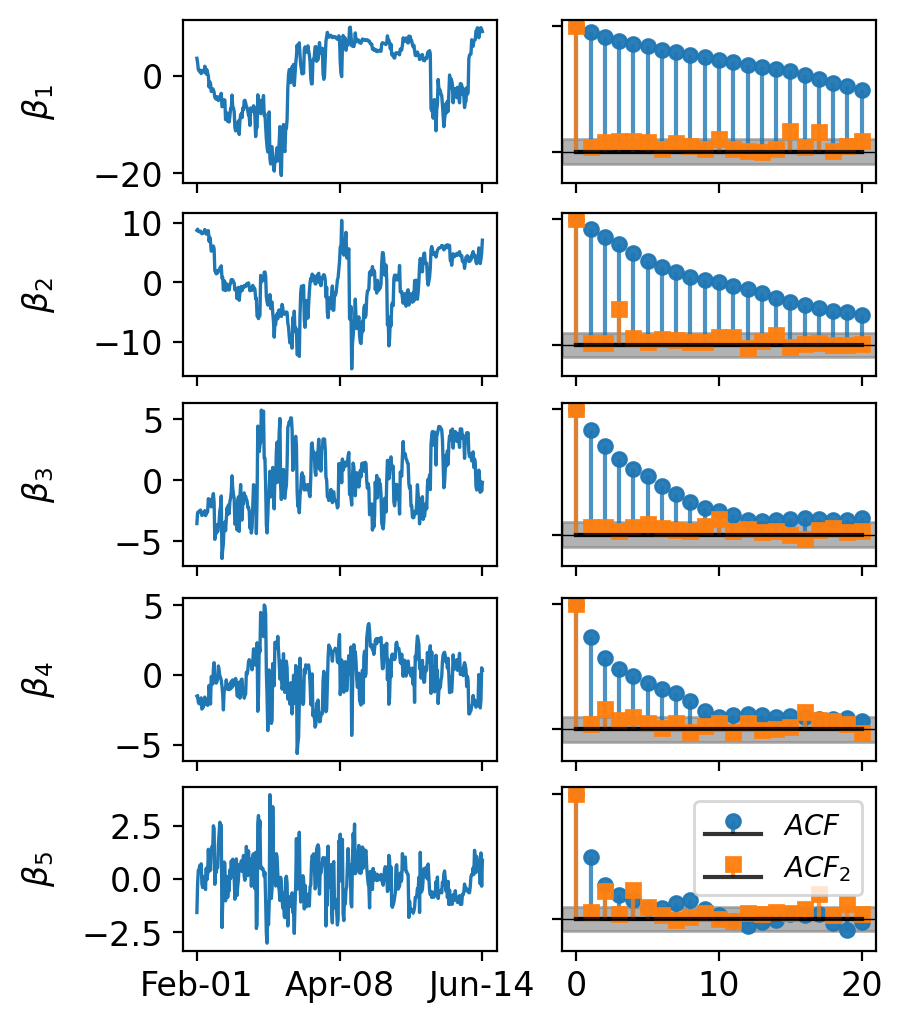} & $\quad$\includegraphics[width=0.31\linewidth]{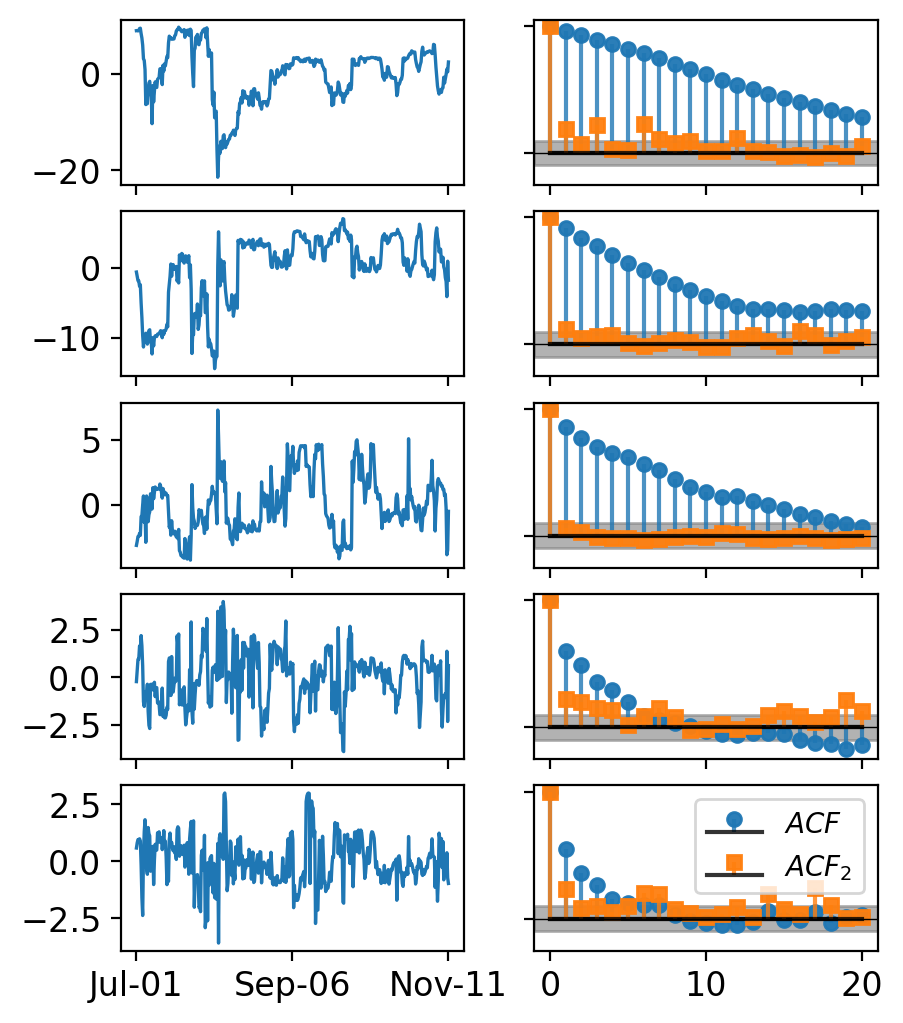}
    \end{tabular}
    \caption{Time series (left) and autocorrelation functions of $\beta_{t,k}$ (right, blue) for ETH5. $ACF_2$ (orange) refers to the autocorrelation of the squared AR(1) residuals $\hat{e}^2_t$. The gray bar indicates the 95\% pointwise cutoff for significance from zero autocorrelation. Each ACF equals 1 at lag 0.}
    \label{fig:time-series-ETH5}
\end{figure}

\begin{figure}[!ht]
    \centering
    \begin{tabular}{ccc}
        $\qquad$Window 1 & Window 2 & Window 3 \\
        \includegraphics[width=0.31\linewidth]{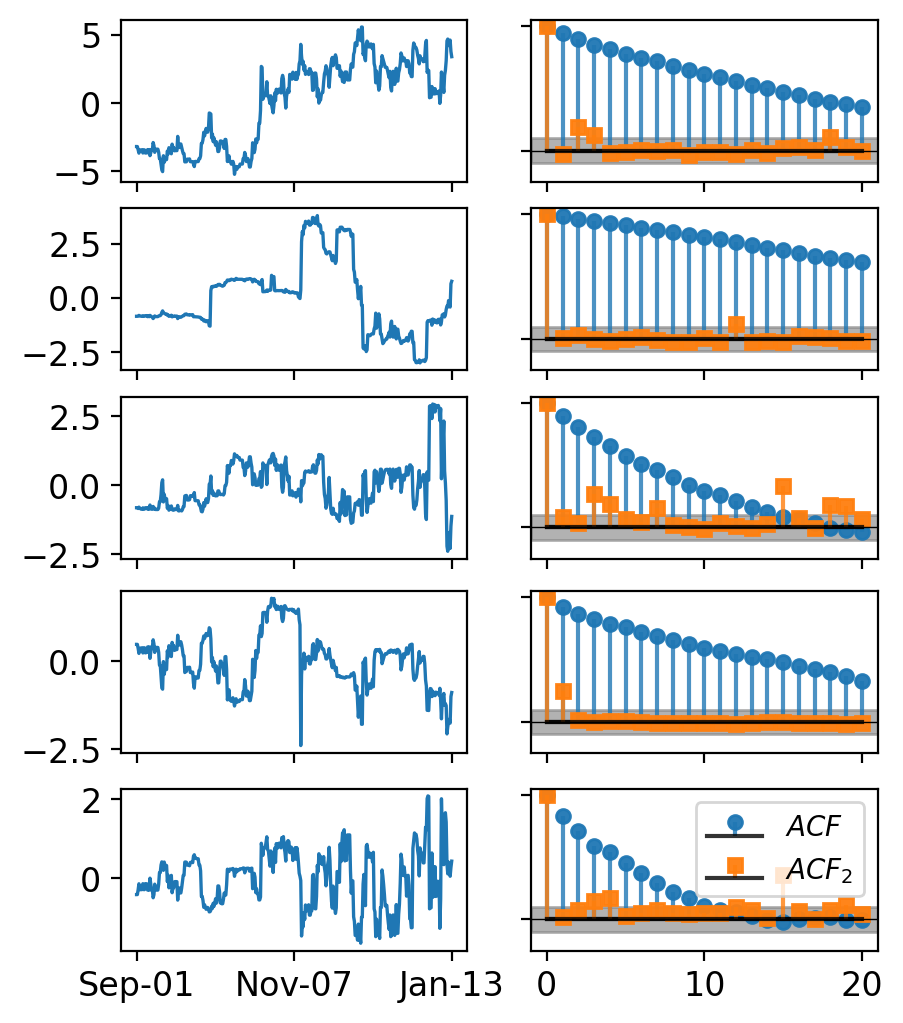} & $\quad$\includegraphics[width=0.31\linewidth]{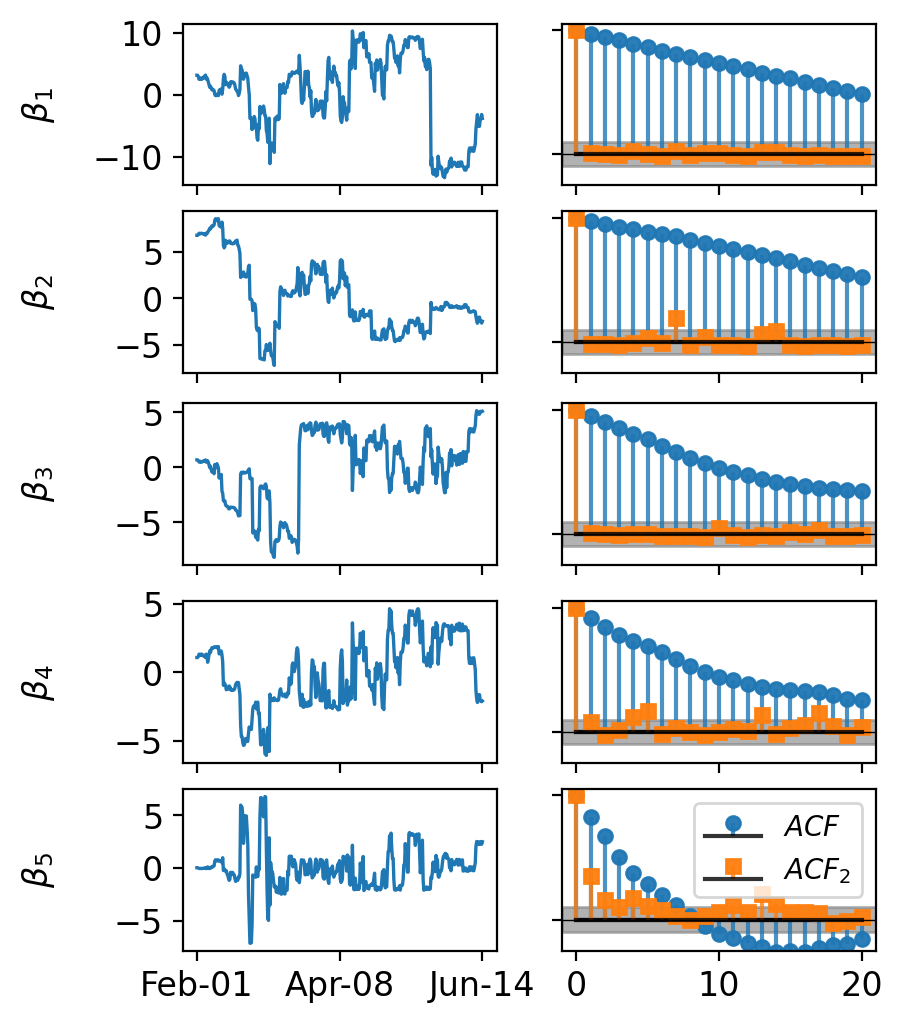} & $\quad$\includegraphics[width=0.31\linewidth]{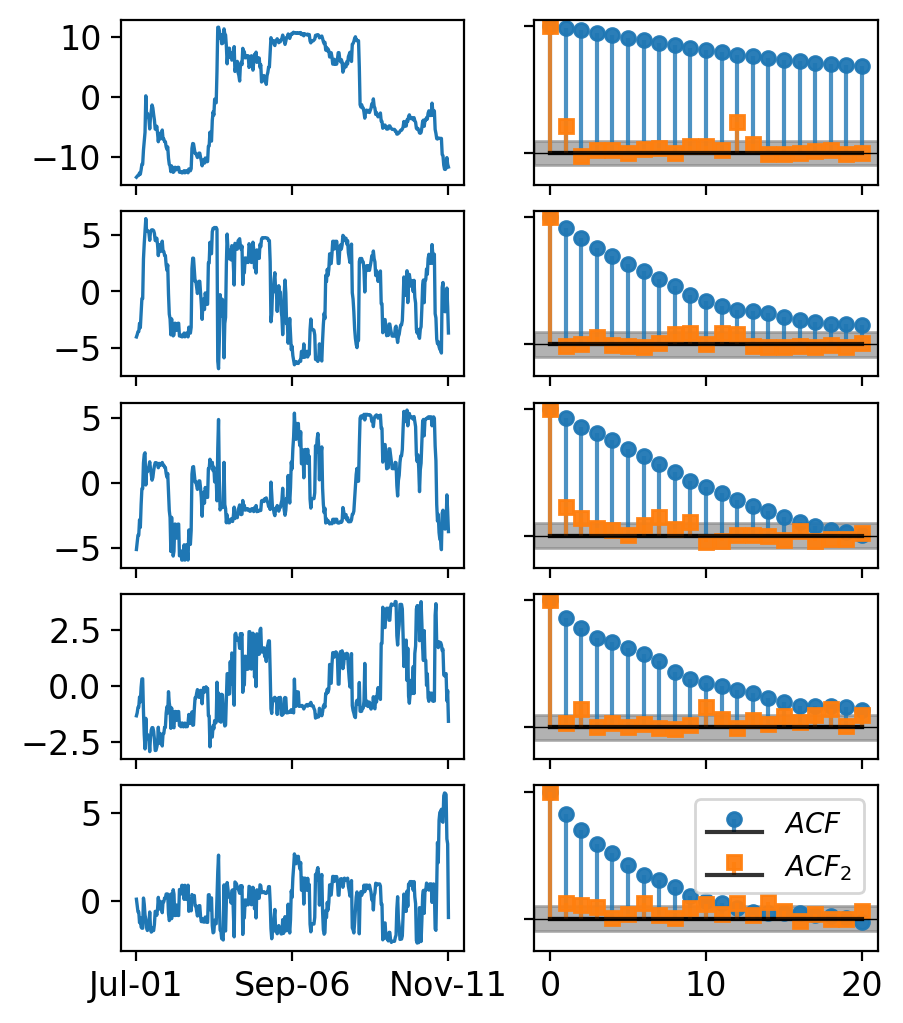}
    \end{tabular}
    \caption{Time series (left) and autocorrelation functions of $\beta_{t,k}$ (right, blue) for ETH30. $ACF_2$ (orange) refers to the autocorrelation of the squared AR(1) residuals $\hat{e}^2_t$. The gray bar indicates the 95\% pointwise cutoff for significance from zero autocorrelation. Each ACF equals 1 at lag 0.}
    \label{fig:time-series-ETH30}
\end{figure}

Figures \ref{fig:time-series-ETH5} (ETH5) and \ref{fig:time-series-ETH30} (ETH30) and Tables \ref{tab:ETH5-time-series} (ETH5) and \ref{tab:eth30-time-series} are displayed here.  The figures show time series, and ACF and ACF of squared residuals analogous to that of Figure \ref{fig:time-series-ARB5} for ARB5.  The tables show various summary statistics, analogous to Table \ref{tab:arb5-time-series} for ARB5.  A summary of takeaways is in Section \ref{sec:time-series-analysis-PCA}.

\begin{table}
\begin{center}\footnotesize
\begin{tabular}{l|rrrr|rrrr|rrrr}
\toprule
\multirow{2}{*}{$k$} & \multicolumn{4}{c}{Window 1} & \multicolumn{4}{c}{Window 2} & \multicolumn{4}{c}{Window 3} \\
\cmidrule{2-5} \cmidrule{6-9} \cmidrule{10-13}
 & PVE$_k$ & ADF $p$ & $\text{sd}(\beta_{t,k})$ & $\tau$ & PVE$_k$ & ADF $p$ & $\text{sd}(\beta_{t,k})$ & $\tau$ & PVE$_k$ & ADF $p$ & $\text{sd}(\beta_{t,k})$ & $\tau$ \\
\midrule
1 & 0.708 & 0.333 & 6.676 & 8.360 & 0.563 & 0.262 & 7.157 & 22.285 & 0.474 & 0.012 & 5.762 & 27.096 \\
2 & 0.125 & 0.064 & 2.808 & 20.500 & 0.247 & 0.003 & 4.743 & 14.349 & 0.342 & 0.193 & 4.892 & 12.638 \\
3 & 0.050 & 0.023 & 1.778 & 10.718 & 0.066 & 0.000 & 2.457 & 5.627 & 0.073 & 0.024 & 2.264 & 6.540 \\
4 & 0.021 & 0.000 & 1.163 & 2.560 & 0.032 & 0.000 & 1.709 & 3.256 & 0.027 & 0.000 & 1.367 & 1.973 \\
5 & 0.015 & 0.000 & 0.982 & 1.912 & 0.012 & 0.000 & 1.049 & 1.428 & 0.015 & 0.000 & 1.019 & 1.658 \\
\bottomrule
\end{tabular}
\end{center}
\caption{Summary statistics for principal component time series of ETH5 for each window. $\tau$ is mean reversion time $-1/\log|\phi|$ from AR(1) fit.}
\label{tab:ETH5-time-series}
\end{table}

\begin{table}
\begin{center}\footnotesize
\begin{tabular}{l|rrrr|rrrr|rrrr}
\toprule
\multirow{2}{*}{$k$} & \multicolumn{4}{c}{Window 1} & \multicolumn{4}{c}{Window 2} & \multicolumn{4}{c}{Window 3} \\
\cmidrule{2-5} \cmidrule{6-9} \cmidrule{10-13}
 & PVE$_k$ & ADF $p$ & $\text{sd}(\beta_{t,k})$ & $\tau$ & PVE$_k$ & ADF $p$ & $\text{sd}(\beta_{t,k})$ & $\tau$ & PVE$_k$ & ADF $p$ & $\text{sd}(\beta_{t,k})$ & $\tau$ \\
\midrule
1 & 0.582 & 0.385 & 3.068 & 17.055 & 0.430 & 0.130 & 6.396 & 29.026 & 0.591 & 0.358 & 7.878 & 123.980 \\
2 & 0.166 & 0.431 & 1.636 & 64.624 & 0.150 & 0.259 & 3.783 & 36.975 & 0.120 & 0.001 & 3.553 & 12.187 \\
3 & 0.049 & 0.000 & 0.890 & 9.819 & 0.110 & 0.100 & 3.239 & 20.934 & 0.087 & 0.009 & 3.030 & 15.286 \\
4 & 0.038 & 0.049 & 0.787 & 11.956 & 0.064 & 0.013 & 2.469 & 13.638 & 0.025 & 0.014 & 1.623 & 7.152 \\
5 & 0.030 & 0.000 & 0.702 & 5.480 & 0.040 & 0.000 & 1.948 & 5.334 & 0.020 & 0.000 & 1.461 & 5.360 \\
\bottomrule
\end{tabular}
\caption{Summary statistics for principal component time series of ETH30 for each window. $\tau$ is mean reversion time $-1/\log|\phi|$ from AR(1) fit.}
\label{tab:eth30-time-series}
\end{center}
\end{table}

\begin{figure}[!ht]
    \centering
    \includegraphics[width=0.95\textwidth]{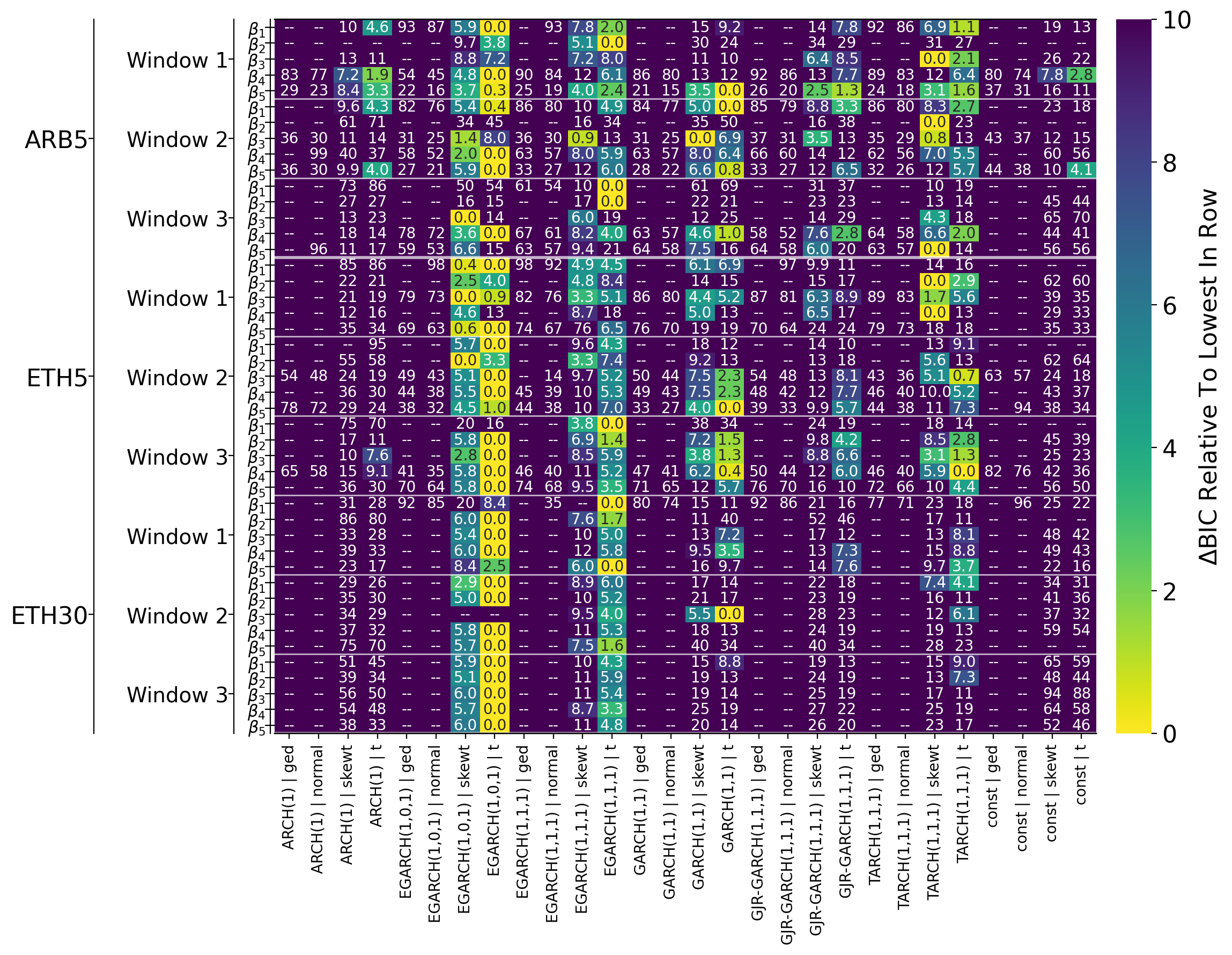}
    \caption{PCA Scores: Heatmap of BIC scores according to heteroskedasticity and distribution assumptions.  Reported is the $\Delta$ BIC relative to the lowest (best) for that series.  All use an AR(1) mean.  Any value of ``--" had $\Delta BIC > 10$.}
    \label{fig:PCA-heteroskedastic-BIC-heatmap}
\end{figure}

\begin{figure}[!ht]
    \centering
    \includegraphics[width=0.95\textwidth]{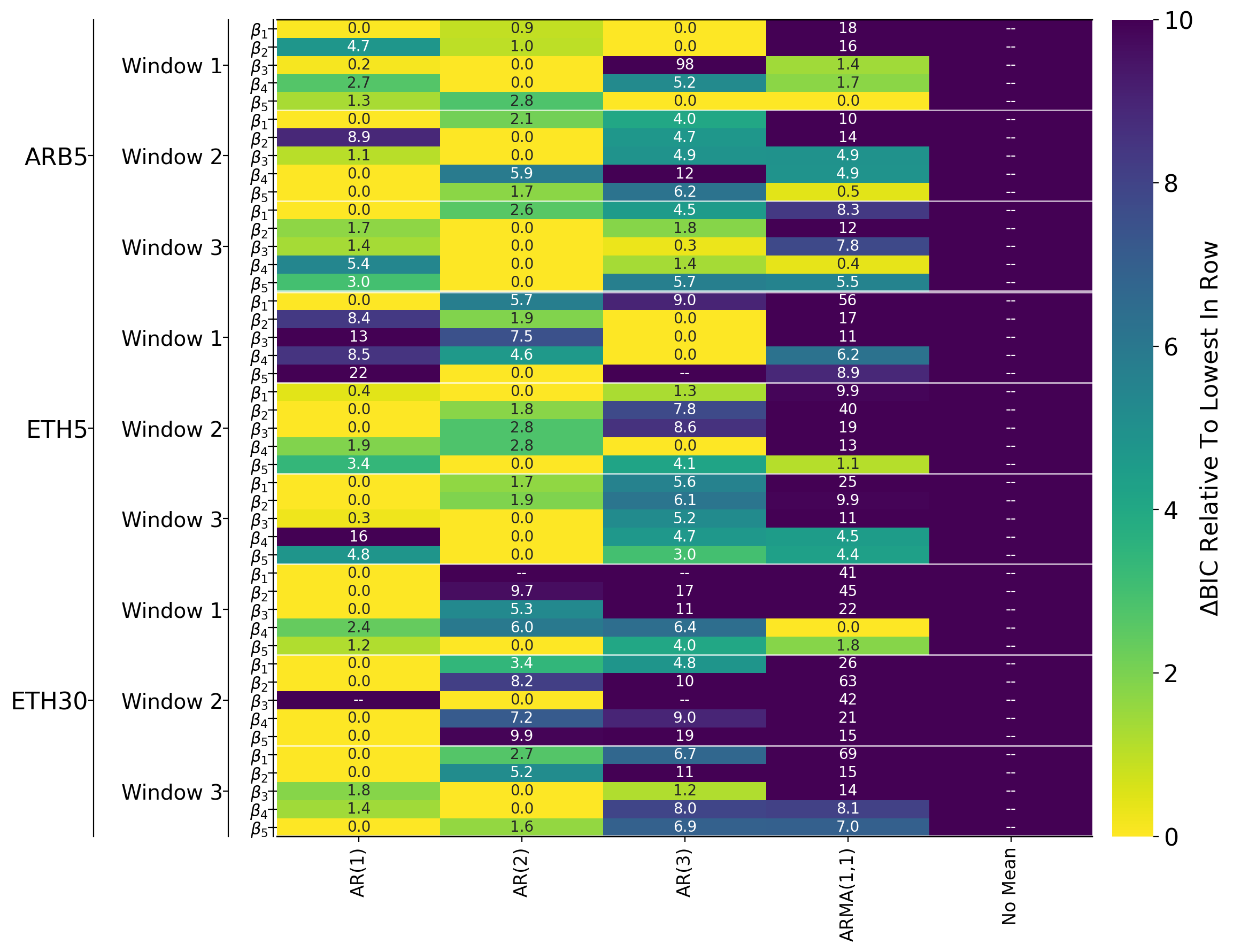}
    \caption{PCA Scores: Heatmap of BIC scores according to mean choice.  Reported is the $\Delta$ BIC relative to the lowest (best) for that series.  All use an EGARCH(1,0,1) volatility and $t$-distributed errors.  Any value of ``--" had $\Delta BIC > 10$.}
    \label{fig:PCA-mean-BIC-heatmap}
\end{figure}

\section{Additional Figures for Legendre Decomposition}\label{sec:tab-fig-legendre-appendix}

\begin{figure}[!ht]
    \centering
    \includegraphics[width=1\linewidth]{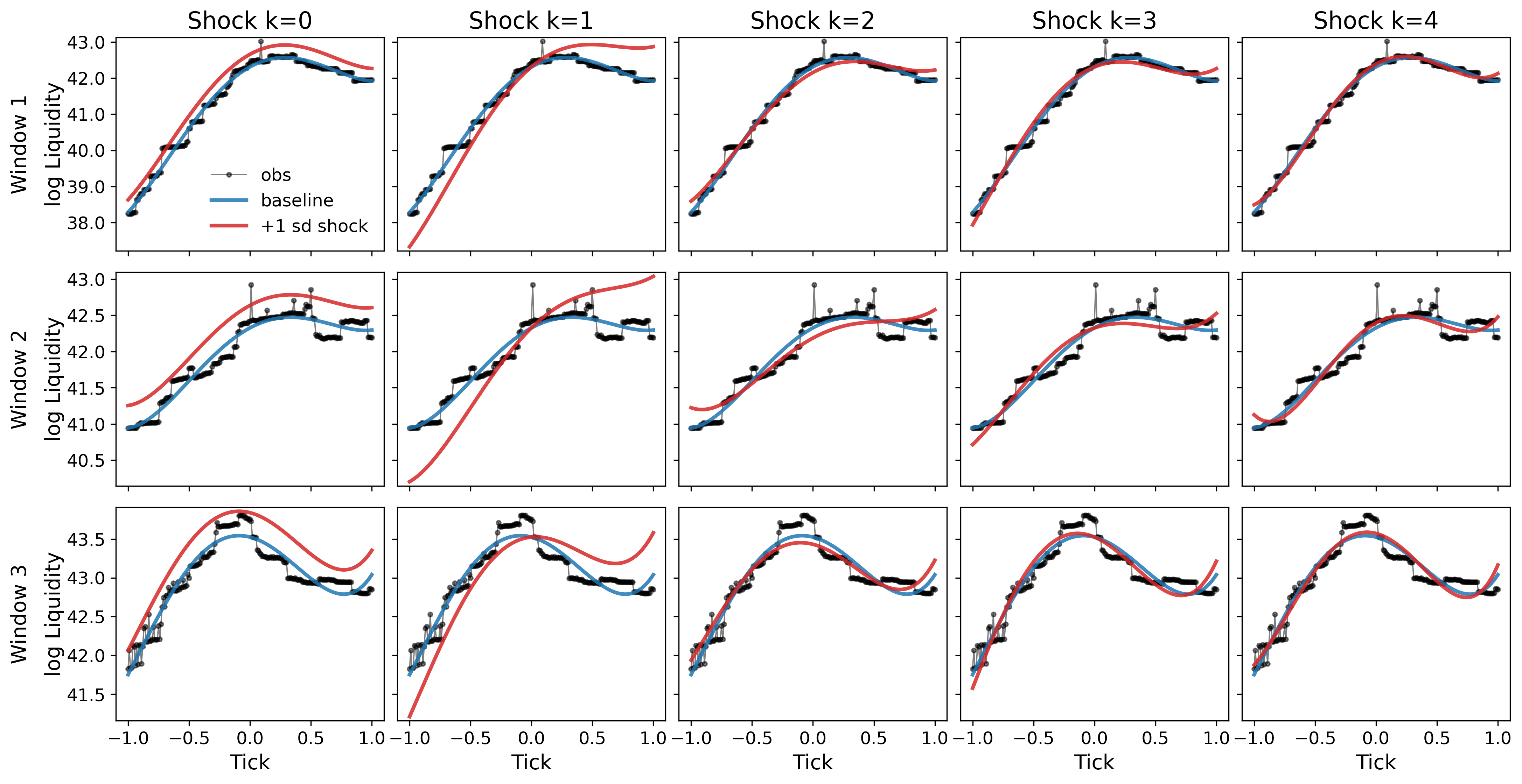}
    \caption{ARB5: effect onto the full cross-section $y_t(x)$ of a 1 standard deviation shock on the $k$th component, $k=1, \ldots, 5$, over the three windows considered.}
    \label{fig:arb5-shock}
\end{figure}

Figure \ref{fig:arb5-shock} gives the analogous version of \ref{fig:eth5-shock} for ARB5.  The shocks have the same effect according to the interpreted basis functions, but are generally less impactful because of lower variance.

\begin{figure}[!ht]
    \centering
    \includegraphics[width=0.95\textwidth]{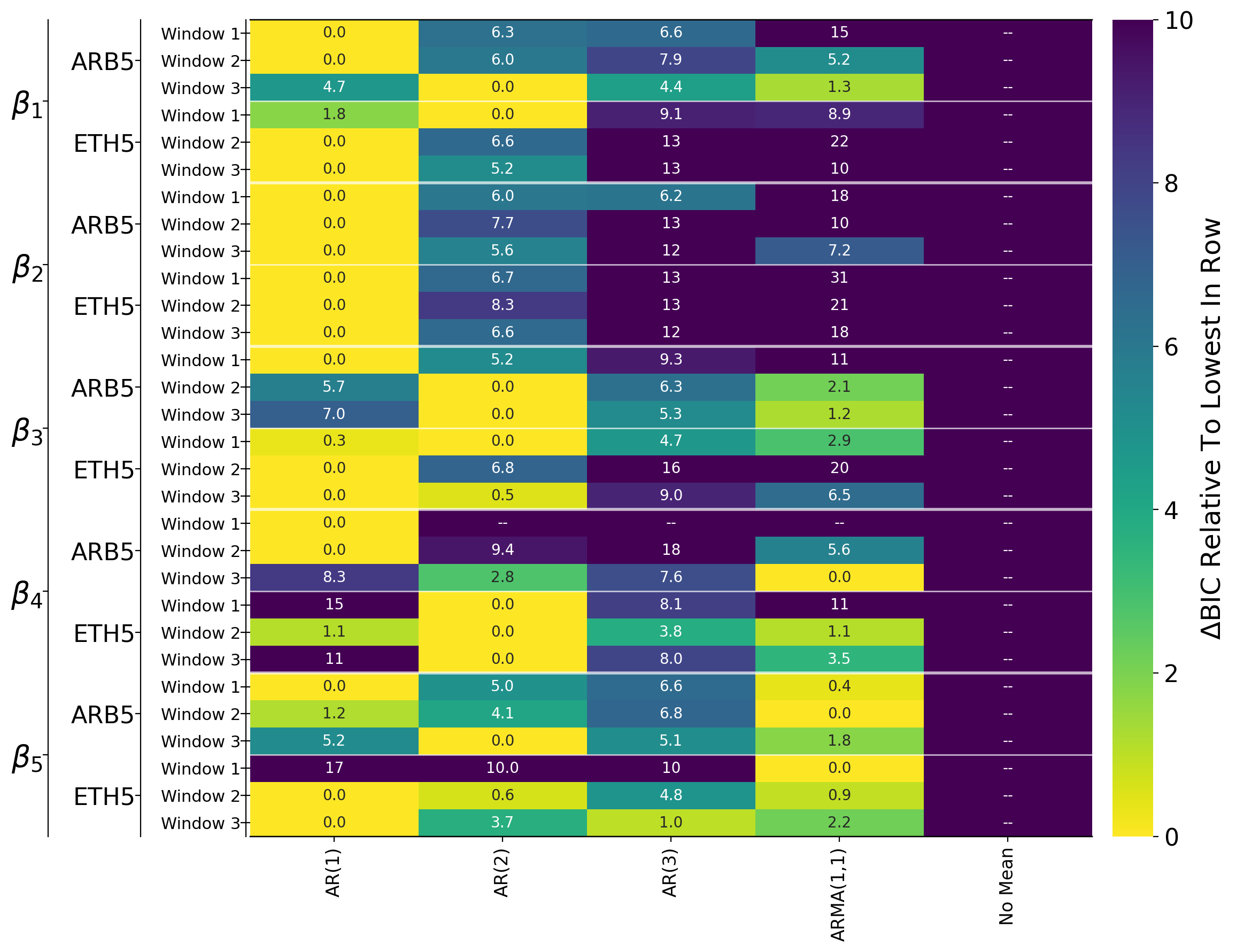}
    \caption{Legendre Coefficients: Heatmap of BIC scores according to mean choice.  Reported is the $\Delta$ BIC relative to the lowest (best) for that series.  All use an EGARCH(1,0,1) volatility and $t$-distributed errors.  Any value of ``--" has $\Delta BIC > 10$.}
    \label{fig:legendre-mean-BIC-heatmap}
\end{figure}

Figure \ref{fig:legendre-mean-BIC-heatmap} gives a BIC heatmap according to mean choice with an EGARCH(1,0,1) volatility and $t$-distributed innovations.  The takeaways are summarized at the end of Section \ref{sec:legendre-time-series}.

\section{Time Series Model Details}\label{sec:appendix-time-series-model-details}

For each series $(\beta_{t,k})_{t=1}^T$ and model $\mathcal M$,
\[
\beta_{t,k}=\mu_t(\mathcal F_{t-1};\mathcal M)+\varepsilon_{t,k},\qquad
\varepsilon_{t,k}=\sigma_{t,k}(\mathcal F_{t-1};\mathcal M)z_{t,k},\qquad
z_{t,k}\overset{\text{iid}}{\sim}F(\mathcal M),\ \mathbb{E} z_{t,k}=0,\ \text{var}(z_{t,k})=1.
\]
An AR(1) without intercept satisfies
\[
\mu_t=\phi\beta_{t-1,k}.
\]
Define standardized shocks $e_{t,k}:= \varepsilon_{t,k}/\sigma_{t,k}$.

\subsection{Volatility Models}
All constraints below are the usual ones ensuring positivity/stationarity. EGARCH does not need positivity constraints because of the log link.

\paragraph{ARCH(1).}
\[
\sigma_{t,k}^2=\omega+\alpha\varepsilon_{t-1,k}^2,\qquad \omega>0,\ \alpha\ge 0.
\]

\paragraph{GARCH(1,1).}
\[
\sigma_{t,k}^2=\omega+\alpha\varepsilon_{t-1,k}^2+\beta\sigma_{t-1,k}^2,\qquad
\omega>0,\ \alpha,\beta\ge 0,\ \alpha+\beta<1\ \text{(sufficient for covariance stationarity)}.
\]

\paragraph{GJR--GARCH(1,1,1) (variance recursion with threshold).}
\[
\sigma_{t,k}^2=\omega+\alpha\varepsilon_{t-1,k}^2+\gamma\varepsilon_{t-1,k}^2\mathbf 1\{\varepsilon_{t-1,k}<0\}
+\beta\sigma_{t-1,k}^2,
\]
with $\omega>0,\ \alpha\ge 0,\ \alpha+\gamma\ge 0,\ \beta\ge 0$.

\paragraph{TARCH(1,1,1) (standard-deviation recursion).}
\[
\sigma_{t,k}=\omega+\alpha|\varepsilon^{\vphantom{2}}_{t-1,k}|+\gamma|\varepsilon^{\vphantom{2}}_{t-1,k}|\mathbf 1\{\varepsilon_{t-1,k}<0\}
+\beta\sigma_{t-1,k},
\]
with $\omega>0,\ \alpha\ge 0,\ \alpha+\gamma\ge 0,\ \beta\ge 0$.

\paragraph{EGARCH(1,0,1) (log-variance, symmetric).}
\[
\log\sigma_{t,k}^2=\omega+\alpha\big(|e_{t-1,k}|-\kappa\big)+\beta\log\sigma_{t-1,k}^2,
\]
where $\kappa=\mathbb E|Z|$ under the reference $Z\sim F$ (e.g., $\kappa=\sqrt{2/\pi}$ for normal).

\paragraph{EGARCH(1,1,1) (log-variance with leverage).}
\[
\log\sigma_{t,k}^2=\omega+\alpha\big(|e_{t-1,k}|-\kappa\big)+\gamma e_{t-1,k}+\beta\log\sigma_{t-1,k}^2.
\]

For GARCH$(p,q)$ (and modifications depending on $(p,o,q)$): increasing $p$ adds additional lagged \emph{shock} terms (squared $\varepsilon$, absolute $|\varepsilon|$, or $|e|$ depending on the model), increasing $o$ adds additional \emph{asymmetry/threshold} lags, and increasing $q$ adds additional \emph{volatility} lags (variance, standard deviation, or log-variance accordingly).

\subsection*{Innovation distributions}
All are standardized to unit variance in estimation:
\begin{itemize}
\item Normal;
\item Student-$t_\nu$ ($\nu>2$);
\item Skew-$t$ (Hansen) with shape $(\nu,\lambda)$;
\item GED with shape $\kappa_{\mathrm{GED}}$, here restricted to $\kappa_{\mathrm{GED}}>2$ to allow platykurtic checks.
\end{itemize}

\section{Alternative Dataset Sizes and Subsampling}\label{sec:alt-preprocessing}
This section motivates some choices used in the paper.    In particular, we consider:
\begin{itemize}
    \item $T=200$ and $T=800$,
    \item an (approximate) subsampling frequency of 4 hours and 16 hours, and
    \item using $M=101, 51, 11$ $x_m \in [-1, 1]$.
\end{itemize}
The main text uses $T=400$, a subsampling frequency of 8 hours, and $M=201$.

\begin{figure}
  \centering
  {%
    \setlength{\tabcolsep}{0pt}
    \begin{tabular}{@{}>{\centering\arraybackslash}m{5.5em} m{\dimexpr\linewidth-3.5em\relax}@{}}
      $M=201$ & \includegraphics[width=\linewidth]{images/pve_windowed_1x3_m4_ntr400_ntrim0_x201_cenmean_lrFalse_lag10_robnone_smooNone.png}\\
      $M=101$ & \includegraphics[width=\linewidth]{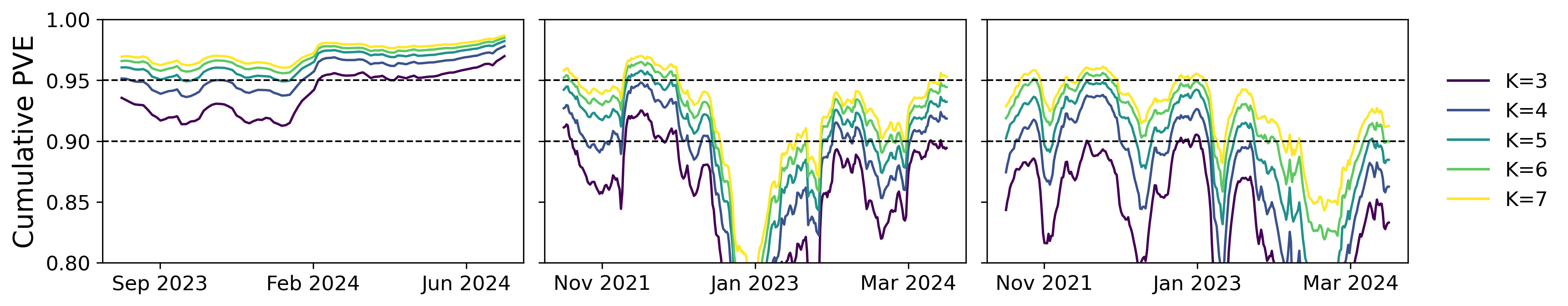}\\
      $M=51$  & \includegraphics[width=\linewidth]{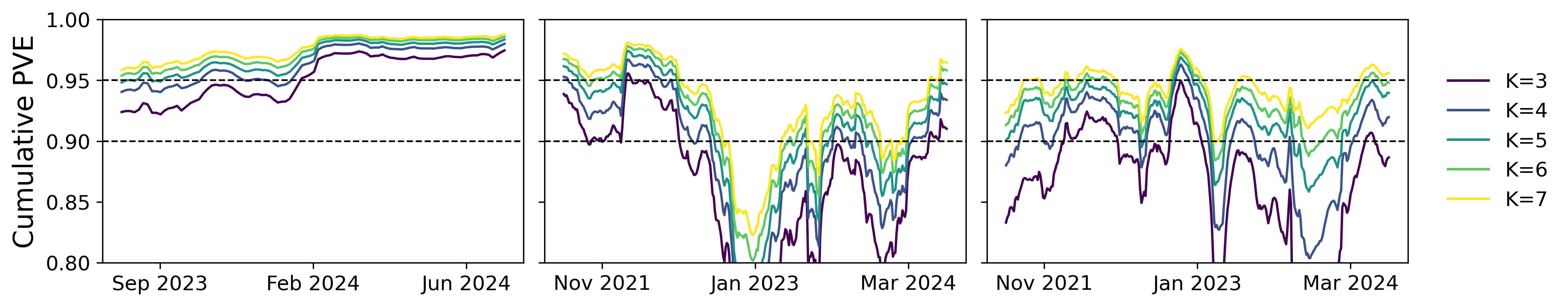}\\
      $M=11$  & \includegraphics[width=\linewidth]{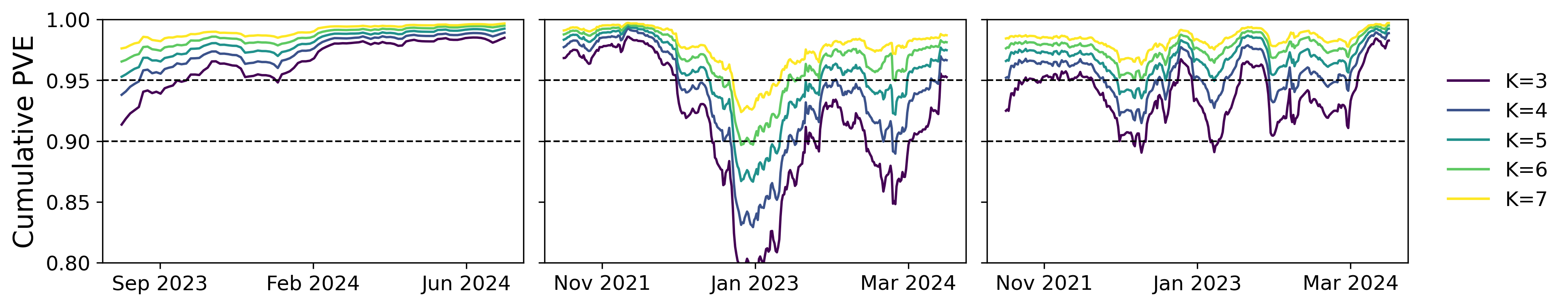}\\
    \end{tabular}%
  }
  \caption{Left to right: ARB5, ETH5, ETH30.  Proportion of variance explained over various choices of $M$ (the number of $x$'s to use centered around the current price).  Analogous to bottom panel of Figure \ref{fig:eigenvalue-CPVE}.}
  \label{fig:prop-var-over-M}
\end{figure}

Figure \ref{fig:prop-var-over-M} illustrates PVE plots analogous to the bottom panel of Figure \ref{fig:eigenvalue-CPVE} for various choices of $M$, with $M=201$ repeated for reference.  The figure yields two major takeaways that hold across all datasets:
\begin{itemize}
    \item PVE is generally higher in the lower $M$ case, particularly $M=11$.  
    \item The change in PVE becomes much less discernible, where the $M=101$ vs.~$M=201$ case is nearly indistinguishable
\end{itemize}

This is makes sense theoretically, since in the discrete data regime the PCA tasks can be thought of as projecting the $M$-dimensional subspace onto a $K$ dimensional one.  For example, $M=11$ means to summarize these 11 variables in terms of $K$ linear combinations.  When the data has some amount of correlation across $x$'s (as we expect it should), there should be little loss of information by truncating to, say, $K=7$.  

The second takeaway can be illustrated by example through comparing $M=201$ and $M=101$.  In particular, the mild difference between the two is effectively saying that the $201-101=100$ prices further away from the center $101$ prices are contributing little information.  Again this is expected since the liquidity curves flatten out and revert to the center price.  Even during periods of high volatility there is little difference between $M=201$ and $M=101$.  Thus, $M=201$ is sufficient for this task.

\begin{figure}
    \centering
\begin{tabular}{cc}
    \includegraphics[width=0.40\linewidth]{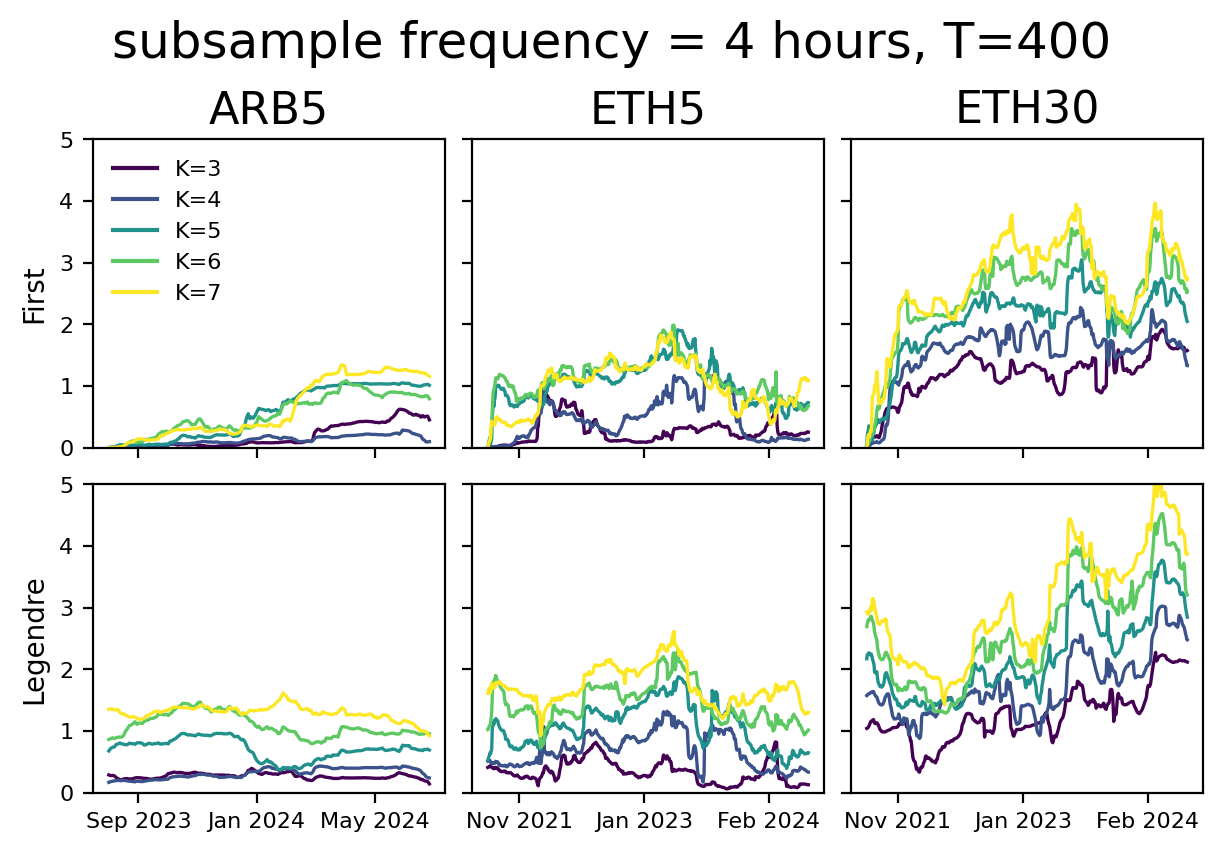} & \includegraphics[width=0.40\linewidth]{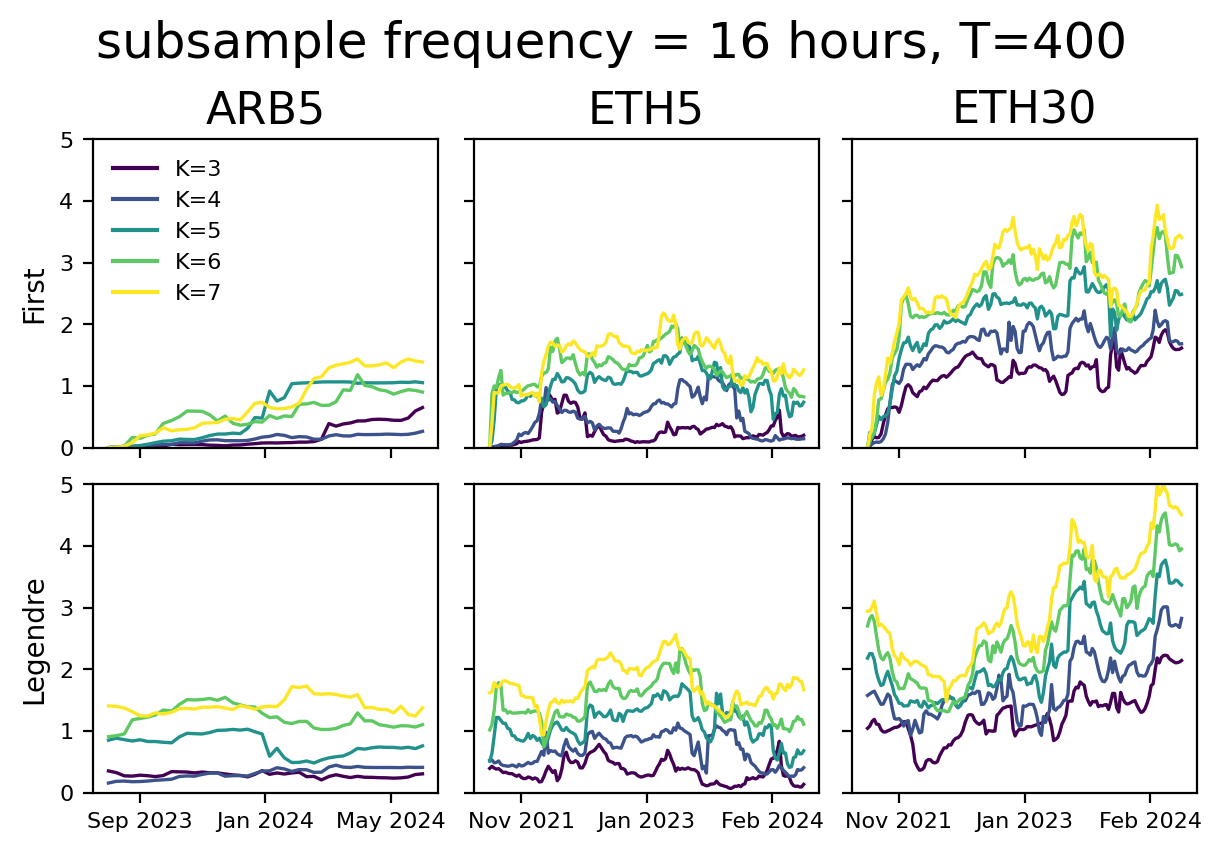} \\
    \includegraphics[width=0.40\linewidth]{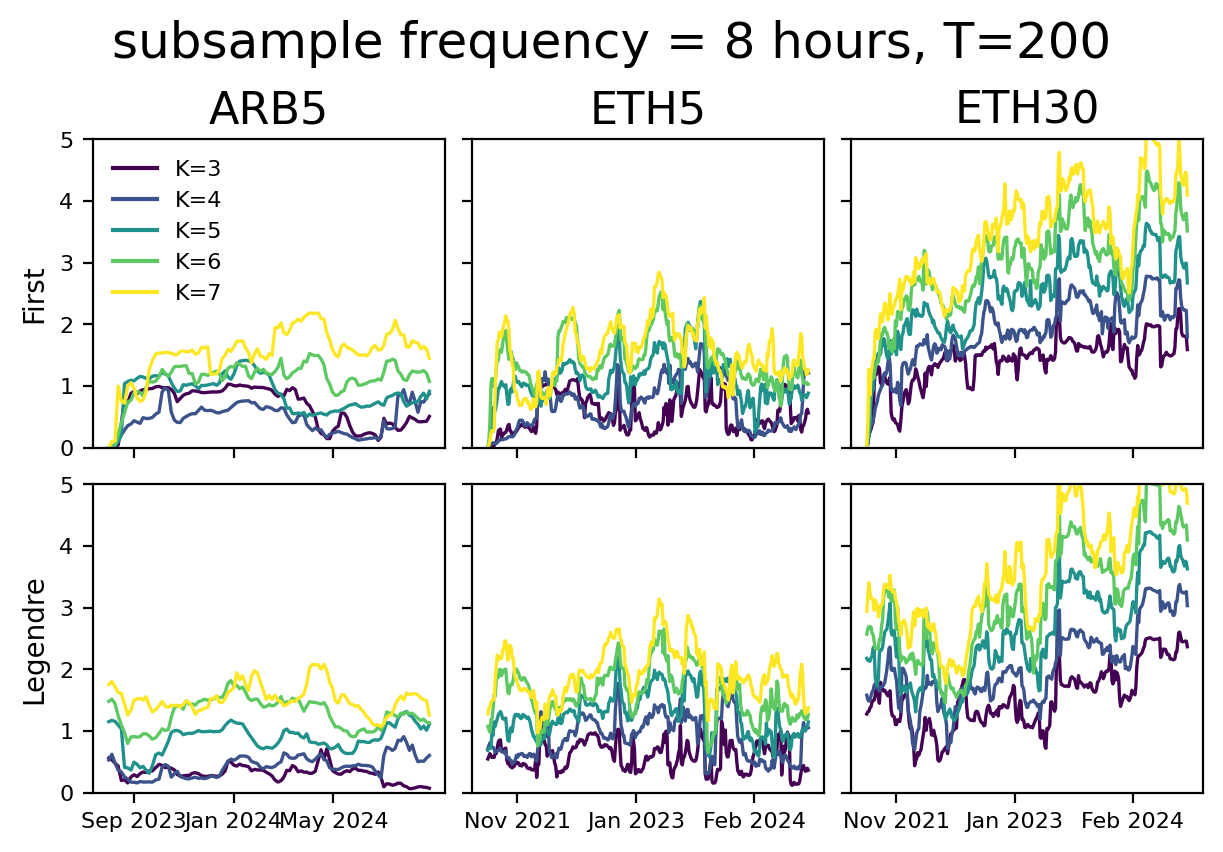} & \includegraphics[width=0.40\linewidth]{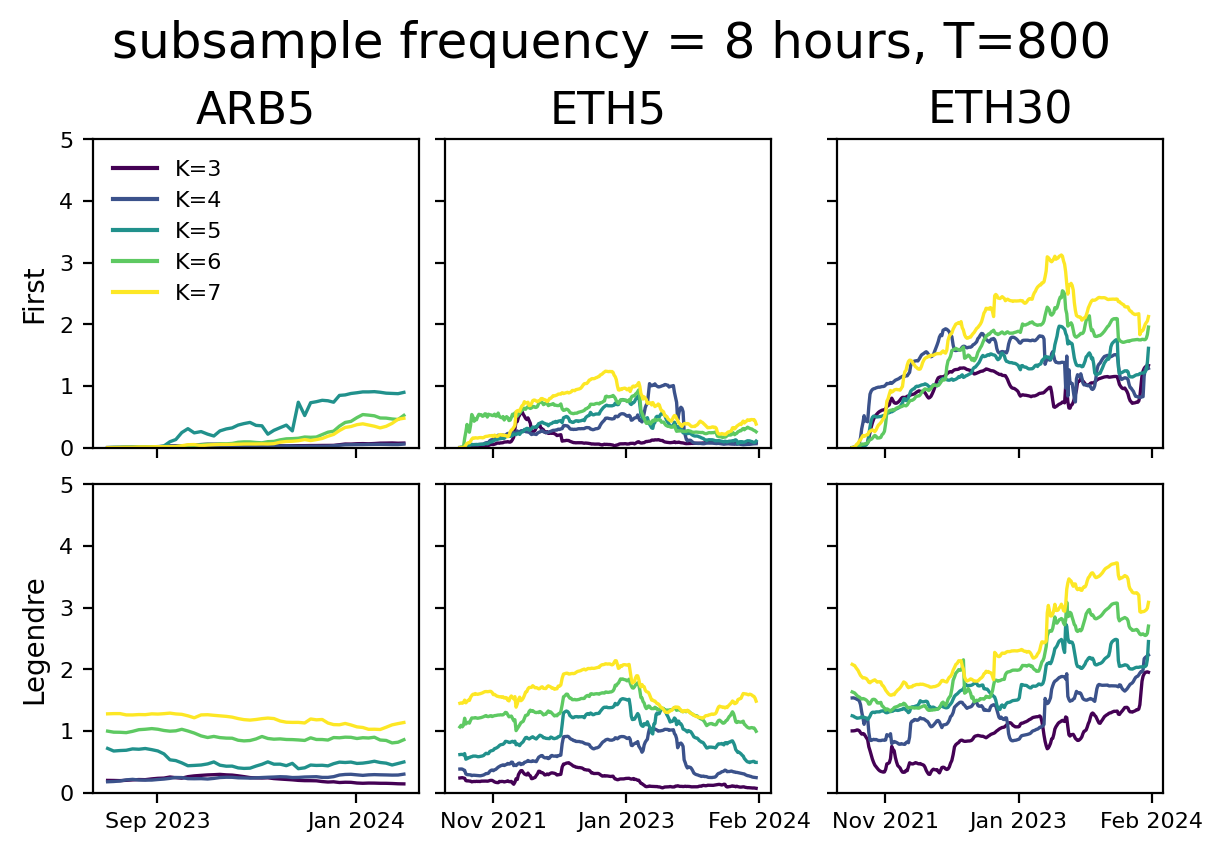} 
\end{tabular}
    \caption{Subspace distances analogous to Figure \ref{fig:eigenmode-stability} over various choices of subsampling frequency and rolling window size $T$.}
    \label{fig:alt-param-subspace-distances}
\end{figure}

Figure \ref{fig:alt-param-subspace-distances} considers various choices of $T$ and subsampling intervals, producing subspace distance plots analogous to Figure \ref{fig:eigenmode-stability}.  The overall takeaways are the same as in the main body ($T=400$, subsample frequency $8$ hours), notably: (1) a period of drift that reverts for ETH5, (2) ETH30 drifting off, and (3) the Legendre basis producing similar results to the PCA basis.  There are no clear differences between subsampling frequencies.  For sample sizes, $T=200$ is much rougher and $T=800$ smoother, appearing to partially subdue some of the drift present in ETH30.